\documentclass[a4paper,11pt]{article}
\pdfoutput=1 
\usepackage{jcappub}
\usepackage{amsmath}
\usepackage{latexsym}
\usepackage{xspace}
\usepackage{color}
\usepackage{hyperref} 
\usepackage{bm}
\usepackage{tabularx}
\usepackage{multirow}
\usepackage{amssymb}
\usepackage[table]{xcolor}
\usepackage{graphicx}
\usepackage{enumitem}
\usepackage{geometry}
\usepackage{xspace}
\usepackage{pdflscape}
\usepackage{dsfont}
\usepackage{tensor}
\usepackage{comment}
\usepackage{soul}
\usepackage{braket}
\usepackage{comment}
\usepackage[utf8]{inputenc}
\usepackage{float}
\usepackage{slashed}

\usepackage{mathtools}

\usepackage{relsize}

\allowdisplaybreaks 

\makeatletter
\gdef\@fpheader{}
\g@addto@macro\bfseries{\boldmath}
\makeatother

\newcommand{\dd}{\mathrm{d}}
\newcommand{\calN}{\mathcal{N}}

\newcommand{\Eq}[1]{Eq.~(\ref{#1})}
\newcommand{\Eqs}[1]{Eqs.~(\ref{#1})}
\newcommand{\Fig}[1]{Fig.~{\ref{#1}}}
\newcommand{\Figs}[1]{Figs.~{\ref{#1}}}
\newcommand{\Refa}[1]{Ref.~{\cite{#1}}}
\newcommand{\Refs}[1]{Refs.~{\cite{#1}}}
\newcommand{\Sec}[1]{Sec.~\ref{#1}}

\newcommand{\App}[1]{Appendix~\ref{#1}}

\newcommand{\ie}{\textsl{i.e.}\xspace}

\newcommand{\eg}{\textsl{e.g.}\xspace}

\newcommand{\etc}{\textsl{etc.}}

\let\oldsqrt\sqrt
\def\sqrt{\mathpalette\DHLhksqrt}
\def\DHLhksqrt#1#2{%
\setbox0=\hbox{$#1\oldsqrt{#2\,}$}\dimen0=\ht0
\advance\dimen0-0.2\ht0
\setbox2=\hbox{\vrule height\ht0 depth -\dimen0}%
{\box0\lower0.4pt\box2}}

\DeclareMathOperator{\erfc}{erfc}

\newcommand{\ee}{e}

\newcommand{\sss}[1]{{\scriptscriptstyle{#1}}}

\newcommand{\uPl}{\mathrm{Pl}}
\newcommand{\uin}{\mathrm{in}}

\newcommand{\uend}{\mathrm{end}}

\newcommand{\uc}{\mathrm{c}}

\newcommand{\usssPl}{\sss{\uPl}}

\newcommand{\calP}{\mathcal{P}}

\newcommand{\Mp}{M_\usssPl}

\newcommand{\efolds}{$e$-folds\xspace}

\newcommand{\beq}{\begin{equation}}
\newcommand{\eeq}{\end{equation}}
\newcommand{\bea}{\begin{equation}\begin{aligned}}
\newcommand{\eea}{\end{aligned}\end{equation}}

\newlength{\wsingfig}
\newlength{\wdblefig}
\newlength{\wquadfig}
\newlength{\wtriplefig}
\newcommand{\PfptV}[1]{P^{\mathrm{V}}_{\sss{\mathrm{FPT}} , #1}}
\newcommand{\Pfpt}[1]{P_{\sss{\mathrm{FPT}} , #1}}
\newcommand{\PfptlV}[1]{P^{\mathrm{V}}_{\sss{\mathrm{FPTL}} , #1}}

\DeclareMathOperator{\arccoth}

\newcommand{\uwell}{\mathrm{well}}

\newcommand{\rph}{r_{\mathrm{ph}}}
\newcommand{\Lfp}{\mathcal{L}_{\mathrm{FP}}}
\newcommand{\vecx}{\vec{x}}

\newcommand{\deflen}[2]{%
    \expandafter\newlength\csname #1\endcsname
    \expandafter\setlength\csname #1\endcsname{#2}%
}

\subheader{}

\title{Clustering of primordial black holes from quantum diffusion during inflation}

\author[]{Chiara Animali,}
\author[]{Vincent Vennin}

\affiliation[]{Laboratoire de Physique de l'\'Ecole Normale Sup\'erieure, ENS, CNRS, Universit\'e PSL, Sorbonne Universit\'e, Universit\'e Paris Cit\'e, F-75005 Paris, France}

\emailAdd{chiara.animali@ens.fr}
\emailAdd{vincent.vennin@ens.fr}

\date{today}

\begin{document}
\sloppy

\abstract{We study how large fluctuations are spatially correlated in the presence of quantum diffusion during inflation. This is done by computing real-space correlation functions in the stochastic-$\delta N$ formalism. We first derive an exact description of physical distances as measured by a local observer at the end of inflation, improving on previous works. Our approach is based on recursive algorithmic methods that consistently include volume-weighting effects. We then propose a ``large-volume'' approximation under which calculations can be done using first-passage time analysis only, and from which a new formula for the power spectrum in stochastic inflation is derived. We then study the full two-point statistics of the curvature perturbation. Due to the presence of exponential tails, we find that the joint distribution of large fluctuations is of the form $P(\zeta_{R_1}, \zeta_{R_2}) = F(R_1,R_2,r) P(\zeta_{R_1})P( \zeta_{R_2})$, where $\zeta_{R_1}$ and $\zeta_{R_2}$ denote the curvature perturbation coarse-grained at radii $R_1$ and $R_2$, around two spatial points distant by $r$. This implies that, on the tail, the reduced correlation function, defined as $P(\zeta_{R_1}>\zeta_\uc, \zeta_{R_2}>\zeta_{\uc})/[P(\zeta_{R_1}>\zeta_\uc) P(\zeta_{R_2}>\zeta_\uc)]-1$, is independent of the threshold value $\zeta_\uc$. This contrasts with Gaussian statistics where the same quantity strongly decays with $\zeta_\uc$, and shows the existence of a universal clustering profile for all structures forming in the exponential tails. Structures forming in the intermediate (\ie not yet exponential) tails may feature different, model-dependent behaviours.
} 



\maketitle

\section{Introduction}
\label{sec:Introduction}

If primordial black holes (PBHs) form in the early universe, they may account for a fraction or all of the dark matter, they may constitute progenitors for some of the black-hole mergers detected by gravitational-wave experiments, they may provide seeds for supermassive black holes in galactic nuclei and be involved in a number of other astrophysical and cosmological processes (for recent reviews see \eg \Refa{Escriva:2022duf, Carr:2023tpt, LISACosmologyWorkingGroup:2023njw}). The detailed role of PBHs in these processes does not only depend on their initial mass distribution, it is also largely affected by the way PBHs are distributed in space. 

For instance, if PBHs are organised in clusters, then the typical distance between two neighbouring black holes is smaller than what it would be if PBHs were uniformly distributed in space, which leads to mergers occurring earlier~\cite{Raidal:2017mfl, Ballesteros:2018swv, Young:2019gfc, Atal:2020igj}. If PBHs form in clusters, this could also alleviate~\cite{Calcino:2018mwh, Gorton:2022fyb, Petac:2022rio} or worsen~\cite{Bringmann:2018mxj, DeLuca:2022uvz} some of the microlensing constraints on their abundance. Clustered PBHs might also explain the early quasars observations~\cite{Dokuchaev:2004kr} and may play a role in early galaxy formation~\cite{Dokuchaev:2008hz}. For these reasons, when investigating PBH formation scenarios, a central question is to characterise their initial clustering~\cite{Chisholm:2005vm, Belotsky:2018wph}, which then determines their subsequent clustering evolution throughout cosmic history~\cite{DeLuca:2020jug}.

In practice, clustering measures how much the spatial locations of PBHs are correlated. At the level of two-point statistics, this can be described by the joint probability $p(M_1,\vec{x}_1;M_2,\vec{x}_2)$ of finding a PBH at location $\vec{x}_1$ with mass $M_1$ and another PBH at location $\vec{x}_2$ with mass $M_2$. When the distance $r=\vert\vec{x}_2 - \vec{x}_1\vert$ is smaller than the size of the black holes, or than a certain exclusion zone, this probability vanishes, since two different black holes cannot exist at the same location. Otherwise, if the positions of PBHs are statistically independent, it factorises as $p(M_1,\vec{x}_1;M_2,\vec{x}_2) = p_{M_1}(\vec{x}_1) p_{M_2}(\vec{x}_2)$, which corresponds to the so-called Poisson distribution. Here, $p_M(\vec{x})$ is the one-point probability of finding a PBH at location $\vec{x}$ with mass $M$. On homogeneous backgrounds, the distribution of black holes is statistically homogeneous, hence $p_M(\vec{x})$ does not depend on $\vec{x}$ and may simply be written as $p_M$. The deviation from the Poisson distribution can thus be characterised by the so-called reduced correlation, defined as~\cite{Kaiser:1984sw}
\bea
\label{eq:clustering:def}
\xi_{M_1,M_2}(r) = \frac{p(M_1,\vec{x}; M_2,\vec{x}+\vec{r}) }{p_{M_1} p_{M_2}} -1 = \frac{p\left(M_2,\vec{x}+\vec{r} \vert M_1,\vec{x} \right)}{ p_{M_2}}-1\,,
\eea
where in the second equality the joint probability has been rewritten as the product of $p_{M_1}$ times the conditional probability $p\left(M_2,\vec{x}+\vec{r} \vert M_1,\vec{x} \right)$.
For statistically homogeneous and isotropic spatial distributions of black holes, $p(M_1,\vec{x}; M_2,\vec{x}+\vec{r})$ only depends on $r=\vert \vec{r}\vert$, hence the notation $\xi_{M_1,M_2}(r)$. 

From this definition, $\xi>0$ indicates positive clustering, \ie the probability to find a PBH at a given location is increased by the presence of surrounding PBHs at distance $r$. This means that PBHs form in clusters. In contrast, $\xi<0$ signals negative clustering, hence the presence of PBHs at distance $r$ makes it less likely to form a PBH at the considered location. In the coincidence limit $r=0$, as mentioned above, $p(M_1,\vec{x};M_2,\vec{x})$ vanishes, hence $\xi_{M_1,M_2}(0)=-1$. Another way to interpret the reduced correlation is to consider the mean number of PBHs with mass $M_2$ contained in a volume $V$ centred on a PBH with mass $M_1$,
\bea
\left\langle N_{M_2}\right\rangle = \bar{n}_{M_2} V +  \bar{n}_{M_2} \int_V \dd^3\vec{r}\, \xi_{M_1,M_2}(r)\, ,
\eea 
where $\bar{n}_{M_2}$ is the mean number density of PBHs with mass $M_2$. In the above expression, the first term is the uniform contribution arising from Poisson fluctuations, while the second term quantifies the excess number of primordial black holes induced by clustering. The latter is of particular relevance when considering the formation of PBH binaries, its ratio with the former may thus be used as a clustering measure~\cite{Desjacques:2018wuu}.  

Usual PBH formation criteria require that the coarse-grained value of a given field (say the density contrast or the compaction function~\cite{Shibata:1999zs, Harada:2015yda, Musco:2018rwt}) is above a certain threshold. If the field under consideration has Gaussian statistics, $p(M_1,\vec{x}_1;M_2,\vec{x}_2) $ can be calculated in the Press-Schechter approach~\cite{Ali-Haimoud:2018dau, Desjacques:2018wuu} or the excursion-set formalism~\cite{Auclair:2024jwj} and one finds that, in the large-threshold limit, $\xi$ is suppressed by the ratio between the squared threshold and the field variance. Therefore, if PBHs are created from the collapse of large Gaussian fluctuations they are born unclustered, as also confirmed by excursion-set calculations~\cite{MoradinezhadDizgah:2019wjf}. 

The presence of non-Gaussianities is nonetheless expected to alter the above conclusion. In particular, primordial non-Gaussianities of the local-type are known to induce correlations between scales, hence this may result in correlating the horizon-size regions over which PBHs form, across larger distances~\cite{Suyama:2019cst, Young:2019gfc, DeLuca:2021hcf, Franciolini:2018vbk}. While this effect has been studied by means of perturbative parametrisations of non-Gaussianities (such as the famed $f_{\mathrm{NL}}$ parametrisation), it has also been shown that PBHs form in the heavy tail of distribution functions, where non-Gaussianities are not under perturbative control~\cite{Pattison:2017mbe, Ezquiaga:2019ftu, Figueroa:2020jkf, Kitajima:2021fpq, Ezquiaga:2022qpw, Cai:2022erk, Hooshangi:2023kss, Kawaguchi:2023mgk, Pi:2022ysn}.

The goal of this article is thus to compute PBH clustering in the presence of non-perturbative non-Gaussianities. To this end, we make use of the stochastic-$\delta N$ formalism~\cite{Enqvist:2008kt, Fujita:2013cna, Vennin:2015hra, Pattison:2017mbe}, which is a non-perturbative framework where the statistics of curvature perturbations is related to first-passage times of stochastic processes, and which we briefly review in \Sec{sec:StocdeltaN}. The calculation of one-point statistics in the stochastic-$\delta N$ formalism has been previously implemented in \Refs{Ando:2020fjm, Tada:2021zzj}, but clustering requires two-point distributions. This is why the main part of this work, \Sec{sec:CoarseGraining}, is devoted to the derivation of two-point statistics in stochastic inflation. When doing so, we clarify how distances measured by a local observer at the end of inflation can be extracted from the graph structure underlying the separate-universe picture on which stochastic inflation is implemented. This generalises the approach presented in \Refs{Ando:2020fjm, Tada:2021zzj}, and allows us to consistently include volume-weighting effects. We then propose a ``large-volume'' approximation, where physical distances of interest are large compared to the Hubble radius at the end of inflation, hence averages over a large number of Hubble patches can be performed. This allows us to compute all quantities of interest in terms of first-passage time distributions only. In \Sec{sec:applications}, we apply our formalism to a couple of toy models and discuss how non-Gaussian heavy tails deeply alter clustering compared to Gaussian statistics. We summarise  our results in \Sec{sec:Conclusion}, where we show that the main conclusions drawn in the toy models are in fact generic. We conclude with a few prospects for future directions that this work opens.

\section{Stochastic-$\delta N$ formalism: a concise review}
\label{sec:StocdeltaN}

At large scales, a gradient expansion can be performed in place of the usual expansion in the size of cosmological perturbations, which allows one to track these fluctuations non perturbatively. This is the so-called $\delta N$ formalism~\cite{Starobinsky:1982ee, Starobinsky:1985ibc, Sasaki:1995aw, Sasaki:1998ug, Lyth:2004gb, Lyth:2005fi}, according to which the comoving curvature perturbation $\zeta$ at super-Hubble scales is related to the amount of \efolds of inflationary expansion $N=\ln{(a)}$ (where $a$ is the scale factor of the Friedmann-Lema\^itre-Robertson-Walker (FLRW) metric) elapsed between an initial flat space-time slice and a final space-time slice of uniform energy density
\bea\label{eq:deltaN}
\zeta(t,\vecx)=N(t,\vecx)-\overline{N}(t) \equiv \delta N\,,
\eea
where $\overline{N}(t)$ is the unperturbed number of \efolds.
This result relies on the separate-universe approach~\cite{Salopek:1990jq, Sasaki:1995aw, Wands:2000dp, Lyth:2003im, Rigopoulos:2003ak, Lyth:2005fi, Artigas:2021zdk, Jackson:2023obv}, which describes super-Hubble sized regions as evolving independently, according to the same field equations as local homogeneous and isotropic FLRW universes. In this setup, $N(t,\vecx)$ in \Eq{eq:deltaN} is the amount of expansion realised in unperturbed, homogeneous universes, and $\zeta$ can thus be inferred from the knowledge of the duration of inflation in a family of such universes. 

The reason why different regions of space inflate for different amounts is because of vacuum quantum fluctuations, which are amplified by gravitational instability during inflation and stretched to large distances. The backreaction of these fluctuations onto the background dynamics can be described through the formalism of stochastic inflation~\cite{Starobinsky:1982ee, Starobinsky:1986fx}. The latter is an effective theory for the long-wavelength part of quantum fields during inflation, which are coarse grained above the Hubble radius. In practice, arranging the fields driving inflation $\phi_{i=1\cdots n}$ and their conjugate momenta $\pi_i=\dd \phi_i/\dd N$ into the field phase-space vector $\bm{\Phi}=(\phi_1, \pi_1,\dots \phi_n,\pi_n)$, the coarse graining is performed over patches of size $(\sigma H)^{-1}$, where $H=\dot{a}/a$ is the Hubble expansion rate, with dots denoting derivatives with respect to cosmic time $t$, and $\sigma\ll 1$ is a fixed parameter that sets the scale at which quantum fluctuations backreact onto the local FLRW geometry,
\bea
\label{eq:general:coarse:graining}
\bm{\Phi}_{\mathrm{cg}}(\vecx)= \left(\frac{a}{R}\right)^3 \int \dd   \vec{y} \,\bm{\Phi}(\vec{y})\, W\left(\frac{a |\vec{y}-\vecx|}{R}\right)\,.
\eea
In the above expression $R=(\sigma H)^{-1}$ and $W$ is a window function that only depends on the distance from $\vecx$ to preserve isotropy, and which satisfies $W(x) \simeq 0$ for $x \gg 1$ and $W(x) \simeq 1$ otherwise. It is normalised such that homogeneous fields are invariant under coarse graining, which implies $ 4 \pi\int_0^\infty x^2 W(x) \dd x=1$\,. In Fourier space such expression can be rewritten as
\bea
\label{eq:general:coarse:graining:Fourier}
\bm{\Phi}_{\mathrm{cg}}(\vecx)=\frac{1}{(2 \pi)^{3/2}}\int \dd \vec{k} \,\widetilde{W}\left(\frac{k}{k_\sigma }\right)\bm{{\Phi}}(\vec{k})e^{-i \vec{k} \cdot \vecx}\, ,
\eea
where $\widetilde{W}$ is the Fourier transform of the real-space window function $W$ and $k_\sigma=\sigma a H$. It is such that $\widetilde{W}\simeq 1$ when $k\ll k_\sigma $ and $\widetilde{W}\ll 1$ when $k\gg k_\sigma$ for the normalisation condition to be satisfied. 

As a result of this coarse-graining procedure, small-wavelength fluctuations act as a random noise on the dynamics of $\bm{\Phi}_{\mathrm{cg}}$ as they cross out the Hubble radius and join the coarse-grained sector. This leads to a stochastic classical theory for $\bm{\Phi}_{\mathrm{cg}}$, which follows a Langevin equation of the form~\cite{Finelli:2010sh}
 \bea\label{eq:Langevin}
 \frac{\dd \bm{\Phi}_{\mathrm{cg}}}{\dd N}=\bm{F}_{\mathrm{cl}}(\bm{\Phi}_{\mathrm{cg}})+\bm{\xi}\,,
 \eea
 where $\bm{F}_{\mathrm{cl}}$ encodes the classical background equation of motion and $\bm{\xi}$ is a white Gaussian noise, with vanishing mean and variance given by
 \bea
 \label{eq:noise:correlator}
 \left\langle \xi_i(\vecx,N) \xi_j(\vecx,N') \right\rangle=\frac{\dd \ln{(k_\sigma)}}{\dd N} \mathcal{P}_{\Phi_i, \Phi_j}[k_\sigma (N),N]\delta_{\mathrm{D}}(N-N')\, .
 \eea
Here, $\mathcal{P}_{\Phi_i, \Phi_j}[k_\sigma (N),N]$ is the reduced cross-power spectrum between the field variables $\Phi_i$ and $\Phi_j$, evaluated at the scale $k_\sigma$ and at time $N$.\footnote{This is valid if coarse graining is performed via a Heaviside window function in Fourier space, which makes the
noises white (\textit{i.e.} uncorrelated at different times).\label{footnote:white}}
From the Langevin equation~\eqref{eq:Langevin} one can derive the Fokker-Planck equation~\cite{risken1989fpe}
\bea
\label{eq:Fokker:Planck:not:adjoint}
\frac{\dd}{\dd N} P(\bm{\Phi},N| \bm{\Phi}_{\mathrm{in}}, N_{\mathrm{in}})=\Lfp(\bm{\Phi}) \cdot P(\bm{\Phi},N| \bm{\Phi}_{\mathrm{in}}, N_{\mathrm{in}})\,,
\eea
which drives the probability to find the system at position $\bm{\Phi}$ in field-phase space at time $N$, knowing that it was at the position $\bm{\Phi}_{\mathrm{in}}$ at time $N_{\mathrm{in}}$. Here, the subscript ``cg'' is dropped for convenience, and $\Lfp(\bm{\Phi})$ is a second-order differential operator in field-phase space (\ie it contains first and second derivatives with respect to the field coordinates $\Phi_i$) called the Fokker-Planck operator.

This sets the stage for the stochastic-$\delta N$ formalism~\cite{Enqvist:2008kt, Fujita:2013cna, Vennin:2015hra}: under the Langevin equation~\eqref{eq:Langevin} the duration of inflation becomes a stochastic variable denoted $\mathcal{N}$, which gives access to the statistics of the curvature perturbation $\zeta$ via \Eq{eq:deltaN}
\bea
\label{eq:zeta:deltaN}
\zeta_{\mathrm{cg}}= \delta N_{\mathrm{cg}}= \mathcal{N}-\overline{N}\, ,
\eea
where $\zeta_{\mathrm{cg}}$ is the curvature perturbation coarse grained at the $\sigma$-Hubble radius at the end of inflation.

Solving for $\mathcal{N}$ is called a first-passage time (FPT) problem, and one can show that the probability density function (PDF) of $\mathcal{N}$ starting from a given field configuration $\bm{\Phi}$, $\Pfpt{\bm{\Phi}}(\mathcal{N})$, satisfies the adjoint Fokker-Planck equation~\cite{Vennin:2015hra, Pattison:2017mbe}
\bea\label{eq:Fokker-Planck}
\frac{\dd \Pfpt{\bm{\Phi}}(\calN) }{\dd \mathcal{N}}=\Lfp^\dagger(\bm{\Phi})\cdot  \Pfpt{\bm{\Phi}}(\mathcal{N})\, .
\eea
Here $\Lfp^\dagger(\bm{\Phi})$ is the adjoint Fokker-Planck operator, \ie it is adjoint to the Fokker-Planck operator in the sense that $\int \dd \bm{\Phi} f_1(\bm{\Phi}) \left[\Lfp(\bm{\Phi})  \cdot f_2(\bm{\Phi})\right]=\int \dd \bm{\Phi}\left[\Lfp^\dagger(\bm{\Phi}) \cdot f_1(\bm{\Phi})\right] f_2(\bm{\Phi})$ for any two functions $f_1(\bm{\Phi})$ and $f_2(\bm{\Phi})$ in field-phase space.
Note that \Eq{eq:Fokker-Planck} needs to be solved with the boundary condition $\Pfpt{\bm{\Phi}}(\mathcal{N})=\delta(\mathcal{N})$ when $\bm{\Phi}$ lies on the end-of-inflation hypersurface, where the number of inflationary \efolds must necessarily vanish.\footnote{If the end-of-inflation hypersurface does not enclose a compact inflationary domain,  additional boundary conditions may be required~\cite{Assadullahi:2016gkk, Vennin:2016wnk}.}

A generic property of \Eq{eq:Fokker-Planck} is that the PDF of $\mathcal{N}$ features heavy non-Gaussian tails at large values of $\cal{N}$, which decay exponentially if the inflating field-phase domain is compact, and even more slowly if not~\cite{Pattison:2017mbe, Panagopoulos:2019ail, Figueroa:2020jkf, Pattison:2021oen, Ezquiaga:2019ftu, Vennin:2020kng, Achucarro:2021pdh, Animali:2022otk, Jackson:2022unc, Briaud:2023eae}. These non-Gaussian tails are non-perturbative modifications of Gaussian falloff, hence they cannot be parametrised by perturbative schemes such as the $\{f_{\mathrm{NL}}$, $g_{\mathrm{NL}}$, \etc$\}$ framework. Since PBHs form when the curvature perturbation is large, they are directly sensitive to these heavy tails, which makes it necessary to employ a non-perturbative framework such as the stochastic-$\delta N$ formalism to derive their expected statistics. 

\section{Coarse graining in stochastic inflation}
\label{sec:CoarseGraining}

In \Sec{sec:StocdeltaN} we have seen how the one-point distribution of the curvature perturbation, coarse grained at the Hubble radius at the end of inflation, can be extracted from the first-passage time statistics of the Langevin equation~\eqref{eq:Langevin} in the stochastic-inflation formalism. This is however still far from the quantities that are required to compute the clustering of PBHs, see \Sec{sec:Introduction}, and three steps are missing.

First, PBHs formation criteria are usually based on the density contrast or the compaction function, rather than the curvature perturbation. In this work, we will not elaborate much on this aspect, on which we further comment in \Sec{sec:Conclusion}. 

Second, in order to study PBHs of arbitrary masses, one must be able to study the statistics of the relevant fields when coarse grained at arbitrary scales, since the mass of the resultant black hole depends on the one contained in the super-threshold region. The goal of this section is thus to show how coarse graining at different scales can be performed within the stochastic-$\delta N$ formalism. 

Third, the reduced correlation~\eqref{eq:clustering:def} involves two-point distributions, and cannot be computed from one-point statistics only. This will also be addressed in this section, see \Sec{sub:two:point}.

\subsection{Forward statistics}
\label{sec:forward:stat}

Although the probabilities relevant for observable quantities have to be defined ``backwards'' in stochastic inflation~\cite{Tada:2016pmk, Ando:2020fjm, Tada:2021zzj}, \ie from the point of view of a local observer sitting on the end-of-inflation hypersurface, let us start by considering the ``forward'' volume produced by a given Hubble patch during inflation. The reason is that it can be calculated from the Langevin equation~\eqref{eq:Langevin} more directly, and thus provides a useful intermediate quantity. 

The problem is the following: at some given time during inflation, consider a $\sigma$-Hubble patch $\mathcal{P}_*$, in which the coarse-grained field $\bm{\Phi}$ takes value $\bm{\Phi_*}$. Since the fields are coarse grained at the $\sigma$-Hubble scale, they can indeed be considered uniform within a $\sigma$-Hubble patch. This patch later expands, and each point within it is therefore mapped to a final point on the end-of-inflation hypersurface. Our goal is to compute the physical volume encompassed by these final points. 

\subsubsection*{Final volume}

Each elementary volume within the patch $\mathcal{P}_*$, labelled by its comoving coordinate $\vecx$, expands by the amount  $\ee^{3 \mathcal{N}_{\mathcal{P}_*}(\vecx)}$, where $\mathcal{N}_{\mathcal{P}_*}(\vecx)$ is the number of \efolds realised on the worldline labelled by $\vecx$ between the initial patch and the end of inflation. Therefore, the final volume $V$ can be expressed as
\bea
\label{eq:vol:from:P*:exact}
\frac{V}{V_*} = \frac{\int_{\mathcal{P}_*}  \ee^{3\mathcal{N}_{\mathcal{P}_*}(\vecx) } \dd \vecx}{\int_{\mathcal{P}_*}  \dd\vecx} = \mathbb{E}_{\mathcal{P}_*}\left[\ee^{3\mathcal{N}_{\mathcal{P}_*}(\vecx)}\right]\, ,
\eea
where $V_* \propto (\sigma H)^{-3}(\bm{\Phi}_*)$ is the volume of the initial patch $\mathcal{P}_*$ and $\mathbb{E}_{\mathcal{P}_*}$ denotes the ensemble average over the worldlines emerging from $\mathcal{P}_*$. The final volume is subject to a distribution function that we denote $P(V \vert \bm{\Phi}_*)$, which only depends on the field value within the initial patch $\bm{\Phi}_*$, since \Eq{eq:Langevin} describes a Markovian process. 

\begin{figure}[H] 
	\centering
	\includegraphics[width=1\hsize]{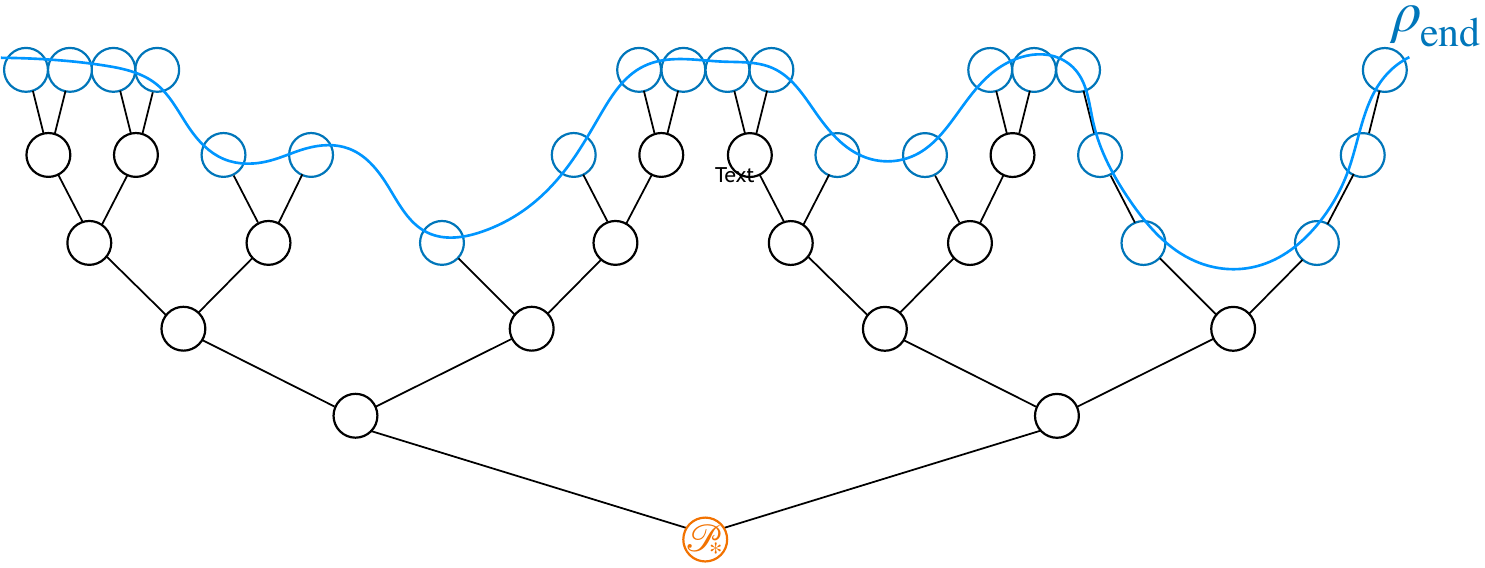}
	\caption{Binary tree associated to the calculation of the final volume. From an initial patch (orange circle), the inflationary expansion creates new patches, with each patch giving rise to two new patches after $\Delta N = \ln(2)/3$ \efolds. This process ends in patches where  $\rho(\bm{\Phi})=\rho_\uend$, where inflation terminates. These final patches are the leaves of the tree (blue circles), and the final volume corresponds to the number of such leaves.}
	\label{fig:sketch:volume}
\end{figure}

For each $\vec{x}$ within the patch $\mathcal{P}_*$, the PDF of $\mathcal{N}_{\mathcal{P}_*}(\vecx)$ is known: it is given by a first-passage time distribution from $\bm{\Phi}_*$ through the end-of-inflation hypersurface, $\Pfpt{\bm{\Phi}_*}(\calN)$, which satisfies the adjoint Fokker-Planck equation~\eqref{eq:Fokker-Planck}. The difficulty comes from the fact that, in \Eq{eq:vol:from:P*:exact}, the worldlines are not independent. The situation is schematically represented in \Fig{fig:sketch:volume}, where the initial patch $\mathcal{P}_*$ is depicted as the orange circle at the bottom. After $\Delta N = \ln(2)/3$ \efolds,\footnote{This is correct if $H$ is constant, which we will assume in this work for simplicity. Otherwise, the time required for the volume to double with respect to the Hubble volume is given by
\bea
\label{eq:DeltaN:SR}
\Delta N = \frac{\ln(2)}{3}+\ln\left[\frac{H(N)}{H(N+\Delta N)}\right] \simeq \frac{\ln(2)}{3}\left(1+\epsilon_1\right) ,
\eea 
where the second expression is valid at first order in slow roll, and $\epsilon_1=-\dd\ln H/\dd N$ is the first Hubble-flow parameter. Implementing \Eq{eq:DeltaN:SR} in the algorithms of \App{app:recursivecode} is straightforward.\label{footnote:slow:roll}} $\mathcal{P}_*$ gives rise to two $\sigma$-Hubble patches, each with a different value of $\bm{\Phi}$, which in turn give rise to new patches as inflation proceeds. This leads to a ``binary tree'' in the language of graph theory, where the ``root'' is the initial patch $\mathcal{P}_*$, and the ``leaves'' are patches that lie on the final hypersurface of uniform energy density, for which $\rho(\bm{\Phi})=\rho_\uend=3\Mp^2 H_\uend^2$. The final volume emerging from $\mathcal{P}_*$ is nothing but the number of leaves, multiplied by the leaf volume $(\sigma H_\uend)^{-3}$. 

This description gives rise to a recursive way to numerically sample realisations of $V$, from which its distribution function can be reconstructed. The  corresponding algorithm is provided in \App{app:recursivecode} where we also account for the fact that inflation may end on a branch between two nodes of the binary tree, \ie after a number of \efolds that is not an integer multiple of $\ln(2)/3$. 

\subsubsection*{Volume-averaged number of \efolds}

Finally, let us consider the volume-averaged number of \efolds realised from the patch $\mathcal{P}_*$, 
\bea
\label{eq:W:def}
W\equiv \mathbb{E}^{\mathrm{V}}_{\mathcal{P}_*} \left[ \mathcal{N}_{\mathcal{P}_*}(\vec{x})\right] ,
\eea
since it will be useful to compute the curvature perturbation below. Here, $\mathbb{E}_{\mathcal{P}_*}^{\mathrm{V}}$ denotes the volume-weighted ensemble average over $\vec{x}$, and can be expressed in terms of the regular ensemble average, $\mathbb{E}_{\mathcal{P}_*}$, as follows
\bea
\label{eq:volMean:N*}
W =\dfrac{\int_{\mathcal{P}_*} \ee^{3 \mathcal{N}_{\mathcal{P}_*}(\vecx)} \mathcal{N}_{\mathcal{P}_*}(\vecx) \dd\vecx}{\int_{\mathcal{P}_*} \ee^{3 \mathcal{N}_{\mathcal{P}_*}(\vecx)} \dd\vecx} =\, \frac{V_*}{V} \mathbb{E}_{\mathcal{P}_*}\left[\ee^{3 \mathcal{N}_{\mathcal{P}_*}(\vecx)} \mathcal{N}_{\mathcal{P}_*}(\vecx) \right] .
\eea
In other words, $\mathbb{E}_{\mathcal{P}_*}$ may be seen as the volume average if the volume between worldlines is measured on the initial patch $\mathcal{P}_*$, while $\mathbb{E}_{\mathcal{P}_*}^{\mathrm{V}}$ is the volume average if the volume between worldlines is measured on the final region. In the language of  \Fig{fig:sketch:volume}, $\mathbb{E}_{\mathcal{P}_*}^{\mathrm{V}}$ is an ensemble average over the leaves of the tree. This implies that the above-mentioned recursive algorithm to sample realisations of the volume can be improved to also sample $\mathbb{E}^{\mathrm{V}}_{\mathcal{P}_*} [ \mathcal{N}_{\mathcal{P}_*}(\vec{x})]$ at the same time, such that the joint distribution function $P(V, W \vert \bm{\Phi}_*)$ can be sampled in this way. The corresponding algorithm is given in \App{app:recursivecode}. 

\subsection{Backward statistics}
\label{sec:backward:stat}
\begin{figure}[H] 
	\centering
	\includegraphics[width=1\hsize]{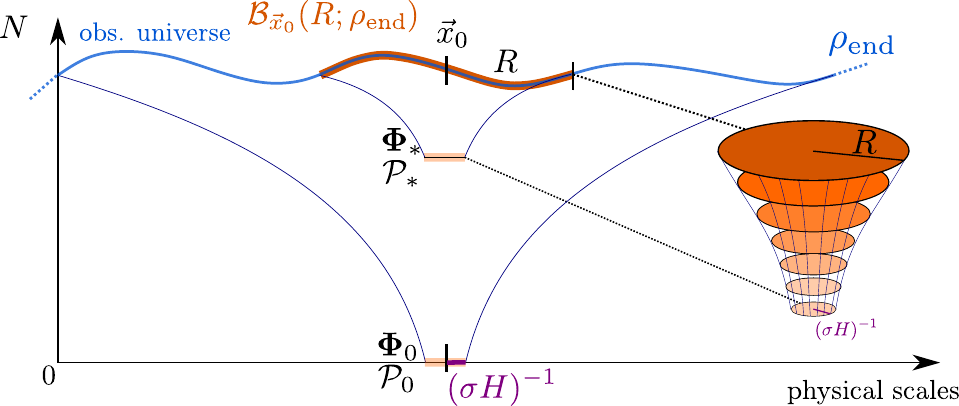}
	\caption{Sketch of the space-time diagram associated to the coarse-graining procedure discussed in \Sec{sec:CoarseGraining}. From a given region $\mathcal{B}_{\vecx_0}(R;\rho_{\mathrm{end}})$ on the final hypersurface, past worldlines can be drawn and $\mathcal{P}_*$ denotes the point at which they are all comprised within the same $\sigma$-Hubble volume. This ``parent patch'' has a volume $[\sigma H(\bm{\Phi}_*)]^{-3}$, where $\bm{\Phi}_*$ denotes the coarse-grained field value~\eqref{eq:general:coarse:graining} within $\mathcal{P}_*$. The final volume of $\mathcal{P}_*$ is nothing but the volume of $\mathcal{B}_{\vecx_0}(R;\rho_{\mathrm{end}})$, denoted $R^3$, and the large-volume approximation~\eqref{eq:P(V):appr} corresponds to $ R^3 \gg (\sigma H_{\mathrm{end}})^{-3}$.}
	\label{fig:sketch1}
\end{figure}
Having described the statistical properties of the final volume emerging from a given patch, let us now adopt the viewpoint of an observer sitting on the hypersurface of uniform energy density $\rho=\rho_{\mathrm{end}}$ where inflation ends, and performing local averages there. We consider the situation depicted in \Fig{fig:sketch1}, where an observer located at the comoving coordinate $\vecx_0$ on the end-of-inflation hypersurface performs coarse graining at a distance $R$ around them. The collection of comoving points $\vecx$ whose physical distance from $\vecx_0$, measured on the final inflation hypersurface, is less than $R$, defines the set
\bea\label{eq:setB}
\mathcal{B}_{\vecx_0}(R;\rho_{\mathrm{end}})=\{\vecx | \rph (\vecx, \vecx_0; \rho_{\mathrm{end}})  \leq R\}\,.
\eea

When the physical distance $R$ matches the size of the observable universe $R_{\mathrm{obs}}$, the set $\mathcal{B}_{\vecx_0}(R_{\mathrm{obs}}, \rho_{\mathrm{end}})$ corresponds to the entire observable universe. Going back in time the physical size of that comoving region decreases, therefore at some point it becomes smaller than the $\sigma$-Hubble volume. This defines the \textit{primeval patch} $\mathcal{P}_0$, which sets the origin of time $N=0$. On the primeval patch the coarse-grained fields $\bm{\Phi}=\bm{\Phi}_0$ are homogeneous across the observable universe. In what follows, all distributions functions will be conditioned to the value of $\bm{\Phi}_0$. If a well-motivated distribution for $\bm{\Phi}_0$ is provided, the final results may be averaged against it, but hereinafter we will keep $\bm{\Phi}_0$ explicitly in the conditioning part of all distribution functions (except when it drops out, see \Sec{sec:single:clock:models}).

Likewise, when $R<R_{\mathrm{obs}}$ is arbitrary, there is a finite time at which all worldlines intersecting $\mathcal{B}_{\vecx_0}(R;\rho_{\mathrm{end}})$ are comprised with the same $\sigma$-Hubble patch, and the corresponding patch is called the  \textit{parent patch} $\mathcal{P}_*$ of $\mathcal{B}_{\vecx_0}(R;\rho_{\mathrm{end}})$. We denote by $\bm{\Phi}=\bm{\Phi}_*$ the coarse-grained fields value within that patch. The final volume that emerges from $\mathcal{P}_*$ is nothing but the volume of $\mathcal{B}_{\vecx_0}(R;\rho_{\mathrm{end}})$, denoted $R^3$. 

\subsubsection*{Backward fields value}

From these considerations, the backward distribution of $\bm{\Phi}_*$, given $V=R^3$, can be expressed using Bayes' theorem as
\bea
\label{eq:backwardP}
P(\bm{\Phi}_*\vert V,\bm{\Phi}_0) = P(V\vert\bm{\Phi_*}) \frac{P(\bm{\Phi}_*\vert\bm{\Phi}_0)}{P_{\bm{\Phi}_0}(V)}\,. 
\eea

The quantities appearing in the right-hand side can be more explicitly defined by considering the following process. 
\begin{itemize}
\item Draw a random time $N$, according to some prior distribution $\pi(N)$
\item Draw a field value $\bm{\Phi}_* =\bm{\Phi}(N)$, according to the Fokker-Planck distribution $P(\bm{\Phi}_*,N \vert \bm{\Phi}_0,0)$
\item Draw a final volume $V$ emerging from $\bm{\Phi}_*$, according to the distribution $P(V\vert \bm{\Phi}_*)$
\end{itemize}
From this process, tuples $(\bm{\Phi}_*,V)$ are generated, hence the joint distribution $P(\bm{\Phi}_*,V \vert \bm{\Phi}_0)$ can be reconstructed, and when applying Bayes' theorem to it one obtains \Eq{eq:backwardP}. In that expression, $P(\bm{\Phi}_*\vert \bm{\Phi}_0) = \int {\rm d} N \pi(N)P(\bm{\Phi}_*,N \vert \bm{\Phi}_0,0)$ thus depends on the choice of prior for $N$, which is unavoidable in a forward formulation. A natural choice consists in a flat prior for $N$, since in that case, up to a normalisation factor, $P(\bm{\Phi}_*\vert \bm{\Phi}_0)$ reduces to the probability that, 
starting from $\bm{\Phi}_0$, the coarse-grained fields intersect $\bm{\Phi}_*$ before inflation ends~\cite{Vennin:2015hra, Noorbala:2018zlv}. It can be formally obtained by solving an adjoint Fokker-Planck equation with vanishing right-hand side, but it will drop out from the final result so its detailed expression is not relevant. Finally, $P_{\bm{\Phi}_0}(V)=\int P(V| \bm{\Phi}_*)P(\bm{\Phi}_*| \bm{\Phi}_0){\rm d} \bm{\Phi}_*$ ensures that the distribution~\eqref{eq:backwardP} is properly normalised. 

\subsubsection*{Volume-averaged number of \efolds}

Along similar lines, using Bayes' theorem  repeatedly, the distribution for the volume-averaged number of \efolds~\eqref{eq:W:def} can be expressed as
\bea
\label{eq:1pt:exact}
P(W \vert V, \bm{\Phi}_0) = & \int\dd\bm{\Phi}_* P(W \vert V, \bm{\Phi}_*, \bm{\Phi}_0) P\left(\bm{\Phi}_* \vert V, \bm{\Phi}_0\right) 
\\ =&  \int\dd\bm{\Phi}_* \frac{P(V,W \vert \bm{\Phi}_*,\bm{\Phi}_0)}{P(V\vert  \bm{\Phi}_*,\bm{\Phi}_0)}  P\left(\bm{\Phi}_* \vert V, \bm{\Phi}_0\right) 
=   \int\dd\bm{\Phi}_*  P(V,W \vert \bm{\Phi}_*) \frac{P(\bm{\Phi}_*\vert\bm{\Phi}_0)}{P_{\bm{\Phi}_0}(V)}\, ,
\eea
where \Eq{eq:backwardP} was also employed to replace $P\left(\bm{\Phi}_* \vert V, \bm{\Phi}_0\right)$ and where we have dropped $\bm{\Phi}_0$ from distributions conditioned to $\bm{\Phi}_*$, since the stochastic process under consideration is Markovian. Regarding this last expression, we already explained how to compute the joint distribution $P(V,W \vert \bm{\Phi}_*) $ in \Sec{sec:forward:stat}, and the other distributions involved in the final result were introduced in the paragraph just above. 

\subsection{One-point coarse-grained curvature perturbation}
\label{sec:one:point}

\subsubsection*{Coarse graining at the Hubble scale}

According to the $\delta N$ formula~\eqref{eq:zeta:deltaN}, at each point in the final region $\mathcal{B}_{\vecx_0}(R;\rho_{\mathrm{end}})$, the curvature perturbation is given by
\bea
\label{eq:zeta:cg:def}
\zeta_{\mathrm{cg}}(\vec{x}) = \mathcal{N}_{\mathcal{P}_0}(\vecx) - \mathbb{E}_{\mathcal{P}_0}^{\mathrm{V}}\left(\mathcal{N}_{\mathcal{P}_0}\right) ,
\eea
where the curvature perturbation is coarse grained at the $\sigma$-Hubble scale at the end of inflation. In this expression, $\mathcal{N}_{\mathcal{P}_0}(\vecx)$ is the number of \efolds realised between the primeval patch and the end of inflation along the worldline labelled by $\vecx$, and $\mathbb{E}_{\mathcal{P}_0}^{\mathrm{V}}(\mathcal{N}_{\mathcal{P}_0})$ is the volume-averaged of $\mathcal{N}_{\mathcal{P}_0}$ across the entire observable universe. Note that a volume average is used here, which as explained above corresponds to an ensemble average over the set of $\sigma$-Hubble patches that  tessellate the end-of-inflation hypersurface (\ie, in the language of \Fig{fig:sketch:volume}, an average over the tree's leaves). This is because these are the statistics available to a local observer located on that hypersurface, with respect to which observable quantities are defined. Recall indeed that in the present formalism, ``volume average'' corresponds to the regular spatial average on the end-of-inflation hypersurface. For instance, a direct consequence of \Eq{eq:zeta:cg:def} is that the \textit{volume} average of $\zeta_{\mathrm{cg}}(\vec{x})$ vanishes, $\mathbb{E}_{\mathcal{P}_0}^{\mathrm{V}}[\zeta_{\mathrm{cg}}(\vecx)]=0$. 

This implies that, at the statistical level, the curvature perturbation must be interpreted with volume-weighted distributions. For $\zeta_{\mathrm{cg}}$, the number of \efolds realised from $\bm{\Phi}_0$ follows the distribution function $\Pfpt{\bm{\Phi}_0}(\mathcal{N}_0)$, which is subject to the adjoint Fokker-Planck equation~\eqref{eq:Fokker-Planck}. From this distribution, one can define its volume-weighted version~\cite{Gratton:2005bi, Winitzki:2008yb},
\bea
\label{eq:statistical:volume:weighting}
\PfptV{\bm{\Phi}_0}(\mathcal{N}_0) = \dfrac{\Pfpt{\bm{\Phi}_0}(\mathcal{N}_0)\ee^{3 \mathcal{N}_0}}{\int_0^\infty \Pfpt{\bm{\Phi}_0}(\mathcal{N})\ee^{3 \mathcal{N}} \dd \mathcal{N} } 
\eea
such that $\zeta_{\mathrm{cg}}$ follows the distribution, neglecting cosmic variance,
\bea
\label{eq:P:zetacg}
P\left(\zeta_{\mathrm{cg}} \vert \bm{\Phi}_0\right) = \Pfpt{\bm{\Phi}_0}^{\mathrm{V}}\left( \left\langle \mathcal{N}_0 \right\rangle_{\mathrm{V}} + \zeta_{\mathrm{cg}} \right),
\eea
where $\langle \mathcal{N}_0 \rangle_{\mathrm{V}} = \int  \mathcal{N}_0\Pfpt{\bm{\Phi}_0}^{\mathrm{V}}(\mathcal{N}_0)\dd\mathcal{N}_0 $ is the volume-weighted stochastic average of $\mathcal{N}_0$. In this work, $\zeta(\vec{x})$ denotes the real-space field, while $\zeta$ (without spatial argument) refers to the random variable that shares its statistics. For instance, through ergodicity, $\mathbb{E}_{\mathcal{P}_0}^{\mathrm{V}}[\zeta_{\mathrm{cg}}(\vecx)]=0$ ensures that $\langle \zeta_{\mathrm{cg}}\rangle_{\mathrm{V}}=0$.

As mentioned in \Sec{sec:StocdeltaN}, depending on the inflationary setup under consideration, the tail of the first-passage time distribution can either be of the exponential type, $\Pfpt{\bm{\Phi}}(\calN)\propto \ee^{-\Lambda \mathcal{N}}$, or of the monomial type $\Pfpt{\bm{\Phi}}(\calN)\propto  \mathcal{N}^{-p}$, at large $\mathcal{N}$. If $\Lambda\leq 3$ in the former case, and for any value of $p$ in the latter case, the integral appearing in \Eq{eq:statistical:volume:weighting} does not converge. In such a situation, the mean final volume is infinite, which signals the occurrence of ``eternal inflation''~\cite{Steinhardt:1982kg, Vilenkin:1983xq, Guth:1985ya, Linde:1986fc}. The problem of defining measures in eternally inflating spacetimes still remains to be solved~\cite{Garcia-Bellido:1994gng, Bousso:2006ev, Linde:2008xf, DeSimone:2008if, Linde:2010xz, Guth:2011ie, Freivogel:2011eg, Garriga:2012bc}, but since it lies outside the scope of our article we will simply restrict to situations where it does not occur. However, we stress that the measure problem would need to be addressed in order to properly compute observable predictions in models where eternal inflation occurs. Note also that the tail profile is independent of the initial field configuration $\bm{\Phi}_0$, since the values of $\Lambda$ or $p$ are characteristics of the Fokker-Planck spectrum~\cite{Pattison:2017mbe, Ezquiaga:2019ftu}. As a consequence, one cannot simply avoid the eternal-inflation problem by assuming that initial conditions are set below the eternally inflating region of field space: if eternal inflation occurs somewhere, \Eq{eq:statistical:volume:weighting} diverges everywhere.

\subsubsection*{Coarse graining at an arbitrary scale}

When coarse graining is performed over the whole final region $\mathcal{B}_{\vecx_0}(R;\rho_{\mathrm{end}})$, the coarse-grained curvature perturbation is noted $\zeta_R(\vecx_0)$. It can be viewed as the volume average of $\zeta_{\mathrm{cg}}(\vec{x})$,\footnote{This can also be conveniently rewritten as
\bea
\label{eq:zetaR:alternative}
\zeta_R (\vecx_0) = \mathbb{E}_{\mathcal{P}_*}^{\mathrm{V}}\left(\mathcal{N}_{\mathcal{P}_0}\right) - \mathbb{E}_{\mathcal{P}_0}^{\mathrm{V}}\left(\mathcal{N}_{\mathcal{P}_0}\right)\, .
\eea} 
\bea
\label{eq:zetaR:interm}
\zeta_R(\vecx_0) = \mathbb{E}^{\mathrm{V}}_{\mathcal{P}_*}\left[\zeta_{\mathrm{cg}}(\vec{x})\right] = 
\dfrac{\int_{\mathcal{P}_*} \ee^{3 \mathcal{N}_{\mathcal{P}_0}(\vecx)} \mathcal{N}_{\mathcal{P}_0}(\vecx) \dd\vecx}{\int_{\mathcal{P}_*} \ee^{3 \mathcal{N}_{\mathcal{P}_0}(\vecx)} \dd\vecx}- \mathbb{E}_{\mathcal{P}_0}^{\mathrm{V}}\left[\mathcal{N}_{\mathcal{P}_0}(\vecx)\right]\, .
\eea
At this point, it is useful to notice that the number of \efolds realised from the primeval patch along a worldline ending in $\mathcal{B}_{\vecx_0}(R;\rho_{\mathrm{end}})$ can always be decomposed into the number of \efolds realised before and after the parent patch $\mathcal{P}_*$
\bea
\label{eq:N:decomp}
\mathcal{N}_{\mathcal{P}_0}(\vecx) = \mathcal{N}_{\mathcal{P}_0\to \mathcal{P}_*}(\vecx)+\mathcal{N}_{\mathcal{P}_*}(\vecx)\, .
\eea
By construction, $\mathcal{N}_{\mathcal{P}_0\to \mathcal{P}_*}(\vecx)$ is the same for all $\vecx$, since all worldlines ending in $\mathcal{B}_{\vecx_0}(R;\rho_{\mathrm{end}})$ lie within the same $\sigma$-Hubble patch prior to $\mathcal{P}_*$. Inserting the decomposition~\eqref{eq:N:decomp} into \Eq{eq:zetaR:interm}, one thus finds
\bea
\label{eq:zetaR:Vol}
\zeta_R(\vecx_0) =\mathcal{N}_{\mathcal{P}_0\to \mathcal{P}_*}(\vecx_0) + W\left(\mathcal{P}_*\right) - \mathbb{E}_{\mathcal{P}_0}^{\mathrm{V}}\left[\mathcal{N}_{\mathcal{P}_0}(\vecx)\right]\, ,
\eea
where we recall that $W(\mathcal{P}_*)=\mathbb{E}_{\mathcal{P}_*}^{\mathrm{V}}\left[\mathcal{N}_{\mathcal{P}_*}(\vecx)\right]$, see \Eq{eq:W:def}. Because of the Markovian nature of the stochastic process, the dynamics before and after the parent patch are independent, hence the joint distribution for $\mathcal{N}_{\mathcal{P}_0\to \mathcal{P}_*}$ and $W$ can be factorised according to
\bea
\label{eq:zetaR:interm:2}
P^{\mathrm{V}}(\mathcal{N}_{\mathcal{P}_0\to \mathcal{P}_*} , W \vert V, \bm{\Phi}_*, \bm{\Phi}_0) = & P^{\mathrm{V}}(\mathcal{N}_{\mathcal{P}_0\to \mathcal{P}_*}  \vert \bm{\Phi}_*, \bm{\Phi}_0) P(W \vert \bm{\Phi}_*, V)\, .
\eea
Here, the superscript ``$\mathrm{V}$'' is a reminder that the distribution should be volume weighted (by definition, the volume emerging from the parent patch is fixed, hence such a superscript would be superfluous for the second distribution in the right-hand side and volume weighting has to be performed over $\mathcal{N}_{\mathcal{P}_0\to \mathcal{P}_*}$ only). Both terms in the right-hand side can be further expanded using Bayes' theorem. The non-volume-weighted version of the first distribution reads
\bea
\label{eq:P(N_P0P*)}
P(\mathcal{N}_{\mathcal{P}_0\to \mathcal{P}_*}  \vert \bm{\Phi}_*,\bm{\Phi}_0) = P_{\mathrm{FP}}\left(\bm{\Phi}_*,\mathcal{N}_{\mathcal{P}_0\to \mathcal{P}_*} \vert \bm{\Phi}_0\right) \frac{P_{\bm{\Phi}_0}\left(\mathcal{N}_{\mathcal{P}_0\to \mathcal{P}_*}\right)}{P\left(\bm{\Phi}_* \vert \bm{\Phi}_0\right)}\,,
\eea
where $P_{\mathrm{FP}}(\bm{\Phi}_*,N \vert \bm{\Phi}_0)$ is the solution of the Fokker-Planck equation~\eqref{eq:Fokker:Planck:not:adjoint}, $P_{\bm{\Phi}_0}(N)=\int_N^\infty \dd \mathcal{N} \Pfpt{\bm{\Phi}_0}(\mathcal{N})$ is the probability that, starting from $\bm{\Phi}_0$, at least $N$ \efolds are realised before inflation ends and $P(\bm{\Phi}_* \vert \bm{\Phi}_0)$ has been already introduced below \Eq{eq:backwardP}. Recall that the volume-weighted distribution can then be obtained from the non-volume weighted one by using \Eq{eq:statistical:volume:weighting}. For the second term in \Eq{eq:zetaR:interm:2} one has
\bea
P(W \vert \bm{\Phi}_*, V) = \frac{P\left(V,W \vert \bm{\Phi}_*\right)}{P\left(V\vert \bm{\Phi}_*\right)}\, ,
\eea
where $P\left(V \vert \bm{\Phi}_*\right)$ and $P\left(V,W\vert \bm{\Phi}_*\right)$ are the forward distributions derived in \Sec{sec:forward:stat}. This leads to
\bea
P^{\mathrm{V}}(\mathcal{N}_{\mathcal{P}_0\to \mathcal{P}_*} , W \vert V, \bm{\Phi}_0)=\int \dd\bm{\Phi}_* P^{\mathrm{V}}(\mathcal{N}_{\mathcal{P}_0\to \mathcal{P}_*} , W \vert V, \bm{\Phi}_*, \bm{\Phi}_0)P(\bm{\Phi}_*\vert V,\bm{\Phi}_0)\, ,
\eea
where $P(\bm{\Phi}_*\vert V,\bm{\Phi}_0)$ is given in \Eq{eq:backwardP}. Combining the above results, one finds
\bea
P(\mathcal{N}_{\mathcal{P}_0\to \mathcal{P}_*} , W \vert V, \bm{\Phi}_0)=
\int \dd\bm{\Phi}_* \frac{P_{\bm{\Phi}_0}\left(\mathcal{N}_{\mathcal{P}_0\to \mathcal{P}_*}\right)}{P_{\bm{\Phi}_0}(V)}P_{\mathrm{FP}}\left(\bm{\Phi}_*,\mathcal{N}_{\mathcal{P}_0\to \mathcal{P}_*} \vert \bm{\Phi}_0\right)P\left(V,W \vert \bm{\Phi}_*\right) ,
\eea
where, again, volume weighting over $\mathcal{N}_{\mathcal{P}_0\to \mathcal{P}_*}$ has to be further performed. 

The distribution of $Z\equiv \mathcal{N}_{\mathcal{P}_0\to \mathcal{P}_*} + W$ can thus be computed by marginalising the above joint distribution, $P(Z \vert V, \bm{\Phi}_0) = \int\dd W  P^{\mathrm{V}}(Z-W, W \vert V, \bm{\Phi}_0) $, and this gives rise to the one-point distribution of $\zeta_R$, see \Eq{eq:zetaR:Vol}. This is one of the key results of the present work, since it allows one to compute the one-point distribution of the curvature perturbation when coarse grained at an arbitrary scale, without any other approximation than those on which the stochastic formalism rests. Note that in this approach, a lattice code is not required and recursive algorithms are sufficient: the spatial arrangement of the leaves and branches is entirely contained in the graph structure of the tree.

\subsection{Large-volume approximation}
\label{subsec:Large:vol:approx}
Although exact, the above procedure may be cumbersome to implement, since it involves the recursive sampling algorithms presented in \App{app:recursivecode}. Each evaluation of the algorithms may be numerically expensive, and a large number of realisations are required for the statistics to converge to a given accuracy.  In order to gain analytical insight, the following approximation may thus be useful. 

\subsubsection*{Final volume} 

The stochastic-inflation formalism is expected to provide a reliable effective theory at large scales, hence it should be employed only in situations where the final volume is large compared to the Hubble volume, \ie when the tree of \Fig{fig:sketch:volume} contains a large number of leaves. Given two leaves in such a tree, consider the ``lowest common ancestor'' (still in the language of graph theory), \ie the latest node they both emerge from.\footnote{In this language, the parent patch is the lowest common ancestor to all the leaves belonging to the final volume $\mathcal{B}_{\vecx_0}$ under consideration, and the primeval patch is the lowest common ancestor of the observable universe.} Prior to this common ancestor, the two paths at the end of which these leaves lie are the same, but they become independent afterwards. As a consequence, the number of \efolds realised along those two paths are only partly correlated. In the case of dependent random variables, the central-limit theorem can be generalised, provided the amount of correlations is bounded (see \eg \Refa{10.1215/S0012-7094-48-01568-3}). In this limit, where the number of leaves is large, the ensemble average may be approximated by the stochastic average of a single element within the ensemble, and one finds
\bea
\label{eq:P(V):appr}
P(V \vert \bm{\Phi}_*) \simeq \delta_{\mathrm{D}}\left(V-V_* \left\langle \ee^{3 \mathcal{N}_{\bm{\Phi}_*}}\right\rangle \right)\, ,
\eea 
where $\delta_{\mathrm{D}}$ is the Dirac distribution and
\bea
\label{eq:mean:e3N}
\left\langle \ee^{3 \mathcal{N}_{\bm{\Phi}_*}}\right\rangle = \int_0^\infty  \Pfpt{\bm{\Phi}_*}(\calN) \ee^{3 \mathcal{N}}\dd \mathcal{N}\, .
\eea

\subsubsection*{Caveat}

It is worth noting that the appearance of Dirac distributions in stochastic processes is usually associated to the ``classical limit'', in which the amplitude of quantum diffusion is small compared to the classical drift. In the present case, we have not assumed that the stochastic noise plays a sub-dominant role, we have simply considered large volumes and used the central-limit theorem. However, the two limits bear some formal analogies (although they are physically distinct) for the following reason. Consider a given path in the binary tree of \Fig{fig:sketch:volume}. As one climbs up the tree along that path, each node gives rise to a final volume that is necessarily smaller than that of the previous node, since the former is comprised within the latter. As a consequence, if \Eq{eq:P(V):appr} holds, $\langle \ee^{3 \mathcal{N}_{\bm{\Phi}_*}} \rangle$ must necessarily be a decreasing function of time. This constraints the stochastic evolution of $\bm{\Phi}_*$, which cannot fluctuate along directions where $\langle \ee^{3 \mathcal{N}_{\bm{\Phi}_*}} \rangle$ increases. In practice, such fluctuations are of course possible but if they occur on short time scales only, then they may not be seen when the dynamics is averaged over long time scales, which the large-volume approximation implicitly assumes. One should thus bear in mind that the large-volume approximation may break down when dynamics over short time scales matters. In this work we use the large-volume approximation only as a way to derive first qualitative results, and we postpone a comparison of this approach to the exact, recursive-based method presented above to future work.

\subsubsection*{Backward fields value}

Let us consider the hypersurface  $\mathcal{S}_*$ made of field-phase space locations $\bm{\Phi}_*$ such that $V_* \left\langle \ee^{3 \mathcal{N}_{\bm{\Phi}_*}}\right\rangle=V$. In the large-volume limit, the field values in the parent patch have to lie on that hypersurface. In practice, the parent patch $\mathcal{P}_*$ can be identified with the first crossing of $\mathcal{S}_*$. One may object that, in the path between $\mathcal{P}_0$ and $\mathcal{P}_*$, the hypersurface  $\mathcal{S}_*$ may be crossed several times so that the parent patch does not have to correspond to a first-crossing event. Although this is correct -and indeed this assumption is not made in \Eq{eq:P(N_P0P*)}-, as mentioned above the large-volume limit assumes that $\langle \ee^{3 \mathcal{N}_{\bm{\Phi}_*}} \rangle$ necessarily decreases, hence $\bm{\Phi}_*$ cannot be crossed more than once when the dynamics is zoomed out on sufficiently long time scales. At the level of the current approximation, the first-crossing assumption is therefore valid. 

We thus introduce the joint distribution for the first-passage time and the first-passage location (FPTL) of  $\mathcal{S}_*$,
\bea
\label{eq:joint:Nfpt:phi*}
\PfptlV{\bm{\Phi}_0\to\mathcal{S}_*}\left(\mathcal{N}_{\bm{\Phi}_0\to \mathcal{S}_*}, \bm{\Phi}_* \vert \bm{\Phi}_0\right) = \PfptV{\bm{\Phi}_0\to\mathcal{S}_*}\left(\mathcal{N}_{\bm{\Phi}_0\to \mathcal{S}_*}\right) P\left(\bm{\Phi}_* \vert \mathcal{N}_{\bm{\Phi}_0\to \mathcal{S}_*}\right) .
\eea
Here, $\Pfpt{\bm{\Phi}_0\to\mathcal{S}_*}$ is nothing but a first-passage time distribution, subject to the adjoint Fokker-Planck equation~\eqref{eq:Fokker-Planck}, and $\PfptV{\bm{\Phi}_0\to\mathcal{S}_*}$ is its volume-weighted version. 

The second distribution, $P(\bm{\Phi}_* \vert \mathcal{N}_{\bm{\Phi}_0\to \mathcal{S}_*})$, corresponds to the first-crossing location, conditioned to a given first-crossing time. This distribution is difficult to reconstruct by direct sampling: amongst all realisations that cross $\mathcal{S}_*$, one must keep only those that produce a certain value for $\mathcal{N}_{\bm{\Phi}_0\to \mathcal{S}_*}$ and compute the statistics of $\bm{\Phi}_*$ only amongst those. Direct sampling methods are thus very inefficient since the vast majority of the simulated realisations have to be discarded. Recently, in \Refa{Tokeshi:2023swe}, a solution to this problem has been proposed, which makes use of the theory of constrained stochastic systems. Namely, modified Langevin and Fokker-Planck equations were derived, that only sample realisations of a given duration. Let us also note that, for the purpose of sampling, it is in fact more efficient to directly sample the joint distribution of $P^{\mathrm{V}}(\mathcal{N}_{\bm{\Phi}_0\to \mathcal{S}_*}, \bm{\Phi}_* \vert \bm{\Phi}_0)$, \ie simulate realisations starting at $\bm{\Phi}_0$ and record the values of $\mathcal{N}_{\bm{\Phi}_0\to \mathcal{S}_*}$ and $\bm{\Phi}_* $ for each of them. The decomposition~\eqref{eq:joint:Nfpt:phi*} is introduced for latter convenience only. 

At this stage, it is worth noting that the large-volume approximation differs from the one proposed in \Refa{Tada:2021zzj}, where, instead of \Eq{eq:P(V):appr}, the replacement 
\bea
\label{eq:Tada:Vennin}
P_{\text{\cite{Tada:2021zzj}}}(V\vert \bm{\Phi}_*) \dd V \simeq \Pfpt{\bm{\Phi}_*} \left(\mathcal{N}\right)\dd \calN
\quad\quad\text{with}\quad\quad
V=V_*\ee^{3\mathcal{N}}
\eea
is rather performed. This rests on the assumption that the final volume can be computed along a representative path, the one that ends at $\vecx_0$. Such an approximation is expected to be valid when there is little variation amongst the set of final leaves, which is the case if the number of leaves is small, and the amount of expansion along a reference trajectory within the region of interest is a more accurate estimate than the averaged expansion. In this sense it may be viewed as describing the opposite regime from \Eq{eq:P(V):appr}. In particular, through Bayes' theorem it gives rise to a backward distribution for $\bm{\Phi}_*$, see Eq.~(3.14) of \Refa{Tada:2021zzj}, which differs from \Eq{eq:joint:Nfpt:phi*} when marginalised over $\mathcal{N}_{\bm{\Phi}_0\to \mathcal{S}_*}$.

\subsubsection*{Coarse-grained curvature perturbation}

In the large-volume limit, the ensemble average in \Eq{eq:volMean:N*} may also be replaced with a stochastic average, $W\simeq  \langle \mathcal{N}_{\bm{\Phi}_*} \ee^{3  \mathcal{N}_{\bm{\Phi}_*}} \rangle/\langle \ee^{3  \mathcal{N}_{\bm{\Phi}_*}}\rangle = \langle \mathcal{N}_{\bm{\Phi}_*} \rangle_{\mathrm{V}} $, and similarly for $ \mathbb{E}_{\mathcal{P}_0}^{\mathrm{V}}(\mathcal{N}_{\mathcal{P}_0}) \simeq \langle \mathcal{N}_{\bm{\Phi}_0}  \rangle_{\mathrm{V}}$.
With these replacements, \Eq{eq:zetaR:Vol} leads to
\bea
\label{eq:zetaR:Vol:2}
\zeta_R \simeq \mathcal{N}_{\mathcal{P}_0\to \mathcal{S}_*} + \left\langle \mathcal{N}_{\bm{\Phi}_*} \right\rangle_{\mathrm{V}} -\left\langle \mathcal{N}_{\bm{\Phi}_0}  \right\rangle_{\mathrm{V}}\, .
\eea
The statistics of $\zeta_R$ is therefore entirely determined by the joint distribution of $\mathcal{N}_{\mathcal{P}_0\to \mathcal{S}_*}$ and $\bm{\Phi}_*$ given in \Eq{eq:joint:Nfpt:phi*}, and reads
\bea
\label{eq:1pt:large:vol:final}
P\left(\zeta_R\vert \bm{\Phi}_0\right) =\int_{\mathcal{S}_*}\dd\bm{\Phi}_*
\PfptlV{\bm{\Phi}_0\to\mathcal{S}_*}\left(\mathcal{N}_{\mathcal{P}_0\to \mathcal{S}_*}=\zeta_R- \left\langle \mathcal{N}_{\bm{\Phi}_*} \right\rangle_{\mathrm{V}} +\left\langle \mathcal{N}_{\bm{\Phi}_0}  \right\rangle_{\mathrm{V}}, \bm{\Phi}_* \vert \bm{\Phi}_0\right)  .
\eea

As a consistency check, one can verify that \Eq{eq:zetaR:Vol:2} implies that $\zeta_R$ is centred, \ie $\langle \zeta_R \rangle_\mathrm{V}=0$. We first consider the case where $\bm{\Phi}_*$ is fixed. In the decomposition $\mathcal{N}_{\bm{\Phi}_0}=\mathcal{N}_{\bm{\Phi}_0\to \bm{\Phi}_*}+\mathcal{N}_{\bm{\Phi}_*}$, $\mathcal{N}_{\bm{\Phi}_0\to \bm{\Phi}_*}$ and $\mathcal{N}_{\bm{\Phi}_*}$ are independent random variables, since the stochastic process is Markovian (see \App{app:sub:vanishing:mean} for a detailed demonstration). One therefore has
\bea
\Pfpt{\bm{\Phi}_0}\left(\mathcal{N}_{\bm{\Phi}_0}\right) = \int_0^{\mathcal{N}_{\bm{\Phi}_0}} \dd \mathcal{N}_{\bm{\Phi}_*} \Pfpt{\bm{\Phi}_0\to \bm{\Phi}_*} \left(\mathcal{N}_{\bm{\Phi}_0}-\mathcal{N}_{\bm{\Phi}_*}\right)\Pfpt{\bm{\Phi}_*}\left(\mathcal{N}_{\bm{\Phi}_*}\right)\, .
\eea
This convolution structure also applies to the volume-weighted statistics, since
\bea
\PfptV{\bm{\Phi}_0}\left(\mathcal{N}_{\bm{\Phi}_0}\right) \propto & \Pfpt{\bm{\Phi}_0}\left(\mathcal{N}_{\bm{\Phi}_0}\right) \ee^{3 \mathcal{N}_{\bm{\Phi}_0}}\\
 = & \int_0^{\calN_{\bm{\Phi}_0}} \dd \mathcal{N}_{\bm{\Phi}_*} \Pfpt{\bm{\Phi}_0\to \bm{\Phi}_*} \left(\mathcal{N}_{\bm{\Phi}_0}-\mathcal{N}_{\bm{\Phi}_*}\right)\Pfpt{\bm{\Phi}_*}\left(\mathcal{N}_{\bm{\Phi}_*}\right)\ee^{3 \mathcal{N}_{\bm{\Phi}_0}}\\
  = & \int_0^{\calN_{\bm{\Phi}_0}} \dd \mathcal{N}_{\bm{\Phi}_*} \Pfpt{\bm{\Phi}_0\to \bm{\Phi}_*} \left(\mathcal{N}_{\bm{\Phi}_0}-\mathcal{N}_{\bm{\Phi}_*}\right) \ee^{3\left(\mathcal{N}_{\bm{\Phi}_0}-\mathcal{N}_{\bm{\Phi}_*}\right)} \Pfpt{\bm{\Phi}_*}\left(\mathcal{N}_{\bm{\Phi}_*}\right)\ee^{3 \mathcal{N}_{\bm{\Phi}_*}}\\
  \propto &
   \int_0^{\calN_{\bm{\Phi}_0}} \dd \mathcal{N}_{\bm{\Phi}_*} \PfptV{\bm{\Phi}_0\to \bm{\Phi}_*} \left(\mathcal{N}_{\bm{\Phi}_0}-\mathcal{N}_{\bm{\Phi}_*}\right)\PfptV{\bm{\Phi}_*}\left(\mathcal{N}_{\bm{\Phi}_*}\right) .
   \label{eq:convolv:vol:weighting}
\eea
This implies that, not only $\langle \mathcal{N}_{\bm{\Phi}_0}\rangle=\langle\mathcal{N}_{\bm{\Phi}_0\to \bm{\Phi}_*}\rangle+\langle\mathcal{N}_{\bm{\Phi}_*}\rangle$, but also $\langle \mathcal{N}_{\bm{\Phi}_0}\rangle_{\mathrm{V}}=\langle\mathcal{N}_{\bm{\Phi}_0\to \bm{\Phi}_*}\rangle_{\mathrm{V}}+\langle\mathcal{N}_{\bm{\Phi}_*}\rangle_{\mathrm{V}}$, hence $\langle \zeta_R \rangle_\mathrm{V}=0$. This is valid for all $\bm{\Phi}_*$, and is thus still valid when marginalising over $\bm{\Phi}_*$.

In practice, the above considerations lead to a simple algorithm to sample the one-point distribution function $P(\zeta_{R} \vert \bm{\Phi}_0)$: from $\bm{\Phi}_0$, simulate a realisation until it crosses $\mathcal{S}_*$ and record the value of $\bm{\Phi}_*$ and $\mathcal{N}_{\bm{\Phi}_0\to \mathcal{S}_*}$. Evaluate $\zeta_{R}$ via \Eq{eq:zetaR:Vol:2}, and weight this outcome by $\ee^{3\mathcal{N}_{\bm{\Phi}_0\to \mathcal{S}_*}}$. Finally, repeat the whole procedure until a sufficiently large number of outcomes is obtained, over which the distribution $P(\zeta_{R} \vert \bm{\Phi}_0)$ can be reconstructed.

\subsection{Two-point coarse-grained curvature perturbation}
\label{sub:two:point}

Let us now study how the curvature perturbations coarse grained at two different locations are correlated. At first sight, the Langevin equation \eqref{eq:Langevin} describes the evolution of independent Hubble-sized patches of the universe, therefore it does not seem to retain any information about the relative spatial position of those patches.  However, as argued in \Refa{Ando:2020fjm}, the relative distance between patches is encoded in the time at which their respective paths become statistically independent. As a result, the stochastic-inflation formalism also allows one to describe the spatial structure of the correlations between the durations of inflation at different points, and therefore it can be extended to reconstruct multiple-point statistics.

\begin{figure}[H] 
	\centering
	\includegraphics[width=1\hsize]{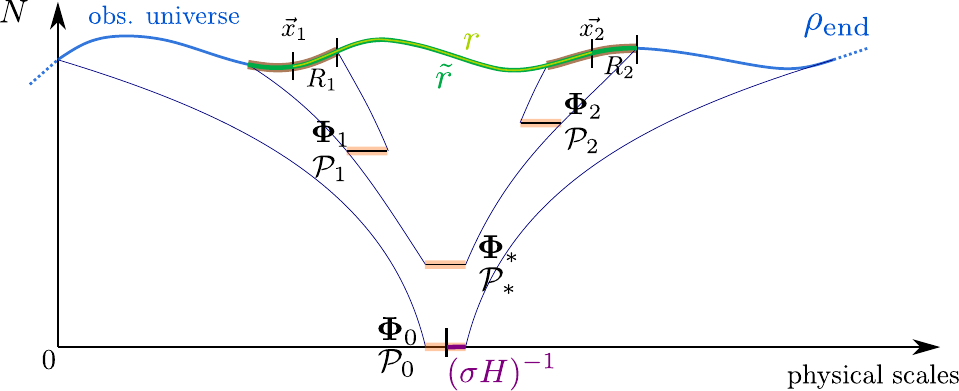}
	\caption{Sketch of the space-time diagram associated to the calculation of the two-point distribution discussed in \Sec{sub:two:point}. The final volume of $\mathcal{P}_i, i=1,2$ is the volume of $\mathcal{B}_{\vecx_i}(R_i;\rho_{\mathrm{end}})$, denoted $R_i^3$, and the final volume of $\mathcal{P}_*$ is given by $(\tilde{r}/2)^3$. }
	\label{fig:sketch2}
\end{figure}

We consider two points of comoving coordinates $\vecx_1$ and $\vecx_2$ on the end-of-inflation hypersurface, at physical distance $r_{\mathrm{ph}}(\vecx_1, \vecx_2; \rho_{\mathrm{end}})=r$.  With $i=1,2$, let $B_{\vecx_i}(R_i;\rho_{\mathrm{end}})$ be the sets of comoving points $\vecx$ around $\vecx_i$ within the physical distance $R_i$ on the hypersurface $\rho=\rho_{\mathrm{end}}$, see \Eq{eq:setB}. These correspond to the two regions over which the curvature perturbation is coarse grained. To each region one can associate a parent patch $\mathcal{P}_i$, see \Sec{sec:backward:stat}. Similarly, let us consider the latest point at which the set of comoving points $B_{\vecx_1}(R_1;\rho_{\mathrm{end}}) \cup B_{\vecx_2}(R_2;\rho_{\mathrm{end}})$ lies within a $\sigma$-Hubble volume. We call this patch the \textit{joint parent patch} $\mathcal{P}_*$, and we denote by $\bm{\Phi}_*$ the field values within that patch. The size of the final region emerging from $\mathcal{P}_*$ is given by $\tilde{r} = r+R_1+R_2$, and the situation is summarised in \Fig{fig:sketch2}.

For each final region one can use \Eq{eq:zetaR:Vol}, namely
\bea\label{eq:two:zeta:gen}
\zeta_{R_i}(\vecx_i) =\mathcal{N}_{\mathcal{P}_0\to \mathcal{P}_i}(\vecx_i) + W\left(\mathcal{P}_i\right) - \mathbb{E}_{\mathcal{P}_0}^{\mathrm{V}}\left[\mathcal{N}_{\mathcal{P}_0}(\vecx)\right]\, ,
\eea
where one can further expand
\bea\label{eq:two:zeta:dec}
\mathcal{N}_{\mathcal{P}_0\to \mathcal{P}_i}(\vecx_i)  = \mathcal{N}_{\mathcal{P}_0\to \mathcal{P}_*}+\mathcal{N}_{\mathcal{P}_*\to\mathcal{P}_i}(\vecx_i)\, .
\eea
The above decomposition holds for $i=1,2$, but the first term in the right-hand side is independent of $i$, since all trajectories ending in $B_{\vecx_1}(R_1;\rho_{\mathrm{end}}) \cup B_{\vecx_2}(R_2;\rho_{\mathrm{end}})$ share the same path prior to the joint parent patch. Let us notice that this last feature is indeed the source of correlation between the curvature perturbations $\zeta_{R_1}(\vecx_1)$ and $\zeta_{R_2}(\vecx_2)$ on the end-of-inflation hypersurface.

A similar calculation as in \Sec{sec:one:point} can then be performed, which allows one to express the joint distribution $P(\zeta_{R_1},\zeta_{R_2})$ in terms of distribution functions that can be sampled by solving stochastic first-passage time problems. We do not write the corresponding expressions here for conciseness and since the derivation is very similar to that presented in \Sec{sec:one:point}. 

Let us instead move to the large-volume approximation, where \Eqs{eq:two:zeta:gen} and~\eqref{eq:two:zeta:dec} lead to
\bea
\zeta_{R_i} =& \mathcal{N}_{\mathcal{P}_0\to \mathcal{S}_i} + \left\langle \mathcal{N}_{\bm{\Phi}_i} \right\rangle_{\mathrm{V}} -\left\langle \mathcal{N}_{\bm{\Phi}_0}  \right\rangle_{\mathrm{V}}\\
= & \mathcal{N}_{\mathcal{P}_0\to \mathcal{S}_*}+\mathcal{N}_{\bm{\Phi}_*\to \mathcal{S}_i}+ \left\langle \mathcal{N}_{\bm{\Phi}_i} \right\rangle_{\mathrm{V}} -\left\langle \mathcal{N}_{\bm{\Phi}_0}  \right\rangle_{\mathrm{V}}\,,
\label{eq:Pjoint:large-volume}
\eea
where $i=1,2$.
Here $\mathcal{S}_*$ is made of the field-phase space locations $\bm{\Phi}_*$ such that $(\sigma H)^{-3}(\bm{\Phi}_*) \left\langle \ee^{3 \mathcal{N}_{\bm{\Phi}_*}}\right\rangle=(\tilde{r}/2)^3$, and, analogously, $\mathcal{S}_i$ are made of the field-phase space locations $\bm{\Phi}_i$ such that $(\sigma H)^{-3}(\bm{\Phi}_i) \left\langle \ee^{3 \mathcal{N}_{\bm{\Phi}_i}}\right\rangle=R_i^3$.

Moreover, we recall that $\mathcal{N}_{\mathcal{P}_0\to \mathcal{S}_*}$ and $\mathcal{N}_{\bm{\Phi}_*\to \mathcal{S}_i}$ are first-passage times that need to be volume weighted. The reason is that, when decomposing $\mathcal{N}_{\mathcal{P}_0\to \mathcal{S}_i}= \mathcal{N}_{\mathcal{P}_0\to \mathcal{S}_*}+\mathcal{N}_{\bm{\Phi}_*\to \mathcal{S}_i}$, $\mathcal{N}_{\mathcal{P}_0\to \mathcal{S}_*}$ and $\mathcal{N}_{\bm{\Phi}_*\to \mathcal{S}_i}$ are two first-passage times that are statistically independent because of the Markovian nature of the stochastic process, see \App{app:sub:vanishing:mean}. Moreover, as argued in \Sec{sec:one:point}, $\mathcal{N}_{\mathcal{P}_0\to \mathcal{S}_i}$ needs to be volume weighted, and as shown around \Eq{eq:convolv:vol:weighting}, volume weighting and convolution are commuting operations. This is why $\mathcal{N}_{\mathcal{P}_0\to \mathcal{S}_*}$ and $\mathcal{N}_{\bm{\Phi}_*\to \mathcal{S}_i}$ can be volume weighted separately.

In practice, this again gives rise to a simple algorithm to sample the joint distribution $P(\zeta_{R_1},\zeta_{R_2})$: from $\bm{\Phi}_0$, simulate a realisation until it crosses $\mathcal{S}_*$ and record the value of $\bm{\Phi}_*$ and $\mathcal{N}_{\bm{\Phi}_0\to \mathcal{S}_*}$. Then, from $\bm{\Phi}_*$, simulate one realisation until it crosses $\mathcal{S}_1$, and record the value of $\bm{\Phi}_1$ and $\mathcal{N}_{\bm{\Phi}_*\to \mathcal{S}_1}$. Likewise, simulate another realisation from $\bm{\Phi}_*$ until it crosses $\mathcal{S}_2$, and record the value of $\bm{\Phi}_2$ and $\mathcal{N}_{\bm{\Phi}_*\to \mathcal{S}_2}$. From these recorded values, evaluate $\zeta_{R_1}$ and $\zeta_{R_2}$ via \Eq{eq:Pjoint:large-volume}, and weight the tuple $(\zeta_{R_1},\zeta_{R_2})$ by $\ee^{3(\mathcal{N}_{\bm{\Phi}_0\to \mathcal{S}_*}+\mathcal{N}_{\bm{\Phi}_*\to \mathcal{S}_1}+\mathcal{N}_{\bm{\Phi}_*\to \mathcal{S}_2})}$. Finally, repeat the whole procedure until a sufficiently large number of tuples are obtained, over which the joint distribution can be reconstructed.

The result of this procedure can be mathematically expressed as follows. We first note that $P(\zeta_{R_1},\zeta_{R_2})$ can be formally written as
\bea
P\left(\zeta_{R_1},\zeta_{R_2}\vert\bm{\Phi}_0\right) =& \int\dd\bm{\Phi}_*\dd\bm{\Phi}_1\dd\bm{\Phi}_2\dd\mathcal{N}_{\bm{\Phi}_0\to \mathcal{S}_*}\dd\mathcal{N}_{\bm{\Phi}_*\to \mathcal{S}_1}\dd\mathcal{N}_{\bm{\Phi}_*\to \mathcal{S}_2} \\
& \quad P\left(\bm{\Phi}_*, \bm{\Phi}_1,\bm{\Phi}_2,\mathcal{N}_{\bm{\Phi}_0\to \mathcal{S}_*},\mathcal{N}_{\bm{\Phi}_*\to \mathcal{S}_1},\mathcal{N}_{\bm{\Phi}_*\to \mathcal{S}_2}\vert\bm{\Phi}_0\right)\\
& \quad\delta_{\mathrm{D}}\left[\calN_{\bm{\Phi}_0 \rightarrow \mathcal{S}_*}+\calN_{\bm{\Phi}_* \rightarrow \mathcal{S}_1}-\langle \calN_{\bm{\Phi}_0} \rangle_{\mathrm{V}}+\langle \calN_{\bm{\Phi}_1} \rangle_{\mathrm{V}}-\zeta_{R_1}\right]\\
& \quad \delta_{\mathrm{D}}\left[\calN_{\bm{\Phi}_0 \rightarrow \mathcal{S}_*}+\calN_{\bm{\Phi}_* \rightarrow \mathcal{S}_2}-\langle \calN_{\bm{\Phi}_0} \rangle_{\mathrm{V}} +\langle \calN_{\bm{\Phi}_2} \rangle_{\mathrm{V}}-\zeta_{R_2}\right]\, ,
\label{eq:2pt:gen}
\eea
where the Dirac distributions implement \Eq{eq:Pjoint:large-volume}. Now, making use of the fact that, once $\bm{\Phi}_*$ is fixed, $\bm{\Phi}_1$ and $\bm{\Phi}_2$ are statistically independent, and once all three are fixed, $\mathcal{N}_{\bm{\Phi}_0\to \mathcal{S}_*}$, $\mathcal{N}_{\bm{\Phi}_*\to \mathcal{S}_1}$ and $\mathcal{N}_{\bm{\Phi}_*\to \mathcal{S}_2}$ are also independent, one has
\bea
P\left(\bm{\Phi}_*, \bm{\Phi}_1,\bm{\Phi}_2,\mathcal{N}_{\bm{\Phi}_0\to \mathcal{S}_*},\mathcal{N}_{\bm{\Phi}_*\to \mathcal{S}_1},\mathcal{N}_{\bm{\Phi}_*\to \mathcal{S}_2}\vert\bm{\Phi}_0\right)=&\\ &  \kern-21em
\PfptlV{\bm{\Phi}_0\to \mathcal{S}_*}\left(\mathcal{N}_{\bm{\Phi}_0\to \mathcal{S}_*},\bm{\Phi}_*\right) \PfptlV{\bm{\Phi}_*\to \mathcal{S}_1}\left(\mathcal{N}_{\bm{\Phi}_*\to \mathcal{S}_1},\bm{\Phi}_1\right)\PfptlV{\bm{\Phi}_*\to \mathcal{S}_2}\left(\mathcal{N}_{\bm{\Phi}_*\to \mathcal{S}_2},\bm{\Phi}_2\right) ,
\eea
where $P^{\mathrm{V}}_{\sss{\mathrm{FPTL}}}$ is the first-passage time and first-passage location distribution introduced in \Eq{eq:joint:Nfpt:phi*}. Inserting this relation into \Eq{eq:2pt:gen}, one can write 
\bea
P\left(\zeta_{R_1},\zeta_{R_2}\vert\bm{\Phi}_0\right) =& \int\dd\bm{\Phi}_*\dd\bm{\Phi}_1\dd\bm{\Phi}_2\dd\mathcal{N}_{\bm{\Phi}_0\to \mathcal{S}_*} 
\ \PfptlV{\bm{\Phi}_0\to \mathcal{S}_*}\left(\mathcal{N}_{\bm{\Phi}_0\to \mathcal{S}_*},\bm{\Phi}_*\right) 
\\ &  
\PfptlV{\bm{\Phi}_*\to \mathcal{S}_1}\left(\zeta_{R_1}-\calN_{\bm{\Phi}_0 \rightarrow \mathcal{S}_*}+\langle \calN_{\bm{\Phi}_0} \rangle_{\mathrm{V}}-\langle \calN_{\bm{\Phi}_1} \rangle_{\mathrm{V}} ,\bm{\Phi}_1\right)\\ &
\PfptlV{\bm{\Phi}_*\to \mathcal{S}_2}\left(\zeta_{R_2}-\calN_{\bm{\Phi}_0 \rightarrow \mathcal{S}_*}+\langle \calN_{\bm{\Phi}_0} \rangle_{\mathrm{V}} -\langle \calN_{\bm{\Phi}_2} \rangle_{\mathrm{V}},\bm{\Phi}_2\right) .
\label{eq:2pt:largevol}
\eea

This formula is one of the other main results of the present work. Let us stress that for the large-volume approximation to be valid, not only $R_1^3,R_2^3\gg (\sigma H_\uend)^{-3}$ are required, but also $r\gg R_1,R_2$ because of the caveat mentioned below \Eq{eq:mean:e3N}.

One can check that, upon marginalising over $\zeta_{R_2}$, one readily recovers the one-point distribution~\eqref{eq:1pt:large:vol:final}, using the fact that
\bea
\kern-0.2em \PfptlV{\bm{\Phi}_0\to \mathcal{S}_1}\left(\mathcal{N}_{\bm{\Phi}_0\to \mathcal{S}_1},\bm{\Phi}_1\right)
= &\\ & \kern-10em
\int\dd\bm{\Phi_*}\dd\mathcal{N}_{\bm{\Phi}_0\to \mathcal{S}_*}
\PfptlV{\bm{\Phi}_0\to \mathcal{S}_*}\left(\mathcal{N}_{\bm{\Phi}_0\to \mathcal{S}_*},\bm{\Phi}_*\right)\PfptlV{\bm{\Phi}_*\to \mathcal{S}_1}\left(\mathcal{N}_{\bm{\Phi}_0\to \mathcal{S}_1}-\mathcal{N}_{\bm{\Phi}_0\to \mathcal{S}_*} ,\bm{\Phi}_1\right) .
\eea
This again translates the fact that, when $\bm{\Phi}_*$ is fixed, $\mathcal{N}_{\bm{\Phi}_0\to \mathcal{S}_*}$ and $\mathcal{N}_{\bm{\Phi}_*\to \mathcal{S}_1}$ are statistically independent.

\section{Single-clock models}
\label{sec:single:clock:models}

In this section, we specialise the large-volume approximation introduced above to ``single-clock'' scenarios, where $\bm{\Phi}$ reduces to a single coordinate $\phi$. This corresponds to single-field models of inflation, along a dynamical attractor such as slow roll, where the momentum conjugated to $\phi$ becomes a fixed function of $\phi$ and the stochastic dynamics is effectively one dimensional~\cite{Grain:2017dqa}.  

The main simplification that occurs in this class of models is that hypersurfaces of fixed mean final volume reduce to single points. In the large-volume approximation, the backward field value $\phi_*$ thus becomes a deterministic variable, which is moreover independent of $\phi_0$. In that limit, the one-point distribution~\eqref{eq:1pt:large:vol:final} thus reduces to
\bea
\label{eq:1pt:single:clock}
P\left(\zeta_R\right) = \PfptV{\phi_0\to \phi_*}\left(\zeta_R-\left\langle \mathcal{N}_{\phi_*}\right\rangle_{\mathrm{V}}+\left\langle \mathcal{N}_{\phi_0}\right\rangle_{\mathrm{V}}\right) .
\eea
Likewise, for the two-point distribution, $\phi_1$, $\phi_2$ and $\phi_*$ become deterministic and independent of $\phi_0$, and \Eq{eq:2pt:largevol} reduces to
\bea
\label{eq:2pt:single:clock}
P(\zeta_{R_1},\zeta_{R_2}) = & \int\dd\mathcal{N}_{\phi_0\to\phi_*} \PfptV{\phi_0\to\phi_*}\left(\mathcal{N}_{\phi_0\to\phi_*}\right)
\\ & \quad
\PfptV{\phi_*\to\phi_1}\left(\zeta_{R_1}-\mathcal{N}_{\phi_0\to\phi_*}+\left\langle \mathcal{N}_{\phi_0} \right\rangle_{\mathrm{V}}-\left\langle \mathcal{N}_{\phi_1} \right\rangle_{\mathrm{V}}\right)
\\ & \quad
\PfptV{\phi_*\to\phi_2}\left(\zeta_{R_2}-\mathcal{N}_{\phi_0\to\phi_*}+\left\langle \mathcal{N}_{\phi_0} \right\rangle_{\mathrm{V}}-\left\langle \mathcal{N}_{\phi_2} \right\rangle_{\mathrm{V}}\right) .
\eea
Both expressions only involve first-passage time distributions, subject to the adjoint Fokker-Planck equation~\eqref{eq:Fokker-Planck}, and which can be readily sampled~\cite{Jackson:2022unc}. This is a substantial simplification.

\subsubsection*{Power spectrum}

As a by-product of the above formulas, let us derive the power spectrum of the curvature perturbation. This is nothing but the Fourier transform of the two-point correlation function in real space, which can be readily extracted from the two-point distribution function
\bea
\left\langle \zeta_{R_1}  \zeta_{R_2} \right\rangle = & \int \dd\zeta_{R_1}\int \dd\zeta_{R_2} P(\zeta_{R_1},\zeta_{R_2})\zeta_{R_1}\zeta_{R_2} \\
= &
\int \dd\zeta_{R_1}\int \dd\zeta_{R_2}
 \int\dd\mathcal{N}_{\phi_0\to\phi_*} \PfptV{\phi_0\to\phi_*}\left(\mathcal{N}_{\phi_0\to\phi_*}\right) \zeta_{R_1}\zeta_{R_2}
\\ & \quad
\PfptV{\phi_*\to\phi_1}\left(\zeta_{R_1}-\mathcal{N}_{\phi_0\to\phi_*}+\left\langle \mathcal{N}_{\phi_0} \right\rangle_{\mathrm{V}}-\left\langle \mathcal{N}_{\phi_1} \right\rangle_{\mathrm{V}}\right)
\\ & \quad
\PfptV{\phi_*\to\phi_2}\left(\zeta_{R_2}-\mathcal{N}_{\phi_0\to\phi_*}+\left\langle \mathcal{N}_{\phi_0} \right\rangle_{\mathrm{V}}-\left\langle \mathcal{N}_{\phi_2} \right\rangle_{\mathrm{V}}\right) ,
\eea
where in the second equality \Eq{eq:2pt:single:clock} has been used.
Notice moreover that in the above expression the integration domain is set by the requirement that the arguments of the first-passage time distributions are positive quantities. Performing the change of integration variable $\zeta_{R_1}\to\mathcal{N}_{\phi_*\to\phi_1}=\zeta_{R_1}- \mathcal{N}_{\phi_0\to \phi_*}  + \left\langle \mathcal{N}_{\phi_0}\right\rangle_{\mathrm{V}}  - \left\langle \mathcal{N}_{\phi_1} \right\rangle_{\mathrm{V}}$ and likewise for $\zeta_{R_2}$, leads to
\bea\label{eq:pow:spectrum:dim}
\left\langle \zeta_{R_1}  \zeta_{R_2} \right\rangle = &
\int\dd\mathcal{N}_{\phi_*\to\phi_1}\int \dd\mathcal{N}_{\phi_*\to\phi_2}
\int \dd   \mathcal{N}_{\phi_0\to \phi_*} \\ & \kern-2em
\left(\mathcal{N}_{\phi_*\to\phi_1}+ \mathcal{N}_{\phi_0\to \phi_*}  - \left\langle \mathcal{N}_{\phi_0}\right\rangle_{\mathrm{V}} + \left\langle \mathcal{N}_{\phi_1} \right\rangle_{\mathrm{V}}\right)
 \left(\mathcal{N}_{\phi_*\to\phi_2}+ \mathcal{N}_{\phi_0\to \phi_*}  - \left\langle \mathcal{N}_{\phi_0}\right\rangle_{\mathrm{V}} + \left\langle \mathcal{N}_{\phi_2} \right\rangle_{\mathrm{V}}\right) \\ &  \kern-2em
 \PfptV{\phi_0\to \phi_*}\left( \mathcal{N}_{\phi_0\to \phi_*}\right) 
\PfptV{\phi_*\to \phi_1}\left(\mathcal{N}_{\phi_*\to\phi_1}\right)
\PfptV{\phi_*\to \phi_2}\left(\mathcal{N}_{\phi_*\to\phi_2}\right)\,.
\eea
Since $\phi_*$, $\phi_1$ and $\phi_2$ are fixed in single-clock models, $\mathcal{N}_{\phi_0\to\phi_*}$, $\mathcal{N}_{\phi_*\to\phi_1}$ and $\mathcal{N}_{\phi_*\to\phi_2}$ are independent stochastic variables, hence\footnote{If $X$ and $Y$ are independent random variables with distribution functions $P_X(X)$ and $P_Y(Y)$, their product $Z=XY$ has  distribution $P(Z) =  \int\dd X \dd Y P(X,Y)\delta_{\mathrm{D}}(Z-X Y)= \int\dd X \dd Y P_X(X) P_Y(Y)\delta_{\mathrm{D}}(Z-X Y)= \int\frac{\dd X}{X}  P_X(X) P_Y(Z/X)$. Its mean value thus reads $\langle Z \rangle =  \int P(Z) Z \dd Z = \int \dd Z \int \frac{\dd X}{X} P_X(X) P_Y(Z/X) Z=  \int \dd X P_X(X)  X \int \frac{\dd Z}{X} P_Y(Z/X) \frac{Z}{X}=  \langle X \rangle \langle Y \rangle$.}
\bea
\label{eq:mean:product}
\int & \dd\mathcal{N}_{\phi_*\to\phi_1}\int \dd\mathcal{N}_{\phi_*\to\phi_2} \PfptV{\phi_*\to \phi_1}\left(\mathcal{N}_{\phi_*\to\phi_1}\right)\PfptV{\phi_*\to \phi_2}\left(\mathcal{N}_{\phi_*\to\phi_2}\right)\mathcal{N}_{\phi_*\to\phi_1} \mathcal{N}_{\phi_*\to\phi_2}=\\
&\langle\mathcal{N}_{\phi_*\to\phi_1} \rangle_{\mathrm{V}} \langle\mathcal{N}_{\phi_*\to\phi_2} \rangle_{\mathrm{V}} \,,
\eea
with similar expressions for analogous terms. This gives rise to
\bea
\left\langle \zeta_{R_1}  \zeta_{R_2} \right\rangle = &
\left(\left\langle \mathcal{N}_{\phi_1} \right\rangle_{\mathrm{V}}-\left\langle \mathcal{N}_{\phi_0} \right\rangle_{\mathrm{V}}\right)
\left(\left\langle \mathcal{N}_{\phi_2} \right\rangle_{\mathrm{V}}-\left\langle \mathcal{N}_{\phi_0} \right\rangle_{\mathrm{V}}\right)
\\ &
+\left(\left\langle \mathcal{N}_{\phi_1} \right\rangle_{\mathrm{V}}-\left\langle \mathcal{N}_{\phi_0} \right\rangle_{\mathrm{V}}\right)
\left(\left\langle \mathcal{N}_{\phi_*\to\phi_2}  \right\rangle_{\mathrm{V}}+\left\langle \mathcal{N}_{\phi_0\to\phi_*}  \right\rangle_{\mathrm{V}} \right)\\ &
+\left(\left\langle \mathcal{N}_{\phi_2} \right\rangle_{\mathrm{V}}-\left\langle \mathcal{N}_{\phi_0} \right\rangle_{\mathrm{V}}\right)
\left(\left\langle \mathcal{N}_{\phi_*\to\phi_1}  \right\rangle_{\mathrm{V}}+\left\langle \mathcal{N}_{\phi_0\to\phi_*}  \right\rangle_{\mathrm{V}} \right)\\ &
+ \left\langle \mathcal{N}_{\phi_*\to\phi_1} \right\rangle_{\mathrm{V}} \left\langle \mathcal{N}_{\phi_*\to\phi_2} \right\rangle_{\mathrm{V}}
+\left\langle \mathcal{N}_{\phi_*\to\phi_1} \right\rangle_{\mathrm{V}} \left\langle \mathcal{N}_{\phi_0\to\phi_*} \right\rangle_{\mathrm{V}}\\ &
+ \left\langle \mathcal{N}_{\phi_0\to\phi_*} \right\rangle_{\mathrm{V}} \left\langle \mathcal{N}_{\phi_*\to\phi_2} \right\rangle_{\mathrm{V}}
+ \left\langle \mathcal{N}_{\phi_0\to\phi_*} ^2 \right\rangle_{\mathrm{V}} \,.
\eea
The next step is to use the result shown in \Eq{eq:convolv:vol:weighting}, namely that convolution and volume-weighting operations commute, which allows to write identities of the type $\left\langle \mathcal{N}_{\phi_0\to\phi_*}  \right\rangle_{\mathrm{V}} +\left\langle \mathcal{N}_{\phi_*\to\phi_2}  \right\rangle_{\mathrm{V}}=\left\langle \mathcal{N}_{\phi_0\to\phi_2}  \right\rangle_{\mathrm{V}}$. Using such identities repeatedly, the above expression simplifies into
\bea
\label{eq:two:point:corr}
\left\langle \zeta_{R_1}  \zeta_{R_2} \right\rangle = &
  \left\langle \mathcal{N}_{\phi_0\to\phi_*} ^2 \right\rangle_{\mathrm{V}} - \left\langle \mathcal{N}_{\phi_0\to\phi_*}  \right\rangle_{\mathrm{V}}^2 \\ \equiv &\left\langle \delta\mathcal{N}^2_{\phi_0\to\phi_*}  \right\rangle_{\mathrm{V}}=
\left\langle \delta\mathcal{N}_{\phi_0}^2 \right\rangle_{\mathrm{V}}-\left\langle \delta\mathcal{N}_{\phi_*}^2 \right\rangle_{\mathrm{V}}
\,,
\eea
where we have introduced the variance notation $\langle\delta \mathcal{N}^2_\phi \rangle \equiv \langle \mathcal{N}^2_\phi \rangle-\langle \mathcal{N}_\phi \rangle^2$. 

Interestingly, $\phi_1$ and $\phi_2$ have dropped from the final result, which therefore depends neither on $R_1$ nor on $R_2$. This seemingly surprising fact can be explained as follows. In Fourier space, $\zeta_{R_1}(\vecx_1)$ and $\zeta_{R_2}(\vecx_2)$ can be expanded as
\bea
\label{eq:zetaR:Fourier}
\zeta_{R_i}(\vec{x}_i) & = \int\frac{\dd \vec{k}}{\left(2\pi\right)^{3/2}}\, \zeta_{\vec{k}}\, e^{i \vec{k}\cdot\vec{x}_i}\, \widetilde{W}\left(\frac{kR_i}{a}\right) ,
\eea
see \Eq{eq:general:coarse:graining:Fourier}. This leads to 
\bea
\left\langle \zeta_{R_1}  \zeta_{R_2} \right\rangle = & 
\int\frac{\dd\vec{k}_1}{\left(2\pi\right)^{3/2}}
\int\frac{\dd\vec{k}_2}{\left(2\pi\right)^{3/2}}
\left\langle \zeta_{\vec{k}_1} \zeta_{\vec{k}_2}^\dagger \right\rangle
e^{i \left(\vec{k}_1\cdot\vec{x}_1-\vec{k}_2\cdot\vec{x}_2\right)} \widetilde{W}\left(\frac{k_1R_1}{a}\right)\widetilde{W}\left(\frac{k_2R_2}{a}\right)\, ,
\eea
with $\langle \zeta_{\vec{k}_1} \zeta_{\vec{k}_2}^\dagger \rangle = (2\pi)^2/k_1^3 \calP_\zeta(k_1) \delta_{\mathrm{D}}(\vec{k}_1-\vec{k_2})$, where $\calP_\zeta(k_1)$ is the reduced power spectrum of the comoving curvature perturbation $\zeta$. The Dirac distribution allows one to integrate out $\vec{k}_2$, and after integrating over the angular degrees of freedom contained in $\vec{k_1}$ one obtains
\bea
\label{eq:zetaR1:zetaR2:WWsinc}
\left\langle \zeta_{R_1}  \zeta_{R_2} \right\rangle = &
 \int_0^\infty \dd \ln k\,  \calP_\zeta(k)\widetilde{W}\left(\frac{kR_1}{a}\right)\widetilde{W}\left(\frac{kR_2}{a}\right) \mathrm{sinc}\left(\frac{kr}{a}\right) .
\eea

At this point, a few words must be said about the window functions. As mentioned in footnote~\ref{footnote:white}, for the noise to be white in the Langevin equation, the window function has been assumed to be of the Heaviside type in Fourier space. Moreover, the noise correlator~\eqref{eq:noise:correlator} can be computed between two distinct spatial locations, and one finds $\langle \xi_i(\vec{x}_1,N)\xi_j(\vec{x}_2,N)\rangle = \langle \xi_1(\vec{x}_1,N)\xi_2(\vec{x}_2,N)\rangle \mathrm{sinc}(k_\sigma r/a )$. This means that, in principle, the noises acting on two separate patches are correlated. In practice, these correlations are not accounted for in the stochastic-inflation formalism, which amounts to approximating the cardinal function by a Heaviside function. Performing these replacements in the above expression one can write
\bea
\left\langle \zeta_{R_1}  \zeta_{R_2} \right\rangle 
= &
 \int_0^\infty \dd \ln k \, \calP_\zeta(k)\theta\left(1-\frac{kR_1}{a}\right)\theta\left(1-\frac{kR_2}{a}\right) \theta\left(1-\frac{kr}{a}\right)\\ 
 = &
 \int_0^{a/r} \dd \ln k \, \calP_\zeta(k)
 \, ,
 \label{eq:two:pt:correl:Fourier}
\eea
where $\theta(x)=1$ if $x\geq 0$ and $0$ otherwise is the Heaviside function. In the second line, we have used that $r>R_1,R_2$, and the resulting expression makes it clear why the two-point correlation function does not depend on $R_1$ and $R_2$, once the filtering functions implicitly employed in the stochastic formalism are implemented. 

By differentiating both sides of \Eq{eq:two:pt:correl:Fourier} with respect to $r$ one also obtains
\bea\label{eq:pow:spec:interm}
\calP_\zeta(k) = & -\left.\frac{\partial}{\partial \ln r}\left\langle \zeta_{R_1}  \zeta_{R_2} \right\rangle  \right\vert_{r=a_\uend/k}
=  \left.\frac{\partial}{\partial \ln r}\left\langle \delta\mathcal{N}^2_{\phi_*} \right\rangle_{\mathrm{V}}\right\vert_{r=a_\uend/k}\,,
\eea
where \Eq{eq:two:point:corr} has been inserted in the last equality. The derivative with respect to $r$ can be expressed in terms of a derivative with respect to $\phi_*$, recalling that in the large-volume approximation these two are related through $\langle \ee^{3 \mathcal{N}_{\phi_*}} \rangle= [\tilde{r} \sigma H(\phi_*)/2]^3$, see \Eq{eq:P(V):appr}. Recalling that $\tilde{r}=r+R_1+R_2$, this leads to
\bea
\label{eq:power:spectrum:single:clock:large:volume}
\mathcal{P}_\zeta(k)=\frac{r}{\tilde{r}}\left. \left[\frac{1}{3}\frac{\partial}{\partial\phi_*}\ln\left\langle \ee^{3\mathcal{N}_{\phi_*}} \right\rangle-\frac{\partial}{\partial\phi_*} \ln H(\phi_*)\right]^{-1} \frac{\partial}{\partial\phi_*}\left\langle \delta\mathcal{N}^2_{\phi_*} \right\rangle_{\mathrm{V}} \right\vert_{\langle \ee^{3 \mathcal{N}_{\phi_*}} \rangle^{1/3}= \frac{1}{2}\frac{\tilde{r}}{r}\frac{a_\uend \sigma H(\phi_*)}{k}}\, .
\eea
In the large-volume limit, as explained in \Sec{sub:two:point}, the condition $r\gg R_1,R_2$ must be fulfilled, hence one shall replace $r/\tilde{r}\simeq 1$ in the above expression. In single-field slow-roll inflation, one may also replace $\partial\ln H/\partial\phi \simeq \sqrt{\epsilon_1/2}/\Mp$, hence this term provides a slow-roll suppressed correction.

Note that this expression differs from the one established in \Refa{Ando:2020fjm} for the reason highlighted around \Eq{eq:Tada:Vennin}, namely the fact that in \Refa{Ando:2020fjm} the final volume is computed along a reference trajectory while in the large-volume approximation it is replaced by its stochastic average. Incidentally, \Eq{eq:power:spectrum:single:clock:large:volume} is closer to the early expressions derived in \Refs{Fujita:2013cna, Vennin:2015hra}, with which it would coincide at leading order in slow roll if volume weighting had been omitted, and if $\mathcal{S}_*$ were defined as fixed-$\langle \mathcal{N} \rangle$ hypersurfaces instead of fixed-$\langle \ee^{3\mathcal{N}} \rangle$ hypersurfaces.
Finally, it is worth mentioning that the power spectrum can also be obtained from the knowledge of the coarse-grained one-point distribution function only. This calculation is performed in \App{app:power:spectrum:one:point}, where we show that the same result as above is obtained, which confirms the consistency of our approach.

\section{Applications }
\label{sec:applications}

Let us now apply the formalism developed so far to a few toy models, in order to illustrate our approach concretely, and to get a first insight into the amount of clustering that is to be expected from quantum diffusion. Our goal is not to provide a comprehensive parameter-space exploration of these models, but rather to obtain qualitative results that will allow us to identify relevant directions that require further exploration. 

In practice, we consider single-field slow-roll models of inflation, where the Langevin equation~\eqref{eq:Langevin} reduces to 
\bea
\label{eq:Langevin:SR}
\frac{\dd\phi}{\dd N} = -\frac{V'}{3H^2}+\frac{H}{2\pi} \xi
\eea 
where $V(\phi)$ is the potential energy stored in the inflaton field $\phi$ and $V'$ is its derivative with respect to $\phi$, and the Hubble rate is given by the Friedmann's equation $H^2=V/(3\Mp^2)$ at leading order in slow roll. Here, the white Gaussian noise $\xi$ is normalised to unit variance, so $\langle \xi(N) \xi(N')\rangle=\delta_{\mathrm{D}}(N-N')$. With the It\^o prescription this gives rise to the adjoint Fokker-Planck operator~\cite{risken1989fpe}
\bea
\Lfp^\dagger(\phi) = -\Mp^2 \frac{v'}{v} \frac{\partial}{\partial\phi}  + v\frac{\partial^2}{\partial\phi^2}\, ,
\eea
where $v\equiv V/(24\pi^2\Mp^4)$ denotes the rescaled inflationary potential.
\subsection{Quantum well}
\label{sec:quantum:well}
\begin{figure}[H] 
	\centering
	\includegraphics[width=0.7\hsize]{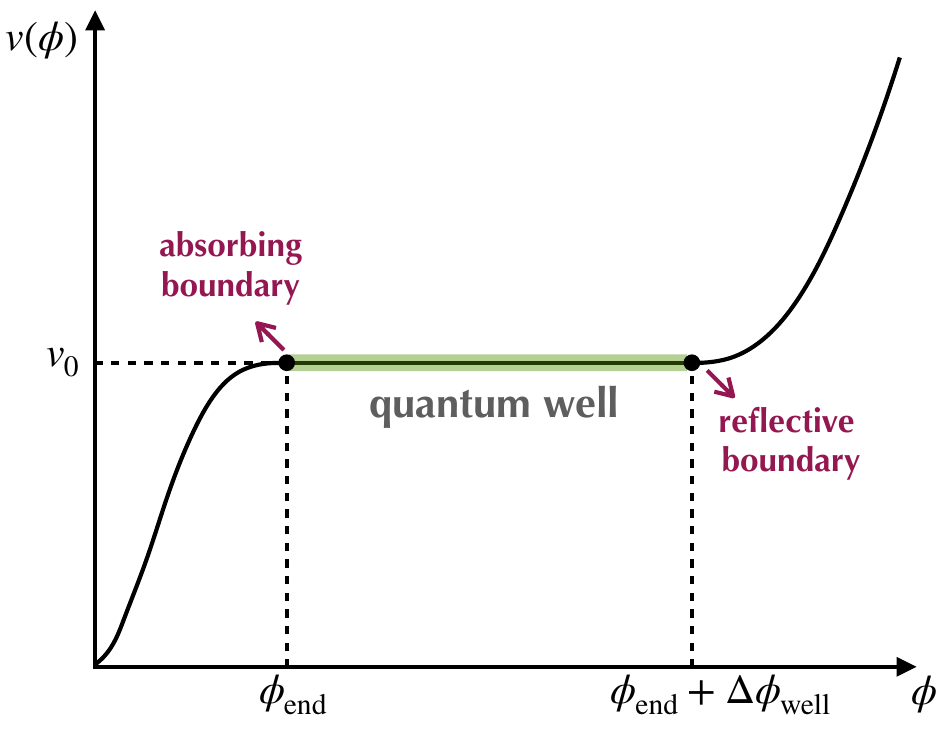}
	\caption{Quantum-well toy model.}
	\label{fig:quantum:well}
\end{figure}

We start by considering the case where the inflationary potential is constant, $v=v_0$, between $\phi=\phi_\uend$ (where inflation ends, hence an absorbing boundary is placed) and $\phi=\phi_\uend+\Delta\phi_{\uwell}$ (where the potential is assumed to become classical-drift dominated, hence a reflective boundary is placed). A sketch of this  potential is displayed in \Fig{fig:quantum:well}.
\noindent This model is simple enough to yield analytical results for the first-passage time problem, while displaying key features of more generic setups~\cite{Pattison:2017mbe, Ezquiaga:2019ftu}, which makes it natural to consider first. 

The solution to the adjoint Fokker-Planck equation~\eqref{eq:Fokker-Planck}, for the first-passage time through the absorbing boundary at $\phi_\uend$, is given by~\cite{Pattison:2017mbe}
\bea
\label{eq:Pfpt:quantum:well}
\Pfpt{\phi}(\mathcal{N}) &= -\frac{\pi}{2\mu^2} \vartheta_2'\left(\frac{\pi}{2} x,\ee^{-\frac{\pi^2}{\mu^2} \mathcal{N}}\right) .
\eea
Here $x=(\phi-\phi_\uend)/\Delta\phi_\uwell$ is the rescaled field value ($x$ varies between $0$ and $1$ within the well) and we have introduced 
\bea
\label{eq:mu:def}
\mu^2=\frac{\Delta\phi_\uwell^2}{v_0\Mp^2}\, .
\eea
One thus sees that the distribution of the random variable $\mathcal{N}/\mu^2$ has a universal profile that does not depend on any other parameter. Finally, $\vartheta_2$ is the second elliptic theta function, $\vartheta_2(z,q)=2\sum_{n=0}^\infty q^{(n+1/2)^2}\cos[(2n+1)z]$, and $\vartheta_2'(z,q)$ is the derivative of $\vartheta_2(z,q)$ with respect to $z$. If the first-passage time through another field value than $\phi_\uend$ needs to be computed, say through $\phi_*$, \Eq{eq:Pfpt:quantum:well} can still be employed, with the rescaling $x\to (x-x_*)/(1-x_*)$, and $\mu\to (1-x_*)\mu$.

From \Eq{eq:Pfpt:quantum:well}, the mean number of \efolds elapsed in the quantum well, when starting from $x$, is given by $\langle \mathcal{N}\rangle = \mu^2/2x(2-x)$, and the mean volume by 
\bea 
\label{eq:flat:well:mean:vol}
\langle \ee^{3\mathcal{N}}\rangle = \frac{\cos\left[\sqrt{3}\mu(1-x)\right]}{\cos\left(\sqrt{3}\mu\right)}\, .
\eea
Notice that this latter quantity is well defined only for $\mu<\mu_\uc\equiv\pi/(2\sqrt{3})$. The reason is that, when $\mu\geq \mu_\uc$, the tail in \Eq{eq:Pfpt:quantum:well} decays more slowly than $\ee^{-3\mathcal{N}}$, hence the mean volume does not converge. This is the regime of eternal inflation discussed below \Eq{eq:P:zetacg}, which we will not consider. Moreover, if $\mu\ll\mu_\uc$, then the mean volume is of order one (in $\sigma$-Hubble units) and the large-volume approximation developed in \Sec{subsec:Large:vol:approx} does not apply. This is why, for the mean volumes to be large though finite, in what follows we work with values of $\mu$ close to (but smaller than) $\mu_\uc$. 

To get \Eq{eq:flat:well:mean:vol}, one can integrate \Eq{eq:Pfpt:quantum:well} against $\ee^{3\mathcal{N}}$ term by term, using the definition of the elliptic function, and resum, but a more direct route is to use the characteristic function, defined as
\bea\label{eq:def:chi:Fourier}
\chi_\calN (t,\phi)=\left\langle \ee^{it\mathcal{N}} \right\rangle = \int_{-\infty}^{\infty} \dd \calN e^{i t \calN} \Pfpt{\phi}(\mathcal{N})\, .
\eea
The characteristic function depends on the dummy parameter $t$ and, from the above definition, it is nothing but the Fourier transform of the first-passage time distribution. From \Eq{eq:Fokker-Planck}, one can readily show that it satisfies $\Lfp^\dagger(\phi)\chi_\calN (t,\phi)=-it \chi_\calN (t,\phi)$, which, contrary to \Eq{eq:Fokker-Planck}, is an ordinary differential equation. It needs to be solved with the boundary conditions $\chi(t,\phi_{\mathrm{end}})=1$ and 
$\frac{\partial}{\partial \phi} \chi(t,\phi)|_{\phi_\uend+\Delta\phi_\uwell}=0$,
which in the present case leads to
\bea\label{eq:chi:flat:cos}
\chi_\calN(t,\phi)=\frac{\cos{\left[(i t)^{1/2} \mu (x-1)\right]}}{\cos{\left[(i t)^{1/2} \mu\right]}}\, .
\eea
Performing the partial fraction decomposition of $\chi_\calN(t,\phi)$, the Fourier transform can be obtained term by term using the residue theorem, and this is in fact how  \Eq{eq:Pfpt:quantum:well} was obtained~\cite{Ezquiaga:2019ftu}.

As far as the mean volume is concerned, it is simply given by $\langle \ee^{3\mathcal{N}_{\phi}} \rangle = \chi_\mathcal{N}(-3i,\phi)$, hence \Eq{eq:flat:well:mean:vol} directly follows from \Eq{eq:chi:flat:cos}. Moreover, from the definition~\eqref{eq:def:chi:Fourier}, the characteristic function of the volume-weighted first-passage time distribution is simply related to the characteristic function of the non-volume-weighted one by
\bea
\chi_\calN^{\mathrm{V}}(t,\phi) = \frac{\chi_\calN(t-3i,\phi)}{\chi_\calN(-3i,\phi)}\, .
\eea
This result implies that the volume-weighted moments of the first-passage time can be derived as follows. First, from Taylor-expanding the exponential function in \Eq{eq:def:chi:Fourier}, one can see that
\bea
\label{eq:meanN:from:char}
\left\langle \mathcal{N}^n_\phi \right\rangle = i^{-n}\left.\frac{\partial^n}{\partial t^n}\chi_\calN(t,\phi)\right\vert_{t=0}\, ,
\eea
which, for $n=1$, gives the formula $\langle \mathcal{N}\rangle = \mu^2/2x(2-x)$ given above. Applying this identity to the volume-weighted characteristic function, one finds
\bea
\label{eq:meanN:vol:weighted:from:char}
\left\langle \mathcal{N}^n_\phi \right\rangle_{\mathrm{V}} = \frac{i^{-n}}{\chi_\calN(-3i,\phi)}\left.\frac{\partial^n}{\partial t^n}\chi_\calN(t,\phi)\right\vert_{t=-3i}\, .
\eea
With \Eq{eq:chi:flat:cos} in the quantum-well model, for $n=1$ this leads to
\bea\label{eq:flat:N:weight}
\left\langle \mathcal{N} \right\rangle_{\mathrm{V}} = \frac{\mu}{2\sqrt{3}}\left\lbrace\tan\left(\sqrt{3}\mu\right)-\left(1-x\right)\tan\left[\sqrt{3}\mu\left(1-x\right)\right]\right\rbrace .
\eea
Again, one recovers that this expression is well-defined only when $\mu<\mu_{\mathrm{c}}$.

Lastly, we notice that \Eq{eq:flat:N:weight} can also be obtained from \Eq{eq:volMean:N*} in the large-volume approximation, namely, $\left\langle \mathcal{N} \right\rangle_{\mathrm{V}}=\left \langle \calN e^{3 \calN} \right \rangle/\left \langle e^{3\calN} \right \rangle$, where $\left \langle \calN e^{3 \calN} \right \rangle=\int_0^\infty \Pfpt{x}(\calN) \calN e^{3 \calN} \dd \calN$ and $\left \langle e^{3\calN} \right \rangle=\int_0^\infty \Pfpt{x}(\calN) e^{3 \calN} \dd \calN$, where one can use the FPT distribution given in \Eq{eq:Pfpt:quantum:well}. Of course, the results obtained following the two different procedures agree.

\subsubsection*{One-point curvature perturbation}

We are now in a position to evaluate the one-point distribution of the coarse-grained curvature perturbation, in the large-volume limit. Setting $\phi_0$ at the reflective boundary of the quantum well, inserting the above results into \Eq{eq:1pt:single:clock} leads to
\bea
\label{eq:1pt:quantum:well}
P\left(\zeta_R\right) =&-\frac{\pi \cos\left[\sqrt{3}(1-x_*)\mu\right]}{2(1-x_*)^2\mu^2} \vartheta_2'\left(\frac{\pi}{2} ,\ee^{-\frac{\pi^2}{(1-x_*)^2\mu^2} \left\lbrace\zeta_R+\frac{\mu}{2\sqrt{3}}(1-x_*)\tan\left[\sqrt{3}\mu\left(1-x_*\right)\right]\right\rbrace}\right) \\
& \times \ee^{3 \left\lbrace\zeta_R+\frac{\mu}{2\sqrt{3}}(1-x_*)\tan\left[\sqrt{3}\mu\left(1-x_*\right)\right]\right\rbrace} .
\eea

In the above expression, $x_*=(\phi_*-\phi_{\mathrm{end}})/\Delta \phi_{\mathrm{well}}$ is related to $R$ through $\langle\ee^{3\mathcal{N}_{\phi_*}}\rangle = (\sigma R H)^3$, see \Eq{eq:P(V):appr}, which reduces to
\bea\label{eq:rel:xstar:R:flat}
x_*(R) =1- \frac{1}{\sqrt{3} \mu}\mathrm{arccos}\left[ (\sigma R H)^3 \cos\left(\sqrt{3}\mu\right)\right] .
\eea

\begin{figure}[H] 
	\centering
	\includegraphics[width=0.7\hsize]{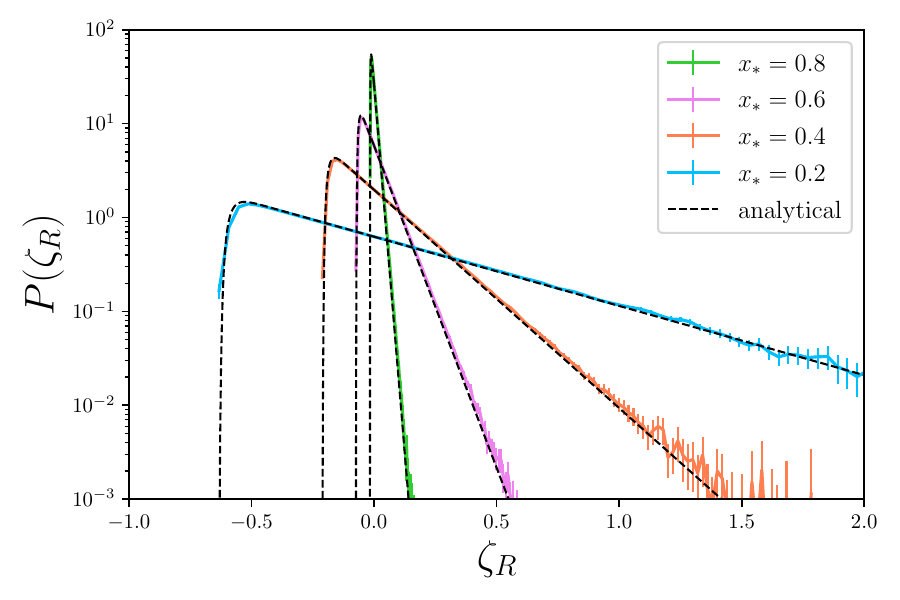}
	\caption{One-point distribution function $P(\zeta_R)$ in the quantum-well potential, for $\mu=\mu_{\mathrm{c}}-0.001$.
  The solid coloured lines are reconstructed from
$10^7$ numerical simulations of the stochastic process described
by the Langevin equation~\eqref{eq:Langevin}, whereas dashed lines stand for the analytical formula~\eqref{eq:1pt:quantum:well}.
Different colours indicate different values of the (rescaled-) field value $x_*$ in the parent patch (see legend), hence
different coarse-graining scales, according to \Eq{eq:rel:xstar:R:flat}.  Error-bars correspond to
the $2\sigma$-estimates of the statistical error obtained via the bootstrap resampling technique with
resampling size $N_{\mathrm{s}} = 1000$.\protect\footnotemark}
	\label{fig:OnePointFlat}
\end{figure}

Using the procedure outlined below \Eq{eq:convolv:vol:weighting}, the one-point distribution function of $\zeta_R$ can be numerically sampled,\footnote{In practice, when reconstructing $P(\zeta_R)$ from a finite sample of Langevin realisations, we use kernel density estimation (we make use of the estimate $\texttt{scipy.stats.gaussian\_kde}$ available in python) with automatic bandwidth determination (or manual determination when necessary) using non-uniform weights.\label{footnote:kernel_density}} and the result of this procedure is compared to the analytical result~\eqref{eq:1pt:quantum:well} in \Fig{fig:OnePointFlat}, where one can check that the agreement is excellent. In particular, one recovers that $P(\zeta_R)$ possesses exponential tails. These are much heavier than Gaussian tails, which implies that the probability to from PBHs is strongly enhanced. 
\footnotetext{The bootstrap method~\cite{Efron:1979bxm}
consists in creating multiple  ``bootstrap samples'' of a given data set, by randomly selecting data points, allowing some to be selected more than once and others not at all. This process is done to mimic the variability in the population and estimate the sampling distribution of a given statistic. In practice, $N_{\mathrm{s}}$ resamples of the same size as the original dataset are generated, and $N_{\mathrm{s}}=10^3-10^4$ is usually considered to be sufficient to provide accurate estimates of the statistical error.}

One also notices that as $x_*$ decreases, hence as $R$ decreases, see \Eq{eq:rel:xstar:R:flat}, the distribution becomes wider, and the tail becomes heavier.\footnote{At small values of $x_*$ the distribution also peaks at rather large, negative values of $\zeta_R$. This is because we work with values of $\mu$ close to $\mu_\uc$, for which the tails are almost flat at small $x_*$. This model thus has to be considered with care, at the qualitative level only.\label{footnote:large:negative:zeta}} This implies that PBHs mostly form at the scales that emerge close to the end point of the quantum well, \ie at the smallest scales produced in the well, in agreement with the finding of \Refa{Tada:2021zzj}. Although the detailed derivation of the PBH mass fraction goes beyond the scope of this work, we thus expect it to be tilted towards smaller masses, in the region where it peaks.

In the far tail, the statistical error becomes substantial, since the statistics is sparser. Let us stress that, although this is always the case, the volume-weighting procedure makes this issue worse. Indeed, the realisations of the Langevin equations are sampled in a non-volume-weighted way, since the dynamics is described forwards, and volume-weighting is only performed at the post-processing level. As a consequence, the tails are up-lifted, and the number of realisations sampled in the tails is much smaller than what the value of the probability density suggests in \Fig{fig:OnePointFlat}. Alternative sampling methods based on importance sampling~\cite{Jackson:2022unc} or stochastic excursions~\cite{Tokeshi:2023swe} might allow one to alleviate this issue, but we leave this possibility for future investigations.

When $\zeta_R-\left\langle \mathcal{N}_{\phi_*}\right\rangle_{\mathrm{V}}+\left\langle \mathcal{N}_{\phi_0}\right\rangle_{\mathrm{V}}\gg\mu^2$, a tail expansion of \Eq{eq:1pt:quantum:well} can be performed, where the elliptic function is dominated by its first term. This gives rise to
\bea
\label{eq:1pt:quantum:well:tail}
P\left(\zeta_R\right) \simeq &\frac{\pi \cos\left[\sqrt{3}(1-x_*)\mu\right]}{(1-x_*)^2\mu^2} 
\ee^{\left[3-\frac{\pi^2}{4(1-x_*)^2\mu^2}\right] \left\lbrace\zeta_R+\frac{\mu}{2\sqrt{3}}(1-x_*)\tan\left[\sqrt{3}\mu\left(1-x_*\right)\right]\right\rbrace} ,
\eea
where one indeed recovers an exponential-tail profile. One can also check that the exponent decays with $x_*$, hence heavier tails are achieved with smaller values of $x_*$, in agreement with the above discussion.

\subsubsection*{Two-point curvature perturbation}

\begin{figure}[H]
  \centering
  \begin{minipage}{.495\textwidth}
 \includegraphics[width=.9\textwidth]{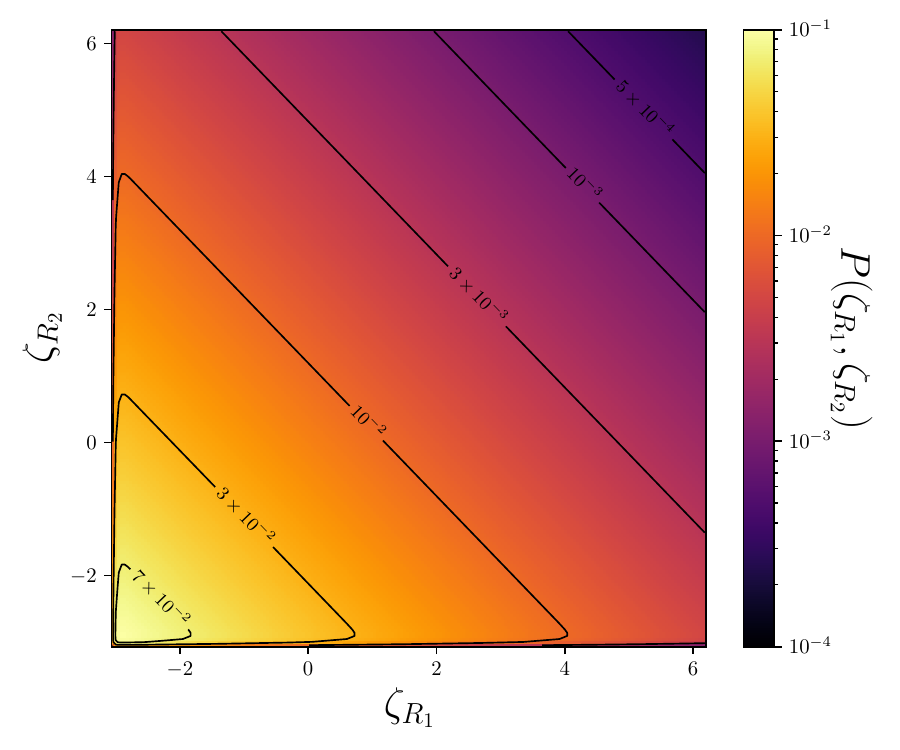}
 \end{minipage}
 \begin{minipage}{.495\textwidth}
 \includegraphics[width=.9\textwidth]{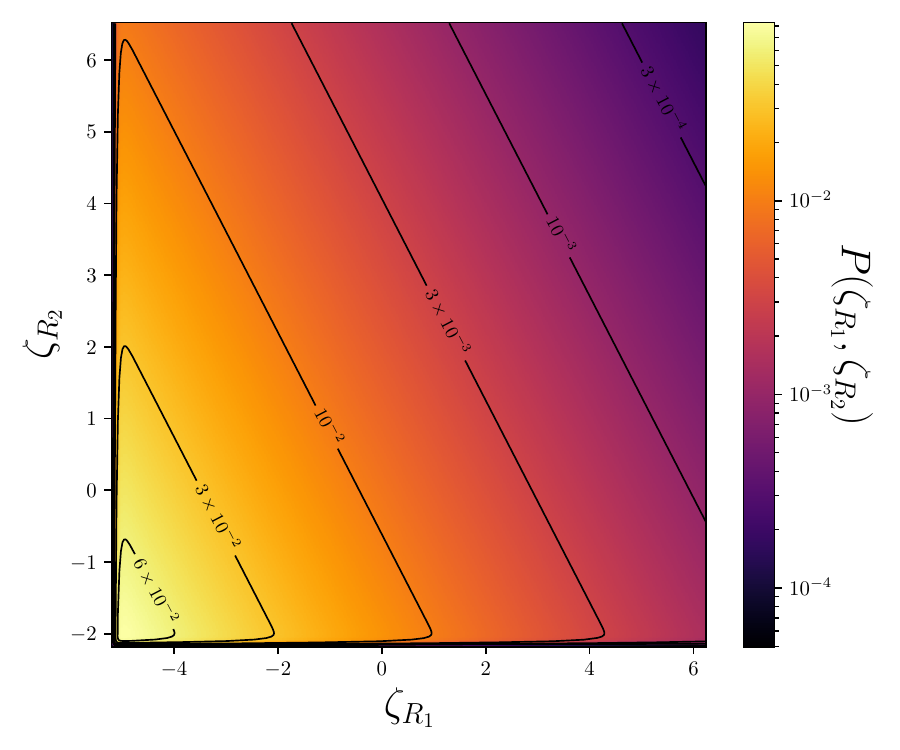}
 \end{minipage}
 \centering
  \begin{minipage}{.495\textwidth}
 \includegraphics[width=.9\textwidth]{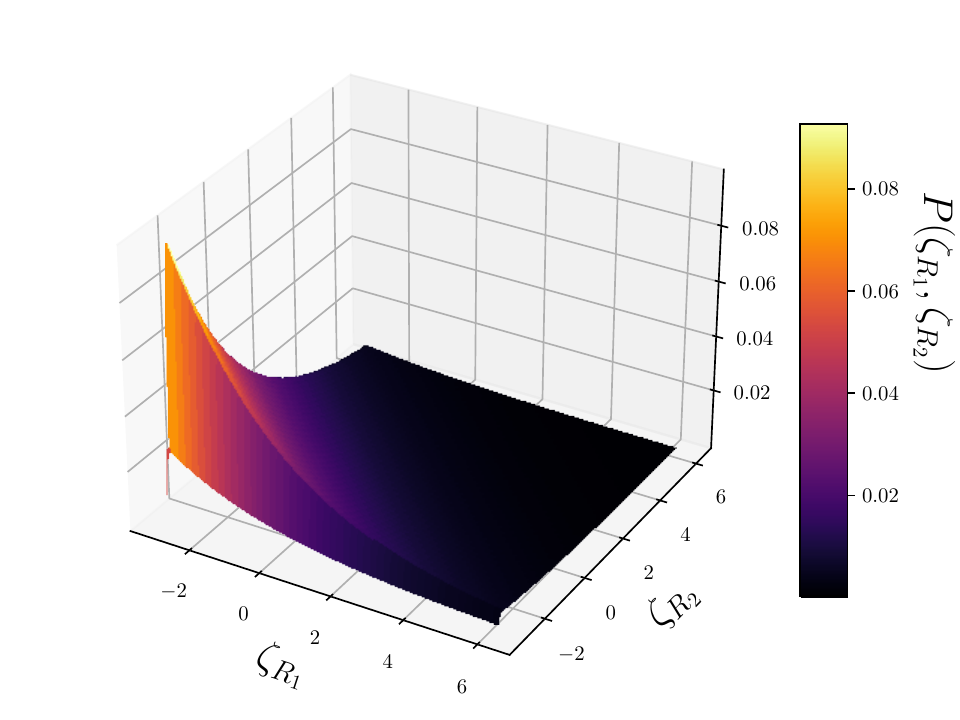}
 \end{minipage}
 \begin{minipage}{.495\textwidth}
 \includegraphics[width=.9\textwidth]{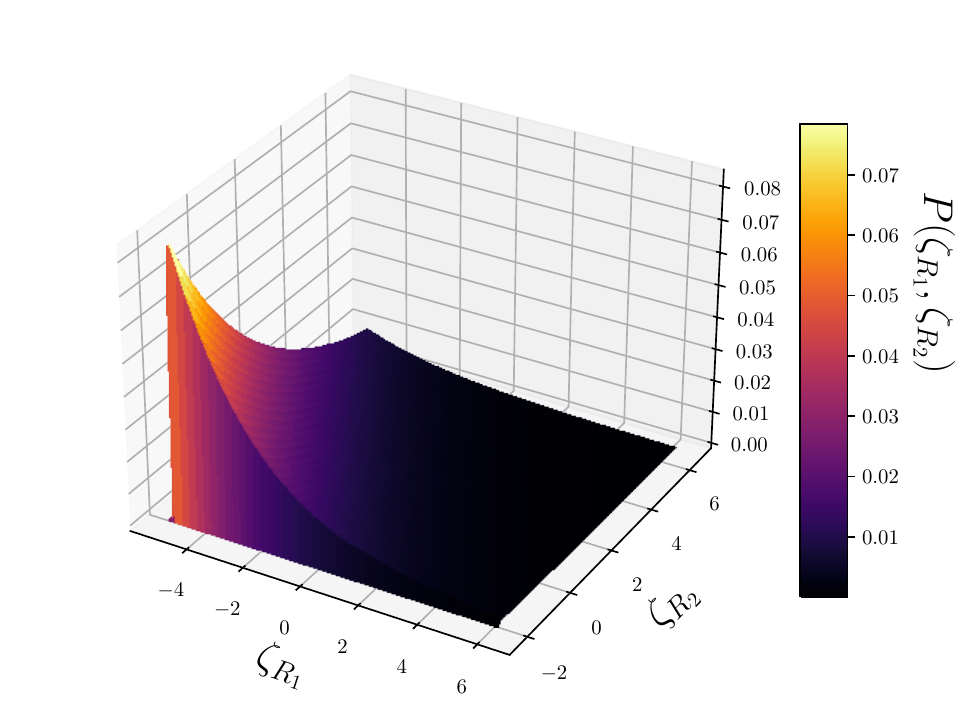}
 \end{minipage}
  \caption{Two-point distribution $P(\zeta_{R_1}, \zeta_{R_2})$ in the quantum-well potential obtained from the numerical integration of \Eq{eq:quantum:well:2pt} for $\mu=\mu_{\mathrm{c}}-0.001$ and $x_*=0.8$. In the left column $x_1=x_2=0.05$, corresponding to the symmetric case $R_1=R_2$, whereas in the right column $x_1=0.03$ and $x_2=0.07$, corresponding to the case $R_1 <R_2$.}
 \label{fig:ColorMapFlat}
\end{figure}

Similarly, for the two-point distribution \Eq{eq:2pt:single:clock} gives rise to
\bea
\label{eq:quantum:well:2pt}
P(\zeta_{R_1},\zeta_{R_2}) = & 
-\frac{\pi^3}{8\mu^6(1-x_*)^2(1-x_1)^2(1-x_2)^2} 
\frac{\cos\left[\sqrt{3}\mu(1-x_1)\right]\cos\left[\sqrt{3}\mu(1-x_2)\right]}{\cos\left[\sqrt{3}\mu(1-x_*)\right]}
\\ & \quad
 \int\dd\mathcal{N}_{\phi_0\to\phi_*}
\vartheta_2'\left(\frac{\pi}{2} ,\ee^{-\frac{\pi^2}{\mu^2(1-x_*)^2} \mathcal{N}_{\phi_0\to\phi_*}}\right)
\\ & \quad
\vartheta_2'\left(\frac{\pi}{2} \frac{x_*-x_1}{1-x_1},\ee^{-\frac{\pi^2}{\mu^2(1-x_1)^2} \left(\zeta_{R_1}-\mathcal{N}_{\phi_0\to\phi_*}+\left\langle \mathcal{N}_{\phi_0} \right\rangle_{\mathrm{V}}-\left\langle \mathcal{N}_{\phi_1} \right\rangle_{\mathrm{V}}\right)}\right)
\\ & \quad
\vartheta_2'\left(\frac{\pi}{2} \frac{x_*-x_2}{1-x_2},\ee^{-\frac{\pi^2}{\mu^2(1-x_2)^2} \left(\zeta_{R_2}-\mathcal{N}_{\phi_0\to\phi_*}+\left\langle \mathcal{N}_{\phi_0} \right\rangle_{\mathrm{V}}-\left\langle \mathcal{N}_{\phi_2} \right\rangle_{\mathrm{V}}\right)}\right)
\\ & \quad
\ee^{3\left(\zeta_{R_1}+\zeta_{R_2}-\mathcal{N}_{\phi_0\to\phi_*}+2\left\langle \mathcal{N}_{\phi_0} \right\rangle_{\mathrm{V}}-\left\langle \mathcal{N}_{\phi_1} \right\rangle_{\mathrm{V}}-\left\langle \mathcal{N}_{\phi_2} \right\rangle_{\mathrm{V}}\right)}\, ,
\eea
where $\left \langle \calN_{\phi_0}\right \rangle_{\mathrm{V}}$, $\left \langle \calN_{\phi_1}\right \rangle_{\mathrm{V}}$ and $\left \langle \calN_{\phi_2}\right \rangle_{\mathrm{V}}$ are given by \Eq{eq:flat:N:weight}.
The result is displayed in \Fig{fig:ColorMapFlat} for $R_1=R_2$ (left panels, where the distribution is symmetric) and $R_1 \neq R_2$ (right panels, where it is asymmetric). Note that the large-volume approximation requires $R_1,\, R_2\ll r$, hence $x_1,\, x_2\ll 1$, which is why the distribution peaks at large negative values of the curvature perturbations, see footnote~\ref{footnote:large:negative:zeta}. 

A tail expansion similar to \Eq{eq:1pt:quantum:well:tail} can be performed for the two-point distribution. However, when $\zeta_{R_1}-\left\langle \mathcal{N}_{\phi_1}\right\rangle_{\mathrm{V}}+\left\langle \mathcal{N}_{\phi_0}\right\rangle_{\mathrm{V}}\gg\mu^2$ and $\zeta_{R_2}-\left\langle \mathcal{N}_{\phi_2}\right\rangle_{\mathrm{V}}+\left\langle \mathcal{N}_{\phi_0}\right\rangle_{\mathrm{V}}\gg\mu^2$, one can check that the integrand appearing in \Eq{eq:quantum:well:2pt} peaks in the tail of the last two elliptic functions, but not in the tail of the first elliptic function. The reason is that, although a large value of $\zeta_{R_1}$ requires a large value of $\mathcal{N}_{\phi_0\to\phi_1} =\mathcal{N}_{\phi_0\to\phi_*}+\mathcal{N}_{\phi_*\to\phi_1} $, it does not imply that both $\mathcal{N}_{\phi_0\to\phi_*}$ and $\mathcal{N}_{\phi_*\to\phi_1}$ are large, and in practice we find that only $\mathcal{N}_{\phi_*\to\phi_1}$ is large (and similarly for $\zeta_{R_2}$). As a consequence, only the last two elliptic functions in \Eq{eq:quantum:well:2pt} can be approximated by their first terms, but the integral over $\mathcal{N}_{\phi_0\to\phi_*}$ can nonetheless be still performed analytically by using the formula $\int_0^\infty \vartheta_2'(\pi/2,\ee^{-x})\ee^{-A x} \dd x=-2\pi/\cosh(\pi\sqrt{A})$, see  Ref.~\cite[\href{https://dlmf.nist.gov/20.10.4}{(20.10.4)}]{NIST:DLMF}. This leads to
\bea
\label{eq:quantum:well:2pt:tail:exp}
P\left(\zeta_{R_1},\zeta_{R_2}\right) \simeq  \frac{P\left(\zeta_{R_1}\right)P\left(\zeta_{R_2}\right) \cos\left(\frac{\pi}{2}\frac{1-x_*}{1-x_1}\right)\cos\left(\frac{\pi}{2}\frac{1-x_*}{1-x_2}\right)}{\cos\left[\sqrt{3}\mu\left(1-x_*\right)\right]\cosh\left\lbrace\sqrt{3}\mu\left(1-x_*\right)\sqrt{1-\frac{\pi^2}{12\mu^2}\left[\frac{1}{\left(1-x_1\right)^2}+\frac{1}{\left(1-x_2\right)^2}\right]}\right\rbrace} ,
\eea 
where $P(\zeta_{R_1})$ and $P(\zeta_{R_2})$ are also expanded according to \Eq{eq:1pt:quantum:well:tail}.

\begin{figure}[H] 
	\centering
	\includegraphics[width=0.7\hsize]{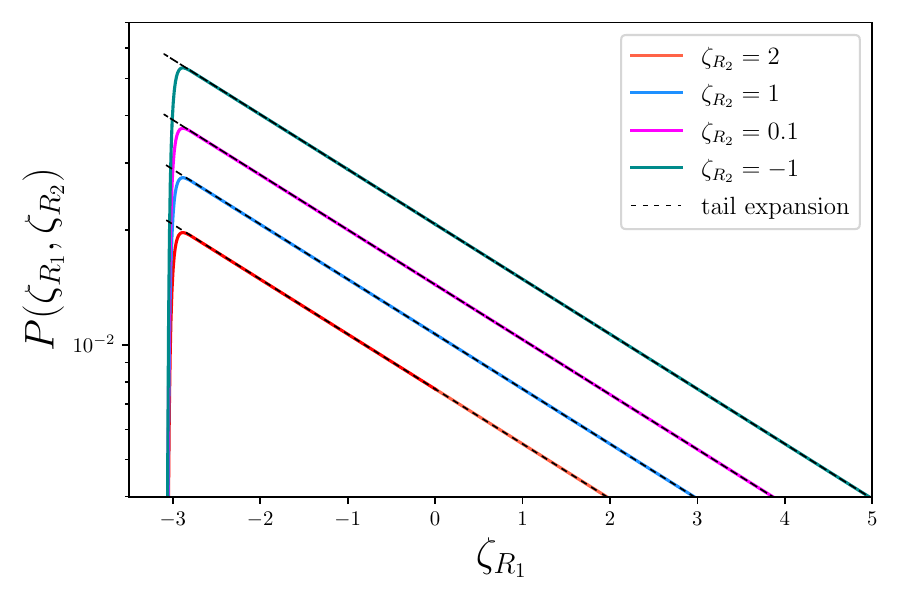}
	\caption{Two-point distributions $P(\zeta_{R_1}, \zeta_{R_2})$ in the quantum-well potential as a function of $\zeta_{R_1}$ for a few fixed values of $\zeta_{R_2}$, with $\mu=\mu_{\mathrm{c}}-0.001$, $x_*=0.8$ and $x_1=x_2=0.05$ (\ie in the same situation as in the left panels of \Fig{fig:ColorMapFlat}). Solid coloured lines stand for the full result~\eqref{eq:quantum:well:2pt}, whereas dashed lines display the tail approximation~\eqref{eq:quantum:well:2pt:tail:exp}.}
	\label{fig:TwoPointFlatSlice}
\end{figure}

One can check that, in the limit $x_*\to 1$, the joint distribution factorises according to $P(\zeta_{R_1},\zeta_{R_2}) = P(\zeta_{R_1}) P(\zeta_{R_1})$, which is expected: any two paths ending in two final regions that do not share any parent node in the tree of \Fig{fig:sketch:volume} cannot be correlated.
The tail expansion~\eqref{eq:quantum:well:2pt:tail:exp} is compared with the full result~\eqref{eq:quantum:well:2pt} in \Fig{fig:TwoPointFlatSlice}, where one-dimensional sections of the two-point distribution are shown as a function of $\zeta_{R_1}$ for a few fixed values of $\zeta_{R_2}$. One can check that the agreement is excellent on the tail.

\subsubsection*{Clustering}

In this work, for simplicity, we assume that primordial black holes form when $\zeta_R>\zeta_\uc$, where $\zeta_\uc$ is a threshold value of order unity,
\bea
\label{eq:pM:zeta:threshold}
p_M = \int_{\zeta_\uc}^\infty P\left(\zeta_R\right)\dd\zeta_R\, ,
\eea 
where $M$ is of the order of the Hubble mass at the time $R$ re-enters the Hubble radius. As mentioned above, this is an oversimplification with respect to more advanced PBH criteria, which rather involve the compaction function and critical scaling, and possible extensions to include these will be discussed in \Sec{sec:Conclusion}. At this stage, this proxy nonetheless allows us to discuss how regions with large curvature perturbations are correlated, and to get a first insight into the amount of clustering to be expected from quantum diffusion. 

The probability to form two primordial black holes with masses $M_1$ and $M_2$ at distance $r$ follows an analogous expression,
\bea
\label{eq:pM1M2:zeta:threshold}
p_{M_1,M_2}(r)=\int_{\zeta_\uc}^\infty P\left(\zeta_{R_1},\zeta_{R_2}\right)\dd\zeta_{R_1}\dd\zeta_{R_2}\, ,
\eea 
from which the reduced correlation defined in \Eq{eq:clustering:def} can be computed.

\begin{figure}[H] 
	\centering
    \begin{minipage}{.495\textwidth}
     \includegraphics[width=.9\textwidth]{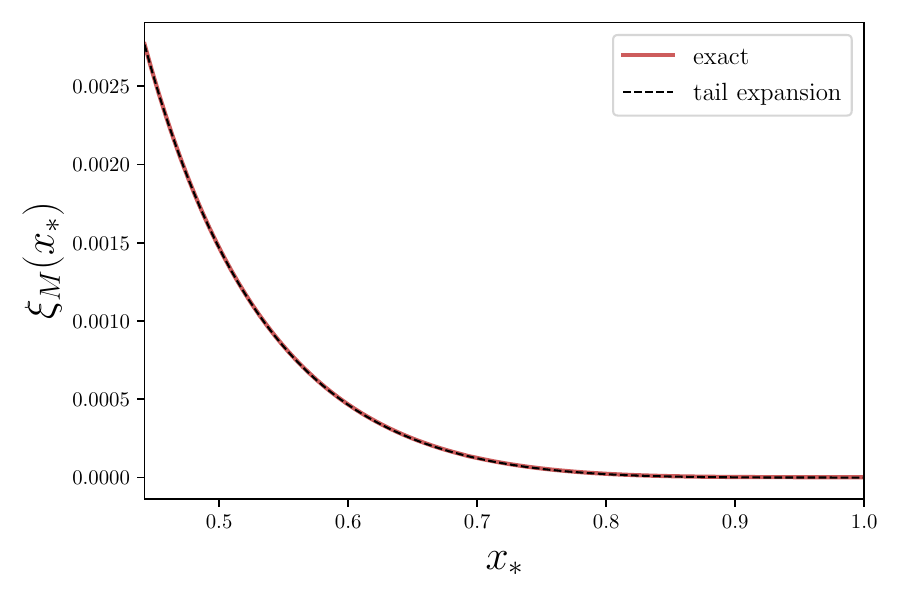}
     \end{minipage}
     \begin{minipage}{.495\textwidth}
    \includegraphics[width=.9\textwidth]{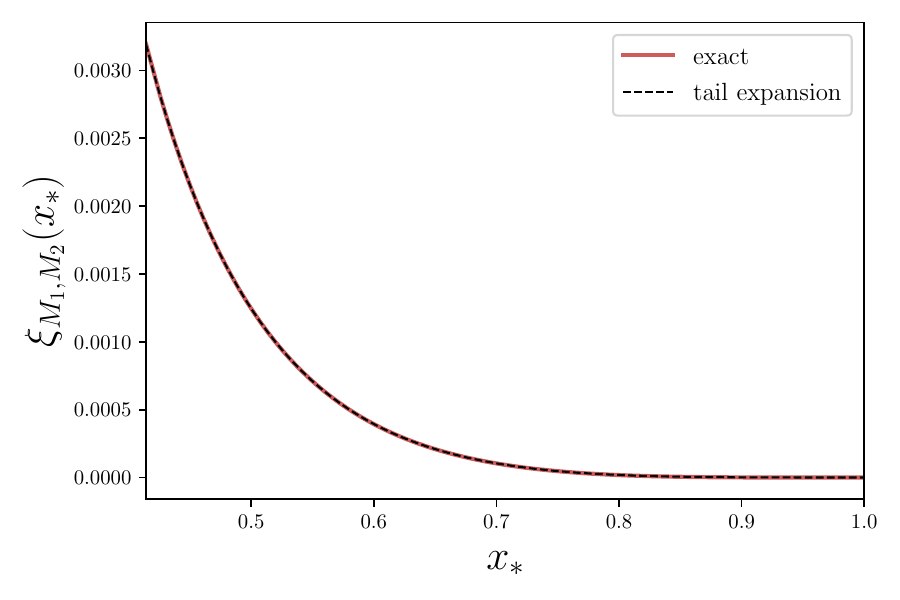}
    \end{minipage}
\caption{Reduced correlation in the quantum-well potential as a function of $x_*$, for $\mu=\mu_{\mathrm{c}}-0.001$ and $x_1=x_2=0.05$ (left panel) and for $\mu=\mu_{\mathrm{c}}-0.001$ and $x_1=0.03$ and $x_2=0.07$ (right panel). We only plot the reduced correlation for values of $x_* $ that satisfy $r\geq R_1+R_2$, where $R_1, R_2$ denote the size of the coarse-graining regions. The solid line stands for the full result obtained by integrating \Eqs{eq:1pt:quantum:well} and~\eqref{eq:quantum:well:2pt} above $\zeta_{\mathrm{c}}=1$, whereas the dashed line stands for the tail-approximation, according to which $1+\xi_{M_1,M_2}$ is nothing but the prefactor in \Eq{eq:quantum:well:2pt:tail:exp}.}
	\label{fig:ClusteringFlat}
\end{figure}

The result is displayed in \Fig{fig:ClusteringFlat} for the parameters corresponding to the left and right panels of \Fig{fig:ColorMapFlat}, where the tail expansion [where $1+\xi_{M_1,M_2}$ is simply given by the prefactor in \Eq{eq:quantum:well:2pt:tail:exp}] is shown to provide an excellent approximation. At small $x_*$, the reduced correlation reaches a maximum when $r\sim R_1+R_2$, below which the exclusion effect discussed in \Sec{sec:Introduction} comes into play. This regime is however not properly described in the large-volume approximation, which requires $r\gg R_1,\, R_2$. In the opposite regime, when $x_*\to 1$, as mentioned below \Eq{eq:quantum:well:2pt:tail:exp} the two-point distribution factorises and $\xi_{M_1,M_2}\to 0$. In practice, if the quantum well is embedded in a full inflationary potential, fluctuations outside the well provide a non-vanishing value for $\xi_{M_1,M_2}$ at $x_*\geq 1$.

\subsection{Tilted quantum well}
\label{sec:tilted:well}

\begin{figure}[H] 
	\centering
	\includegraphics[width=0.7\hsize]{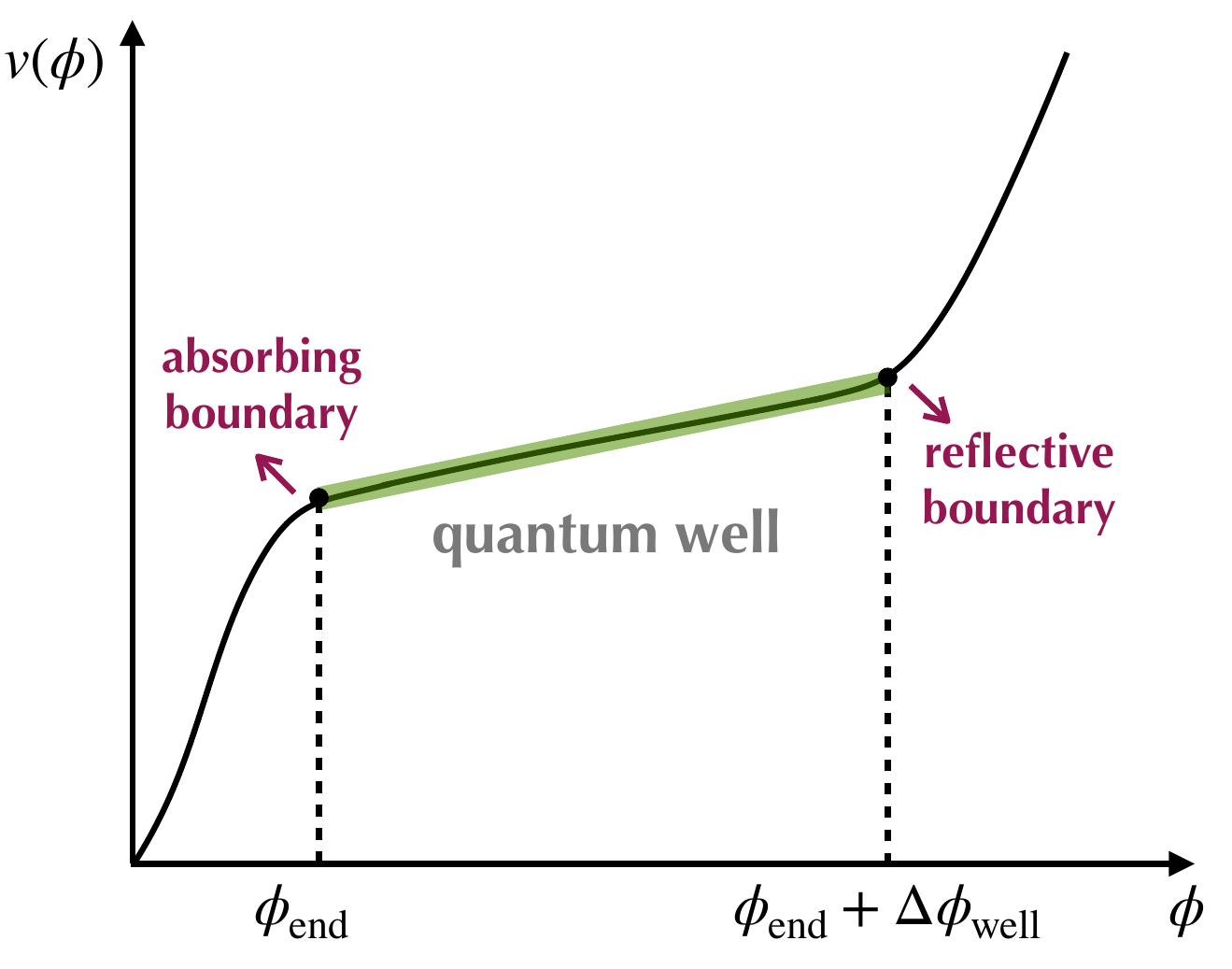}
	\caption{Tilted quantum-well toy model.}
	\label{fig:sketch:tilted}
\end{figure}

The above toy model is useful since it can be described with few parameters and most of the calculation can be performed analytically. However, it does not allow us to compare our results with classical predictions, where quantum diffusion is not accounted for. This is because, in the slow-roll regime, the inflaton never escapes a flat well without quantum diffusion, and the curvature perturbation diverges. The lack of classical counterpart, and our inability to reach a classical limit, makes it impossible to determine whether quantum diffusion enhances or suppresses clustering, and for this reason we now consider a slight modification of the quantum well where the potential function features a linear slope,
\bea
v(\phi)=v_0\left(1+\alpha\frac{\phi}{\Mp}\right) ,
\eea 
where $\alpha$ is a positive parameter.
It is still endowed with an absorbing boundary at $\phi_\uend$ and a reflective boundary at $\phi_\uend+\Delta\phi_\uwell$, and we will consider the regime where the potential is dominated by its constant term, therefore
\bea 
\label{eq:constant:slope:vacuum:dom:cond}
\alpha \frac{\Delta\phi_\uwell}{\Mp}\ll 1\, .
\eea 
Moreover, for the slow-roll approximation~\eqref{eq:Langevin:SR} to apply, one requires $\alpha\ll 1$. A sketch of the potential is displayed in \Fig{fig:sketch:tilted}.
\noindent 
This model has been extensively studied in \Refs{Ezquiaga:2019ftu, Animali:2022otk}, from which we recall the main results. In the regime~\eqref{eq:constant:slope:vacuum:dom:cond}, the characteristic function is given by
\bea\label{eq:char:constant:slope}
\chi_\calN(t,\phi)=e^{\frac{d  \mu^2 x}{2}} \frac{ \sqrt{4 i t - d^2 \mu^2} \cos{\left(\frac{x-1}{2}\sqrt{4 i t- d^2 \mu^2} \mu\right)}- d \mu \sin{\left(\frac{x-1}{2} \sqrt{4 i t- d^2 \mu^2} \mu\right)}}{ \sqrt{4 i t - d^2 \mu^2}\cos{\left(\frac{1}{2} \sqrt{4 i t - d^2 \mu^2}\mu\right)+d \mu \sin{\left(\frac{1}{2} \sqrt{4 i t - d^2 \mu^2}\mu\right)} }}\,,
\eea
where $x$ and $\mu$ have been defined in \Sec{sec:quantum:well} and we have introduced
\bea
d= \alpha\frac{\Mp}{\Delta\phi_\uwell}\, .
\eea 
When $\alpha=d=0$ and one can check that \Eq{eq:chi:flat:cos} is recovered. 

From the characteristic function, the mean volume realised from an initial field location $x$ can be obtained as explained in \Sec{sec:quantum:well}, \ie $\langle \ee^{3\mathcal{N}_{\phi}} \rangle = \chi_\mathcal{N}(-3i,\phi)$, and one finds 
\bea\label{eq:vol:const:slope:char}
\left \langle e^{3 \calN_\phi} \right \rangle =&e^{\frac{d \mu^2 x}{2}}\frac{\sqrt{12-d^2 \mu^2}\cos{\left( \frac{x-1}{2}\sqrt{12-d^2 \mu^2} \mu\right)}-d \mu \sin{\left( \frac{x-1}{2} \sqrt{12-d^2 \mu^2} \mu\right)}}{\sqrt{12-d^2 \mu^2}\cos{\left(\frac{\mu}{2} \sqrt{12-d^2 \mu^2}\right)+ d \mu \sin{\left(\frac{\mu}{2} \sqrt{12-d^2 \mu^2}\right)}}}\,.
\eea
The mean number of \efolds can also be obtained in a compact form using \Eq{eq:meanN:from:char}, and one obtains
\bea
\label{eq:meanN:constant:slope}
\left\langle \mathcal{N}_\phi \right\rangle = \frac{x}{d} +\ee^{-d\mu^2}\frac{1-\ee^{d\mu^2 x}}{d^2\mu^2}\, .
\eea 
In this expression, the first term corresponds to the number of \efolds realised classically, \ie in the absence of quantum diffusion. This makes the interpretation of the parameter $d$ easy: $1/d$ is nothing but the classical duration of inflation across the tilted well. Finally, the volume-weighted number of \efolds can be obtained by means of \Eq{eq:meanN:vol:weighted:from:char}, which gives
\bea
\label{eq:constant:slope:meanN:vol:weighted}
\left\langle \mathcal{N}_\phi \right\rangle_{\mathrm{V}} = & \Bigg\lbrace x \left(d^2 \mu ^2-6\right) \sin \left[\frac{\mu}{2}   (x-2) \sqrt{12-d^2 \mu ^2}\right]+2 (d-3 x+6) \sin \left(\frac{\mu}{2}   x \sqrt{12-d^2 \mu ^2}\right)
 \\ & 
-d^2 \mu ^2 x \sqrt{\frac{12}{d^2 \mu ^2}-1} \cos \left[\frac{\mu}{2}   (x-2) \sqrt{12-d^2 \mu ^2}\right]\Bigg\rbrace 
\\ & 
\Bigg( d^2 \mu^2  \sqrt{12-d^2 \mu ^2} \left[\sin \left(\frac{\mu}{2}   \sqrt{12-d^2 \mu ^2}\right)+\sqrt{\frac{12}{d^2 \mu ^2}-1} \cos \left(\frac{\mu}{2}   \sqrt{12-d^2 \mu ^2}\right)\right]
 \\ & 
\left\lbrace \sqrt{\frac{12}{d^2 \mu ^2}-1} \cos \left[\frac{\mu}{2}   (x-1) \sqrt{12-d^2 \mu ^2}\right]-\sin \left[\frac{\mu}{2}   (x-1) \sqrt{12-d^2 \mu ^2}\right]\right\rbrace\Bigg)^{-1}\, .
\eea 

In order to inverse Fourier transform \Eq{eq:char:constant:slope} and get the first-passage time distribution, one needs to extract the poles of the characteristic function, which requires to solve a transcendental equation.
Analytically, this cannot be done in general, but only in the following two regimes~\cite{Ezquiaga:2019ftu, Animali:2022otk}:
when $\mu^2 d\ll 1$ (the so-called ``narrow-well limit''), where one recovers the flat quantum-well setup studied in \Sec{sec:quantum:well}; and when $\mu^2 d\gg 1$ (the so-called ``wide-well limit''), where one finds
\bea
\label{eq:Pfpt:constant:slope:theta3}
\Pfpt{\phi}(\mathcal{N}) = -\frac{\pi}{2 \mu^2} e^{\mu^2 d \frac{x}{2}-\frac{\mu^2 d^2}{4} \calN}  \vartheta_3^\prime\left(\frac{\pi}{2} x, e^{-\frac{\pi^2}{\mu^2}\calN}\right) ,
\eea 
where $\vartheta_3 (z,q)$ is the third elliptic theta function, $\vartheta_3(z,q)=1+2 \sum_{n=1}^\infty q^{n^2} \cos{(2 n z)}$, and $\vartheta_3^\prime$ denotes its derivative with respect to its first argument $z$. This regime allows one to recover the classical limit when $\mu^2 d \to\infty$, which is why we will now restrict to it. Note that, since the wide-well approximation breaks down for large-index poles, \Eq{eq:Pfpt:constant:slope:theta3} provides a good approximation to the first-passage time distribution only near its maximum and along its upper tail where the first few poles dominate (if $x$ is not too close to $1$), but it fails on its lower tail where all poles equally contribute, \ie at low values of $\mathcal{N}$. In particular, it is not properly normalised, and it should thus be used with care. This is why, in what follows, unless specified otherwise the first-passage distribution is computed from numerically inverse-Fourier transforming \Eq{eq:char:constant:slope}.

At large $\mathcal{N}$, \Eq{eq:Pfpt:constant:slope:theta3} gives $\Pfpt{\phi}(\mathcal{N})\propto \ee^{-(\pi^2/\mu^2+\mu^2 d^2/4)\calN}$, hence the tail of the first-passage time distribution decays more rapidly than $e^{-3\mathcal{N}}$ only if $\alpha^2>12 v_0$, or if $\alpha^2<12 v_0$ and $\mu<\pi/\sqrt{3-\alpha^2/(4v_0)}$.\footnote{In terms of $d$, this condition can be rewritten as $d>3/\pi$ or $d<3/\pi$ and $\mu^2<6/d^2(1-\sqrt{1-d^2\pi^2/9})$ or $\mu^2>6/d^2(1+\sqrt{1-d^2\pi^2/9})$. } Under these conditions the mean volume converges, hence the volume-weighted probabilities are well defined. 

Let us finally note that, when one is interested in the first-passage through a field value $\phi_*$ different from $\phi_{\mathrm{end}}$, \Eqs{eq:char:constant:slope} and \eqref{eq:Pfpt:constant:slope:theta3} can still be employed provided the rescalings $x\to (x-x_*)/(1-x_*)$, $\mu\to (1-x_*)\mu$ and $d \rightarrow d/(1-x_*)$ are performed.

\subsubsection*{One-point curvature perturbation}

\begin{figure}[H]
  \centering
  \begin{minipage}{.495\textwidth}
 \includegraphics[width=.9\textwidth]{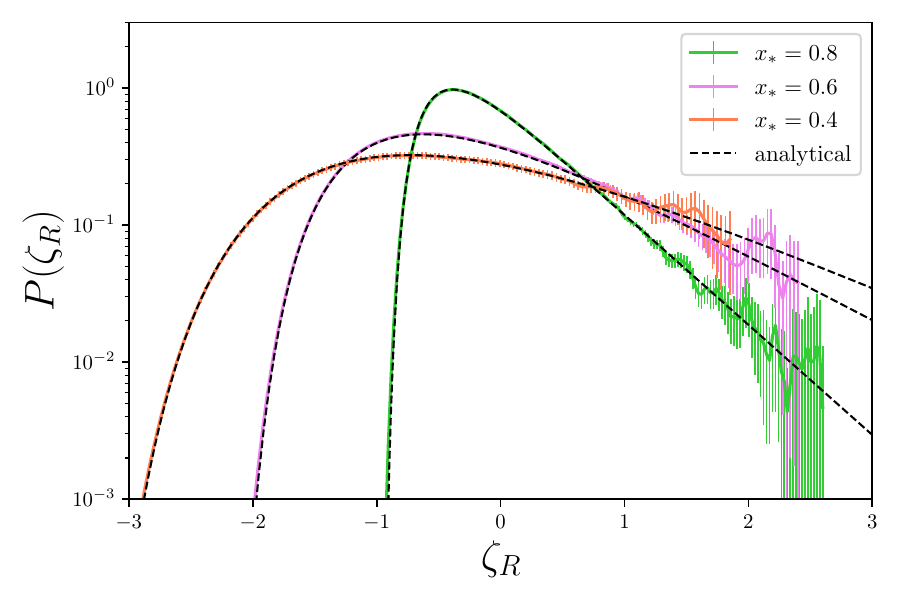}
 \end{minipage}
 \begin{minipage}{.495\textwidth}
 \includegraphics[width=.9\textwidth]{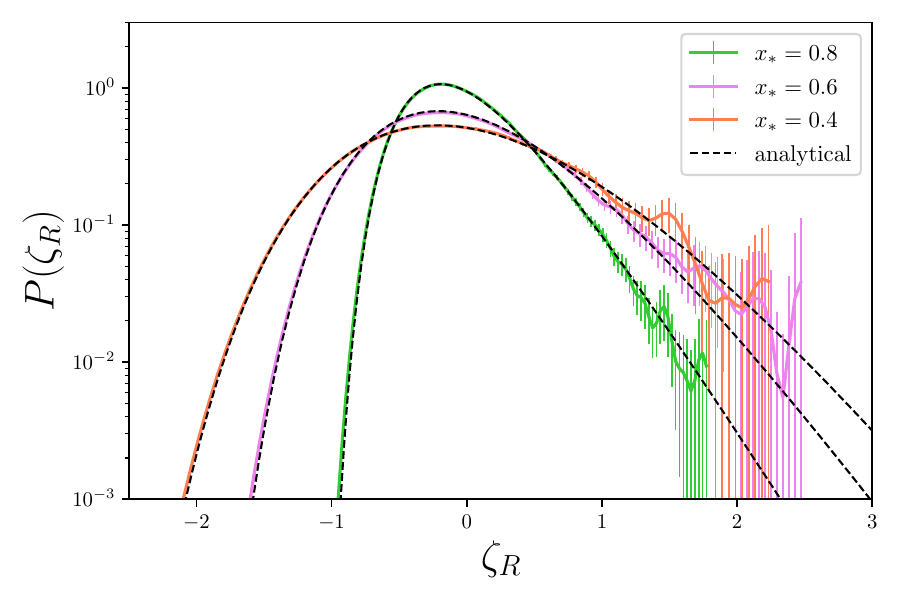}
 \end{minipage}
  \caption{One-point distribution  $P(\zeta_R)$ in the tilted-well potential, for $\mu=13$ and $d=0.3$ (left panel) and for $\mu=25$ and $d=0.2$ (right panel). The solid coloured lines are reconstructed from $10^7$ (left panel) and $10^6$ (right panel) numerical simulations, whereas dashed lines stand for the result obtained by numerically performing the inverse Fourier transform of the volume-weighted version of the characteristic function~\eqref{eq:char:constant:slope}. Different colours indicate different values of the (rescaled) field value $x_*$ in the parent patch, hence different coarse-graining scales. Error bars correspond to the $2\sigma$-estimates of the statistical error obtained via the bootstrap resampling technique with resampling size $N_{\mathrm{s}} = 1000.$}
 \label{fig:OnePointTilted}
\end{figure}

In the large-volume approximation, the one-point distribution for the coarse-grained curvature perturbation can be computed from \Eq{eq:1pt:single:clock}, where the first-passage time distribution is given by the inverse Fourier transform of \Eq{eq:char:constant:slope} and $\langle\mathcal{N}_x\rangle_{\mathrm{V}}$ is given in \Eq{eq:constant:slope:meanN:vol:weighted}. The result is shown in \Fig{fig:OnePointTilted}, where it is compared with a direct sampling reconstruction. Similarly to the flat-well model, it features exponential tails, which are heavier at larger $x_*$ (hence at larger coarse-graining radius). 

One also notices that, in the right panel of \Fig{fig:OnePointTilted}, the distributions look more Gaussian close to their maximum than in the left panel. This is because $\mu^2 d$ is larger in the right panel ($\mu^2 d=125$) than in the left panel ($\mu^2 d=50.7$), and, as argued above, the classical regime is obtained in the limit $d \mu^2 \rightarrow \infty.$ This confirms that the standard, Gaussian results are recovered in that limit.

\subsubsection*{Two-point curvature perturbation}

\begin{figure}[t]
  \centering
  \begin{minipage}{.495\textwidth}
 \includegraphics[width=.9\textwidth]{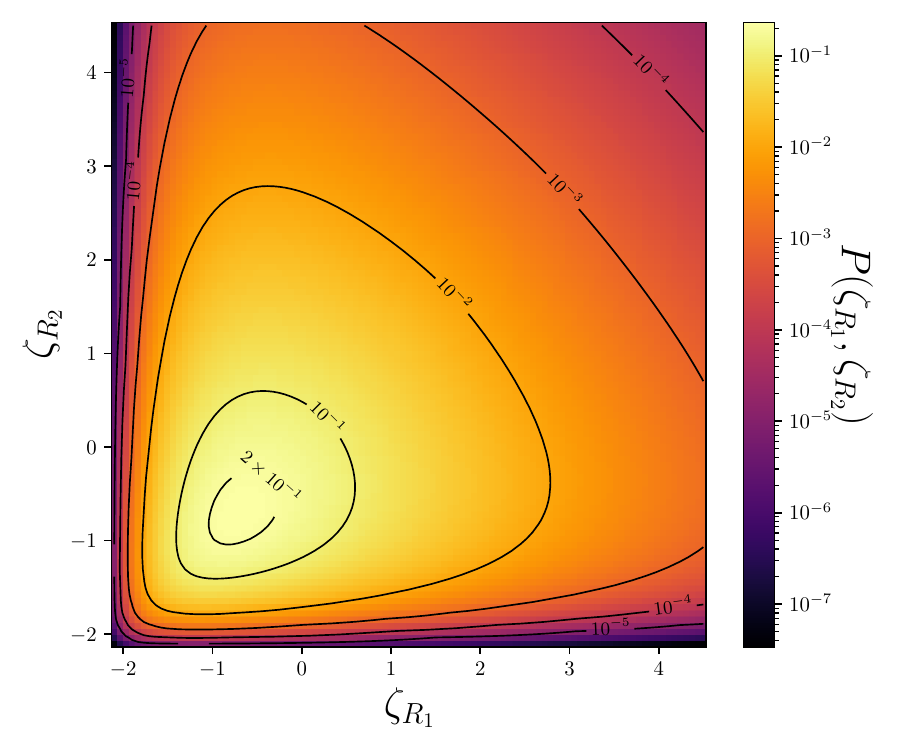}
 \end{minipage}
 \begin{minipage}{.495\textwidth}
 \includegraphics[width=.9\textwidth]{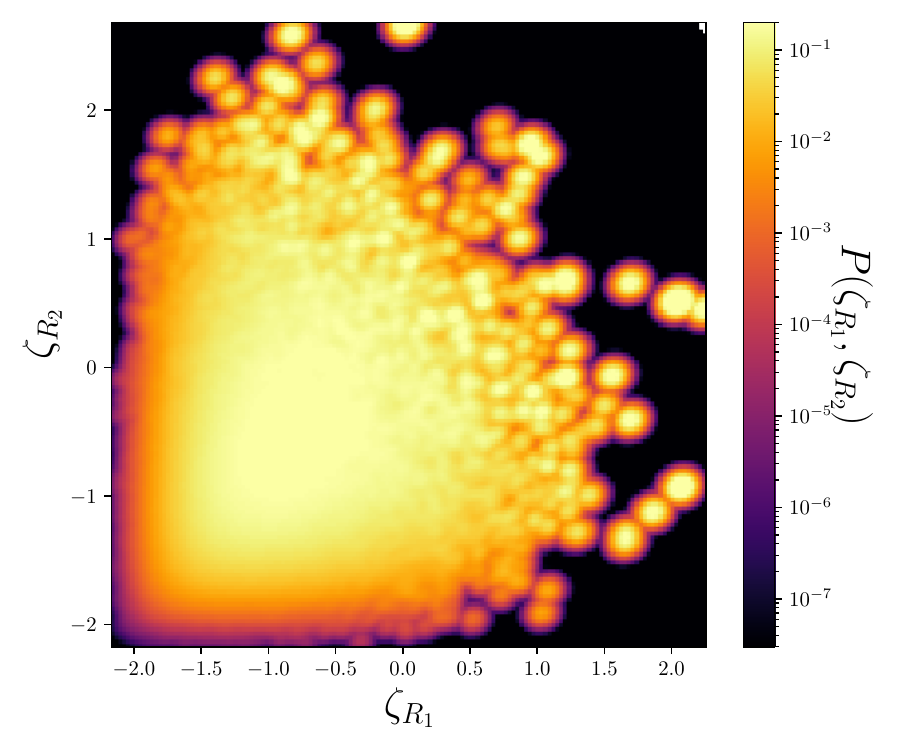}
 \end{minipage}
 \centering
  \begin{minipage}{.495\textwidth}
 \includegraphics[width=.9\textwidth]{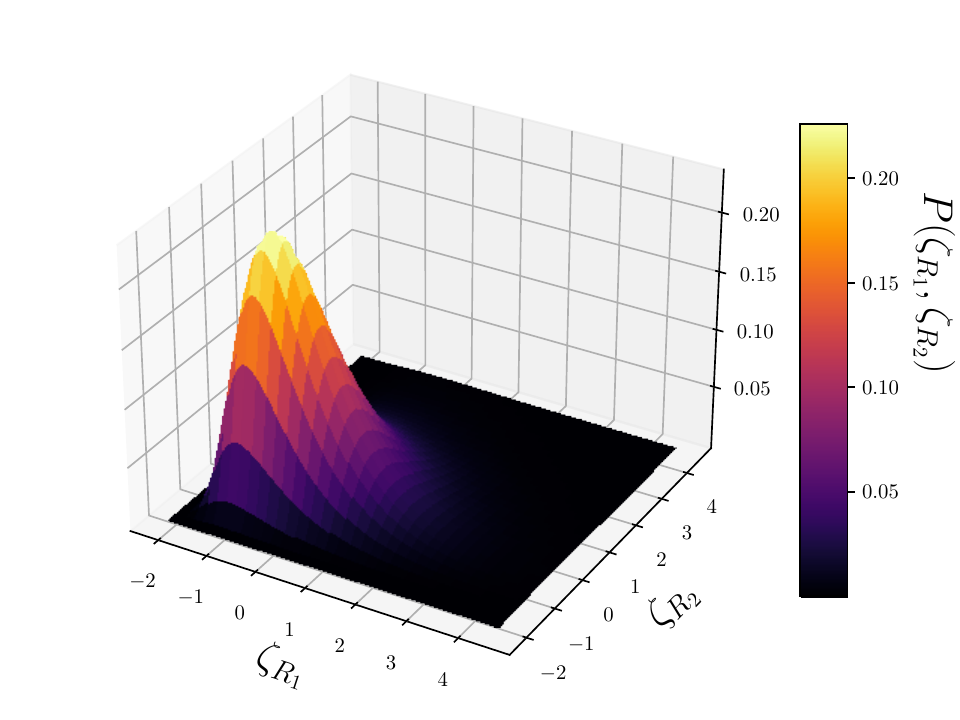}
 \end{minipage}
 \begin{minipage}{.495\textwidth}
 \includegraphics[width=.9\textwidth]{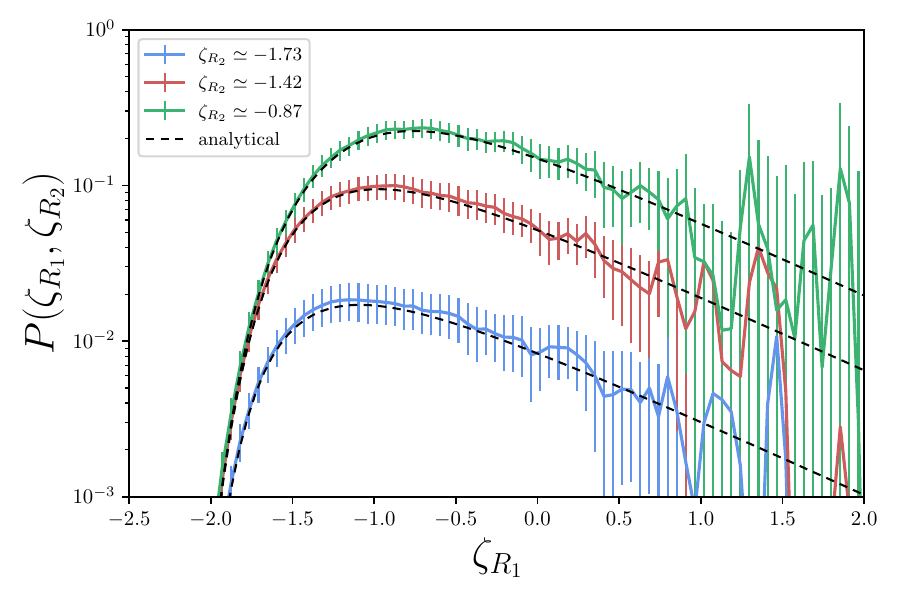}
 \end{minipage}
  \caption{Two-point distribution function $P(\zeta_{R_1},\zeta_{R_2})$ in the tilted-well potential, for $\mu=13$, $d=0.3$, $x_*=0.85$ and $x_1=x_2=0.6$. The left panels show the analytical result~\eqref{eq:2pt:single:clock}, whereas the top-right shows the reconstruction from $10^6$ numerical simulations of the Langevin equation. In the bottom-right we compare the analytical (black dashed lines) and the numerically-sampled (solid coloured lines) distribution as a function of $\zeta_{R_1}$ for a few fixed values of $\zeta_{R_2}$.}
 \label{fig:TwoPointTiltmu13}
\end{figure}

\begin{figure}[t]
  \centering
  \begin{minipage}{.495\textwidth}
 \includegraphics[width=.9\textwidth]{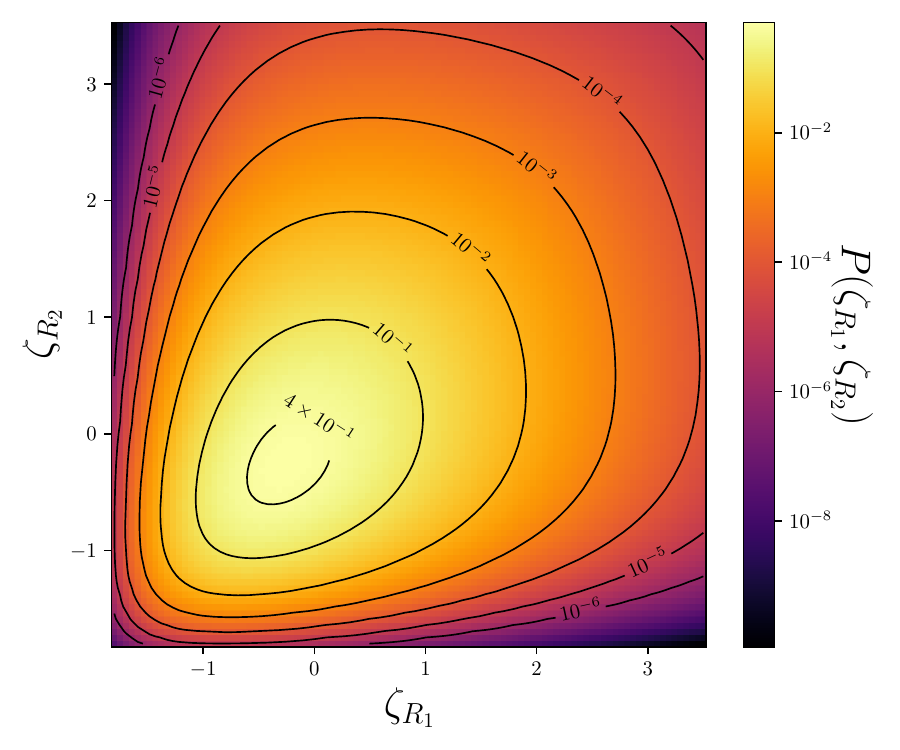}
 \end{minipage}
 \begin{minipage}{.495\textwidth}
 \includegraphics[width=.9\textwidth]{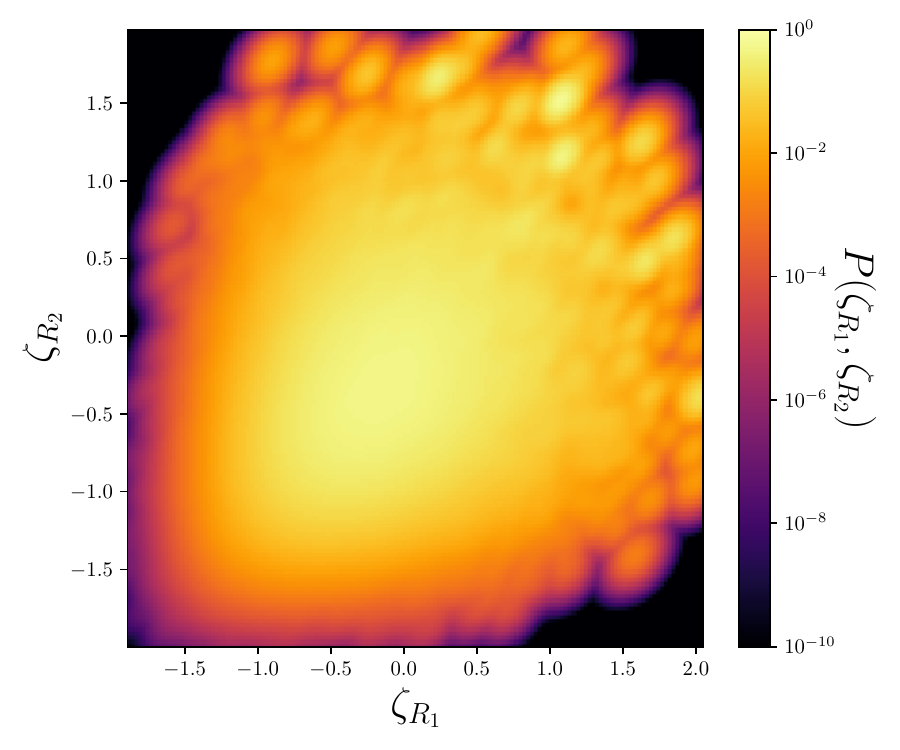}
 \end{minipage}
 \centering
  \begin{minipage}{.495\textwidth}
 \includegraphics[width=.9\textwidth]{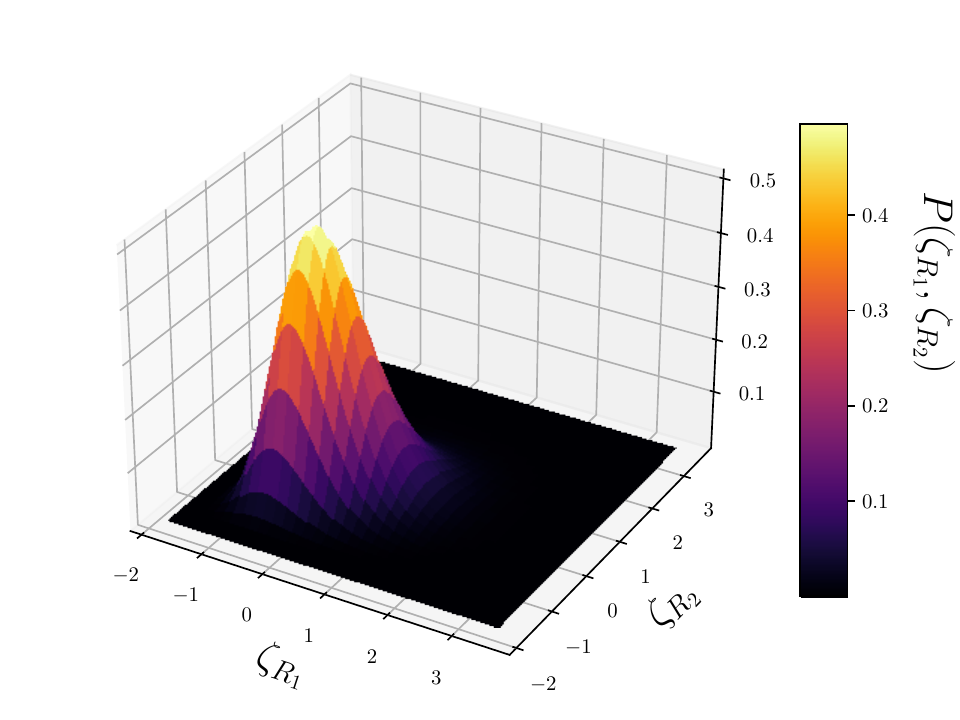}
 \end{minipage}
 \begin{minipage}{.495\textwidth}
 \includegraphics[width=.9\textwidth]{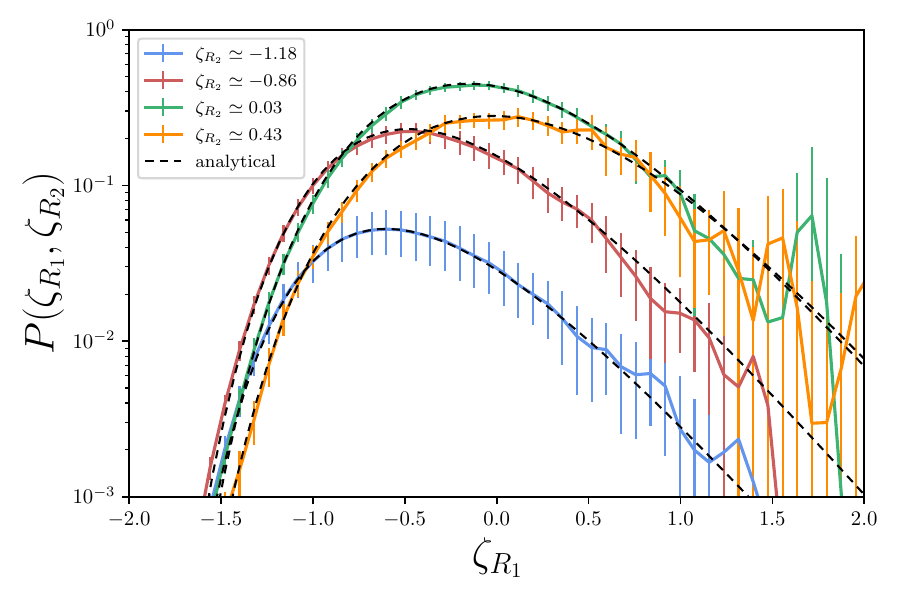}
 \end{minipage}
  \caption{Same as \Fig{fig:TwoPointTiltmu13} for $\mu=25$, $d=0.2$, $x_*=0.85$ and $x_1=x_2=0.6$.}
 \label{fig:TwoPointTiltmu25}
\end{figure}

Similarly, the two-point distribution of the coarse-grained curvature perturbation can be obtained by means of \Eq{eq:2pt:single:clock}, where the first-passage time distribution is given by inverse-Fourier transforming \Eq{eq:char:constant:slope} and the integral in \Eq{eq:2pt:single:clock} is also performed numerically. The result is shown in \Figs{fig:TwoPointTiltmu13} and \ref{fig:TwoPointTiltmu25} for the parameters corresponding to the left and right panels of \Fig{fig:OnePointTilted} respectively. In the right panels of \Figs{fig:TwoPointTiltmu13} and \ref{fig:TwoPointTiltmu25}, the two-point distribution is reconstructed from a direct sampling of the Langevin equations, using the procedure outlined below \Eq{eq:Pjoint:large-volume} and by means of kernel density estimation, see footnote~\ref{footnote:kernel_density}. One can check that the agreement with the analytical expression~\eqref{eq:2pt:single:clock} is excellent, within the statistical error bars.

It is also worth stressing that in \Fig{fig:TwoPointTiltmu25}, the two-point distribution seems more Gaussian than in \Fig{fig:TwoPointTiltmu13}. This can be seen at the level of the contours (top left panels), which are close to ellipses near the maximum in \Fig{fig:TwoPointTiltmu25}, as well as at the level of the one-dimensional slices (bottom-right panels). This confirms the remark made above for the one-point distribution, namely the fact that the Gaussian, classical limit is recovered in the regime $\mu^2 d\gg 1$.

Finally, a tail expansion can be performed, following the same procedure as the one employed in the flat-well model, see \Eq{eq:quantum:well:2pt:tail:exp}. In the tail, the volume-weighted first-passage time distribution decays exponentially as $\PfptV{x}(\mathcal{N}) \simeq a_{\mathrm{V}}(x) e^{-(\Lambda_0+3) \calN}$, where $\Lambda_0 \simeq \mu^2 d^2/4 + \pi^2/ \mu^2$ is the lowest pole of the characteristic function and $a_{\mathrm{V}}(x)=a_0(x)/\langle \ee^{3\mathcal{N}} \rangle_x$, with $a_0(x)$ the lowest residue of the characteristic function.\footnote{In practice, $\Lambda_0$ and $a_0(x)$ are extracted numerically from the characteristic function, using that $\Lambda_0$ is the lowest zero of the denominator in \Eq{eq:char:constant:slope}, and that~\cite{Ezquiaga:2019ftu}
\bea
a_0(x)=-i \left[\frac{\partial}{\partial t} \chi^{-1}(t=-i \Lambda_0,x)\right]^{-1}\,.
\eea 
Indeed, as mentioned below \Eq{eq:Pfpt:constant:slope:theta3}, the wide-well approximation should be used with care, and for the cases of interest below it is insufficiently accurate [its validity condition, $\mu^2 d\gg 1$, is also less verified once $\mu^2 d$ is rescaled by $1-x_1$ or $1-x_2$, according to the scaling rules mentioned below \Eq{eq:Pfpt:constant:slope:theta3}].
\label{footnote:tail_exp_tilt_2}}

This leads to 
\bea
\label{eq:tilted:well:2pt:tail:exp}
P(\zeta_{R_1},\zeta_{R_2}) = & P(\zeta_{R_1})P(\zeta_{R_2}) 
\frac{a_{\mathrm{V}}\left(x_*,x_1\right)}{a_{\mathrm{V}}\left(x_0,x_1\right)}\frac{a_{\mathrm{V}}\left(x_*,x_2\right)}{a_{\mathrm{V}}\left(x_0,x_2\right)}
\\ & 
\int\dd\mathcal{N} \PfptV{x_0\to x_*}\left(\mathcal{N}_{x_0\to x_*}\right)
\ee^{\left[\frac{\mu^2d^2}{2}+\frac{\pi^2}{\mu^2(1-x_1)^2}+\frac{\pi^2}{\mu^2(1-x_2)^2}-6\right]\mathcal{N}_{x_0\to x_*}}\, .
\eea 
Here, $a_{\mathrm{V}}(x,x')$ corresponds to $a_{\mathrm{V}}(x)$ when the absorbing condition is set at location $x'$, \ie it is obtained from the $a_{\mathrm{V}}(x)$ function given above after rescaling parameters according to the rule stated below \Eq{eq:Pfpt:constant:slope:theta3}. The structure of the above equation is similar to the one obtained in the flat-well model, see \Eq{eq:quantum:well:2pt:tail:exp}, and the consequences of this peculiar tail structure will be further discussed in \Sec{sec:Conclusion}.

\subsubsection*{Clustering}

\begin{figure}[t]
  \centering
  \begin{minipage}{.495\textwidth}
 \includegraphics[width=.9\textwidth]{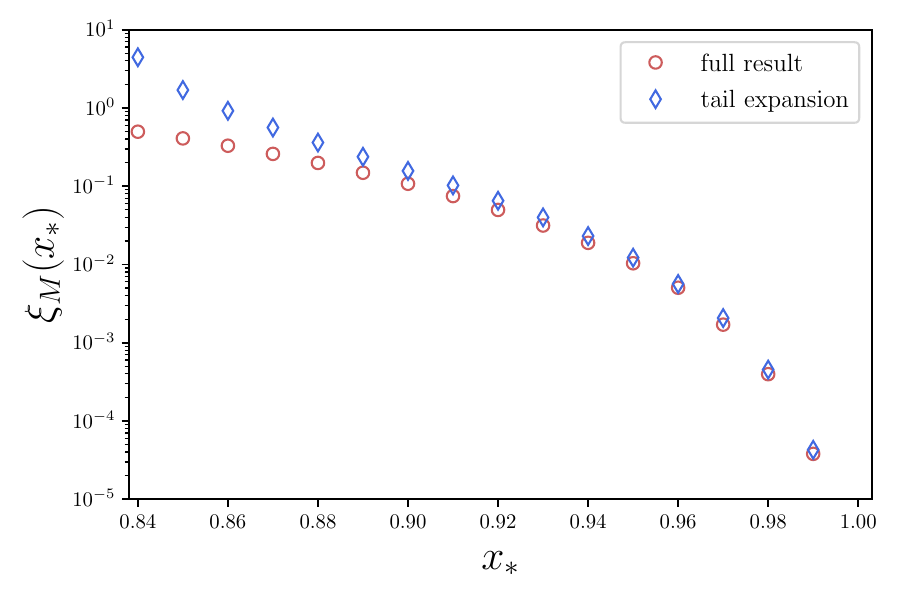}
 \end{minipage}
 \begin{minipage}{.495\textwidth}
 \includegraphics[width=.9\textwidth]{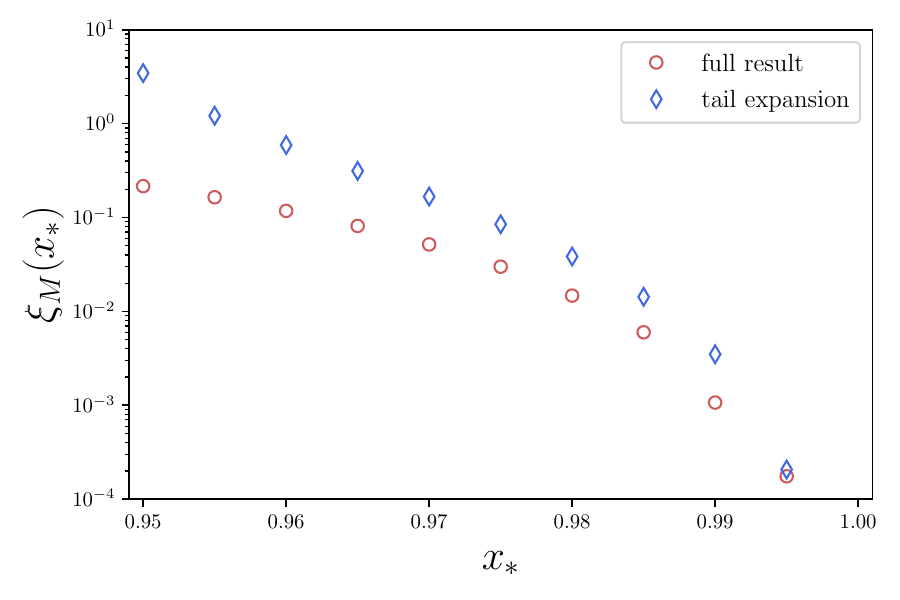}
 \end{minipage}
  \caption{Large-distance behaviour of the reduced correlation function in the tilted-well model, for $d=0.3$, $\mu=13$ (left panel) and for $d=0.2$, $\mu=25$ (right panel), as a function of $x_*$. The coarse-graining size is fixed by $x_1=x_2=0.6$ in both cases. Circles denotes the full stochastic result obtained by numerically integrating the numerically-reconstructed one-point and two-point distributions above $\zeta_{\mathrm{c}}=1$, whereas diamonds denote the tail expansion given by the prefactor in \Eq{eq:tilted:well:2pt:tail:exp}, where the residues and the poles have been found numerically (see footnote~\ref{footnote:tail_exp_tilt_2}) and the integral over $\calN_{x_0\to x_*}$ evaluated numerically.}
 \label{fig:ClusteringTail}
\end{figure}

The reduced correlation function is obtained from integrating the one-point and the two-point distribution functions above the PBH threshold, as explained above, and the result is shown in \Fig{fig:ClusteringTail} for the same two sets of parameters as used in previous figures. At large distance, the tail approximation, according to which $1+\xi$ is given by the prefactor in \Eq{eq:tilted:well:2pt:tail:exp}, provides a good fit to the full result. The reason why this approximation may break down at small distance is that, in that case, $\mathcal{N}_{\phi_0\to\phi_1}=\mathcal{N}_{\phi_0\to\phi_*}+\mathcal{N}_{\phi_*\to\phi_1} $ is dominated by $\mathcal{N}_{\phi_0\to\phi_*}$, hence $\Pfpt{\phi_*\to\phi_1}$ cannot always be tail-expanded in \Eq{eq:2pt:single:clock}, and likewise for $\Pfpt{\phi_*\to\phi_2}$. 

Note that the large-volume approximation, under which the above results were derived, requires that $r^3\gg R_1^3,\,  R_2^3$, hence the large-distance limit is the one of main interest anyway. But it is still important to stress that the tail-expansion formula~\eqref{eq:tilted:well:2pt:tail:exp} may not be sufficient in cases where $\zeta_\uc$ probes the intermediate tail.

One also notices that the validity of the tail expansion declines when $\mu^2 d$ is larger. This is because, as mentioned above, $\mu^2 d\gg 1$ corresponds to the classical limit, in which the first-passage time distribution is quasi-Gaussian except in its very-far tail, hence $\zeta_\uc$ does not fall in the exponential tail unless $x_*$ is very close to one.

\subsubsection*{Comparison with the classical limit}

Let us now compare our results with the classical, standard calculation. As mentioned above, the classical limit corresponds to $\mu^2 d\to \infty$, hence we expect standard results to be recovered in the wide-well regime. For instance, for the mean number of \efolds, the second term in \Eq{eq:meanN:constant:slope} is exponentially suppressed when $\mu^2 d\gg 1$, and the first term precisely corresponds to the duration of inflation when the stochastic noise is switched off.

At leading order in cosmological perturbation theory, the curvature perturbation features Gaussian statistics, and so does its coarse-grained version. Therefore, the one-point distribution $P(\zeta_R)$ is Gaussian, with a variance given by
\bea 
\label{eq:sigmaR2}
\sigma_R^2 \equiv \left\langle \zeta_R^2 \right\rangle = \int_0^{a/R}\dd\ln k\,  \mathcal{P}_{\zeta}(k)\, ,
\eea 
see the discussion around \Eq{eq:zetaR:Fourier}. In the tilted-well model, $\calP_\zeta=v/\epsilon_1=2v/\alpha^2$ at leading order in slow roll and in the regime of \Eq{eq:constant:slope:vacuum:dom:cond}. This makes the integral of \Eq{eq:sigmaR2} diverge, but one should recall that it only applies to those scales exiting the Hubble radius in the tilted well. Ignoring the contribution from larger scales, which were also discarded in the above stochastic-$\delta N$ calculation, the lower bound of the integral~\eqref{eq:sigmaR2} should be replaced with $k_{\mathrm{IR}}= a_\uend H \ee^{-1/d}$. Indeed, this scale matches the Hubble radius at the entry of the tilted well, when $x=1$.
Using the simple PBH formation criterion adopted above, see \Eq{eq:pM:zeta:threshold}, one thus has
\bea\label{P1_class}
p_M = \frac{1}{2}\erfc\left(\frac{\zeta_\uc}{\sqrt{2 \sigma_R^2}}\right) .
\eea

The two-point distribution is also Gaussian. The diagonal entries of its covariance matrix are given by $\sigma_{R_1}^2$ and $\sigma_{R_2}^2$, and its off-diagonal entry is already expressed in terms of the power spectrum in \Eq{eq:two:pt:correl:Fourier}, \ie
\bea 
\tau_r^2 = \left\langle \zeta_{R_1}(\vec{x}) \zeta_{R_2}(\vec{x}+\vec{r})\right\rangle = \int_0^{a/r}\dd\ln k\,  \mathcal{P}_{\zeta}(k)\, ,
\eea 
where the lower bound of the above integral should again be replaced with $k_{\mathrm{IR}}= a_\uend H \ee^{-1/d}$ in the tilted-well model. The probability to form two black holes with masses $M$ at distance $r$ can then be computed using \Eq{eq:pM1M2:zeta:threshold} and one finds~\cite{Ali-Haimoud:2018dau}
\bea\label{P2_class}
p_{M,M}(r)=
\frac{1}{\sqrt{2\pi}} \int_0^\infty \dd x e^{-x^2/2} \erfc{\left[\frac{\zeta_\uc}{\sqrt{\sigma_R^2+\tau_r^2}}\left(1+\sqrt{\frac{\sigma_R^2-\tau_r^2}{2}} \frac{x}{\zeta_\uc}\right)\right]}\, ,
\eea 
where we have restricted the above formula to the case where the two black holes have the same mass for simplicity. 

\begin{figure}[t] 
	\centering
	\includegraphics[width=0.495\hsize]{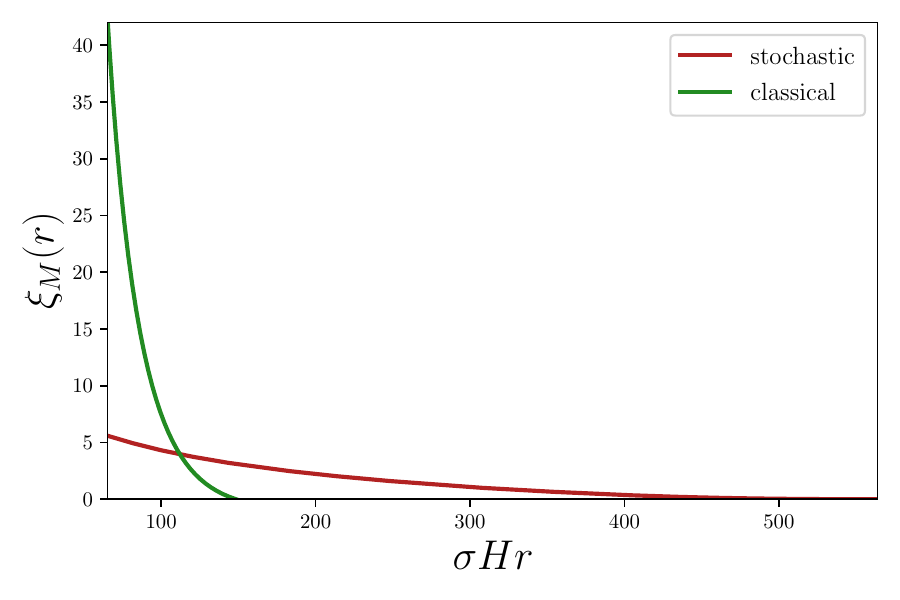}
 \includegraphics[width=0.495\hsize]{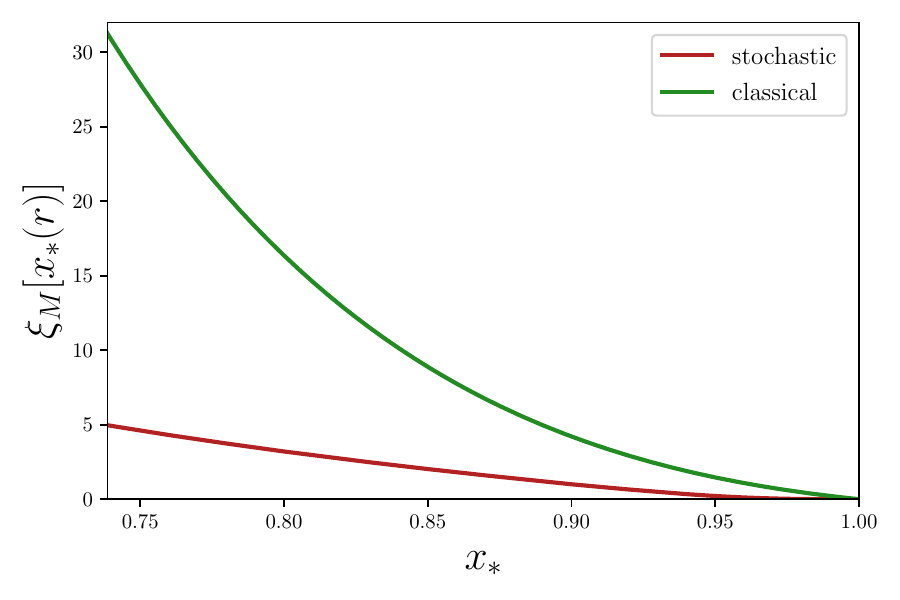}
	\caption{Reduced correlation function in the tilted-well potential, for $\mu=25, d=0.2$. The solid red line stands for the full stochastic calculation with $x_1=x_2=0.6$, while the green line stands for the classical result obtained from Eqs.~\eqref{P1_class} and~\eqref{P2_class}. In the left panel, the reduced correlation function is displayed as a function of $r$, and $R$ in the classical formulas is set to be the same as $R_1=R_2$ of the stochastic setup. The lower bound on $r$ is set by the requirement $r>R_1 + R_2$. In the right panel, the reduced correlation function is displayed as a function of $x_*$, the (rescaled-) field value in the tilted-well, where in the classical formulas $R=e^{x_1/d}=e^{x_2/d}$, where $x_1=x_2=0.6$. The lower bound on $x_*$ is extracted by the requirement $r(x_*)>R_1(x_1)+R_2(x_2).$}
	\label{fig:Clusterigr}
\end{figure}

From these expressions, the reduced correlation~\eqref{eq:clustering:def} can be computed, and it is compared with our stochastic results in \Fig{fig:Clusterigr}. An important observation is that larger distances $r$ are covered in the stochastic calculation than in its classical counterpart. The reason is that, in the stochastic setup, the relationship between scales and field values is given by the mean volume as argued in \Sec{subsec:Large:vol:approx}. Therefore, the largest distance produced in the tilted well is $\tilde{r}=2\langle \ee^{3\mathcal{N}} \rangle_{x=1}^{1/3}$,
where the mean volume is given in \Eq{eq:vol:const:slope:char}. In the classical limit, scales and field values are simply related by the (classical) number of \efolds, hence the largest distance one can probe is given by $r=\ee^{1/d}$, where distances are quoted in $\sigma$-Hubble units. We thus reach the conclusions that PBHs are correlated over longer distances once quantum diffusion is taken into account.

One may wonder if the discrepancy between the classical and the stochastic results in the left panel of \Fig{fig:Clusterigr} could be merely due to this scale-field distorted relationship. To check whether or not this is the case, in the right panel of \Fig{fig:Clusterigr} the reduced correlation is displayed in terms of $x_*$, for PBH masses such that the same values of $x_1$ and $x_2$ are taken in both approaches. We thus compare the $\xi(x_*,x_1,x_2)$ functions rather than the $\xi(r,R_1,R_2)$ functions. One can see that the two clustering profiles are qualitatively similar, which confirms that the scale-field distortion is the main reason for the large difference in the left panel. Nonetheless, substantial differences are still present although $\mu^2 d=125$ and such large values leave exponentially suppressed corrections in the one and two-point functions of $\mathcal{N}$, see for instance \Eq{eq:meanN:constant:slope}. This is because PBHs are sensitive to the intermediate tail, which becomes Gaussian much deeper in the classical regime than the neighborhood of the maximum of the distribution.

\section{Discussion and conclusion}
\label{sec:Conclusion}

In this work, we have computed the amount of clustering between primordial black holes that are seeded by highly non-Gaussian fluctuations. Since PBHs require large fluctuations that can hardly be described by perturbative, quasi-Gaussian approaches, previous results on clustering between Gaussian or quasi-Gaussian peaks had indeed to be extended to fully non-Gaussian statistics. In practice, we considered inflationary models where quantum diffusion is included via the stochastic-inflation formalism, and is known to be responsible for heavy non-Gaussian tails in the statistics of cosmological fluctuations. This first led us to revisiting the way first-passage-time statistics is connected to coarse-grained curvature perturbations in the stochastic-$\delta N$ formalism, improving on previous works with respect to two aspects. 

First, we have clarified how physical distances measured by a local observer on the end-of-inflation hypersurface are related to initial patches during inflation, in a way that is free from any approximation. This relies on computing the final volume emerging from a given patch, and its distribution function conditioned to the field values within that patch. This distribution function is defined operationally via a recursive sampling algorithm, that could lead to explicit numerical implementations in the future. In the limit where the distances of interest are large compared to the Hubble radius, we have argued that the (volume-weighted) mean of that distribution can be used as a proxy, which greatly simplifies the analysis. In practice, most calculations presented in this work were done in this ``large-volume'' limit, although the framework was developed for a full treatment to be carried out in the future. 

Second, we have consistently implemented volume weighting. Different regions of space inflate for different amounts, but those that inflate most contribute a larger volume on the end-of-inflation hypersurface. As a consequence, a local observer located on that final hypersurface and measuring the statistics of coarse-grained fields by recording their values in each coarse-grained patch, encounters more patches in regions that have inflated for longer. This implies that the statistics reconstructed by local observers are volume-weighted. In standard cosmological perturbation theory, all patches inflate for almost the same amount, hence the volume-weighting distortion only arises at higher order. In the presence of large quantum diffusion, this is not the case anymore, hence volume weighting has to be properly included in the formalism and we have proposed a framework to do so.

We have then applied these methods to compute clustering in two toy models, one where the inflationary potential is exactly flat (the ``flat well'') and the first-passage times have highly non-Gaussian distributions, and one where the inflationary potential is linearly tilted (the ``tilted well'') where these distributions are nearly Gaussian around their maximum, and deviate from Gaussianity only in the far tails.

In the flat well, we have checked that direct sampling methods give numerical results that are in agreement with the analytical formulas that can be obtained in this model. Both the one-point and the two-point distributions of the curvature perturbation have exponential tails, and the reduced correlation decays with the distance between the black holes in a way that we were able to characterise. This model is interesting since it is analytically solvable, however it does not have a classical counterpart hence it does not allow one to describe the clustering properties that arise as an effect of quantum diffusion per se.

This is why we then considered the tilted well, which is only semi-analytically tractable, but for which a direct comparison with standard cosmological perturbation theory can be performed. The one- and two-point distributions still have exponential tails, but they are close to Gaussian near their maxima. Due to the volume weighting distorsion, the range of scales (hence the range of PBH masses and distances over which they are correlated) that emerges from the tilted well is very different in the classical and stochastic frameworks, even in the regime where non-Gaussianities only appear in the far tail. The reason is that volume-weighting modulates the one-point distributions $P(\zeta_R)$ by $\ee^{3\zeta_R}$, and the two-point distribution by a similar (although less straightforward) factor, hence it enhances the contribution from the tails. We thus observed that PBHs can be created with larger masses, and with spatial correlations across longer distances, once quantum diffusion is included. 

In both models, the two-point distribution is endowed with a peculiar structure on its tail, which can be seen in \Eqs{eq:quantum:well:2pt:tail:exp} and~\eqref{eq:tilted:well:2pt:tail:exp}. This structure is in fact generic for the following reason. If field space is compact, on its tail the first-passage-time distribution is exponential and can always be approximated by its leading pole 
\bea
\Pfpt{\phi_\uin\to\phi}(\mathcal{N})\simeq a(\phi_\uin,\phi)\ee^{-\Lambda_0(\phi) \mathcal{N}}\, ,
\eea 
where, crucially, $\Lambda_0$ does not depends on the initial condition $\phi_\uin$~\cite{Ezquiaga:2019ftu}. After volume weighting, one thus has
\bea
\PfptV{\phi_\uin\to\phi}(\mathcal{N})\simeq a_{\mathrm{V}}(\phi_\uin,\phi)\ee^{\left[3-\Lambda_0(\phi)\right] \mathcal{N}}\, ,
\eea 
where $a^{\mathrm{V}}$ is a different normalisation factor. The tail expansion of the one-point distribution~\eqref{eq:1pt:single:clock} thus gives
\bea
\label{eq:1pt:single:clock:tail}
P\left(\zeta_R\right) \simeq a_{\mathrm{V}}(\phi_0,\phi_*)\ee^{\left[3-\Lambda_0(\phi_*)\right] \left(\zeta_R-\left\langle \mathcal{N}_{\phi_*}\right\rangle_{\mathrm{V}}+\left\langle \mathcal{N}_{\phi_0}\right\rangle_{\mathrm{V}}\right)}\, .
\eea
As explained above, on the tail of the two-point distribution, the last two first-passage-time distributions appearing in \Eq{eq:2pt:single:clock} can be tail-expanded, which leads to
\bea
\label{eq:2pt:single:clock:tail}
P(\zeta_{R_1},\zeta_{R_2}) \simeq  &
a_{\mathrm{V}}\left(\phi_*,\phi_1\right)a_{\mathrm{V}}\left(\phi_*,\phi_2\right)
\\ & 
\ee^{\left[3-\Lambda_0(\phi_1)\right] \left(\zeta_{R_1}+\left\langle \mathcal{N}_{\phi_0} \right\rangle_{\mathrm{V}}-\left\langle \mathcal{N}_{\phi_1} \right\rangle_{\mathrm{V}}\right)}
\ee^{\left[3-\Lambda_0(\phi_2)\right] \left(\zeta_{R_2}+\left\langle \mathcal{N}_{\phi_0} \right\rangle_{\mathrm{V}}-\left\langle \mathcal{N}_{\phi_2} \right\rangle_{\mathrm{V}}\right)}
\\ & 
\int\dd\mathcal{N}_{\phi_0\to\phi_*} \PfptV{\phi_0\to\phi_*}\left(\mathcal{N}_{\phi_0\to\phi_*}\right)
\ee^{\left[\Lambda_0(\phi_1)+\Lambda_0(\phi_2)-6\right]\mathcal{N}_{\phi_0\to\phi_*}} .
\eea
Combining the two above equations, one obtains
\bea
\label{eq:2pt:almost:factorised}
P(\zeta_{R_1},\zeta_{R_2}) \simeq F\left(R_1,R_2,r\right) P(\zeta_{R_1}) P(\zeta_{R_2})
\eea 
where
\bea
\label{eq:F:def}
F\left(R_1,R_2,r\right) = \frac{a_{\mathrm{V}}\left(\phi_*,\phi_1\right)}{a_{\mathrm{V}}\left(\phi_0,\phi_1\right)}\frac{a_{\mathrm{V}}\left(\phi_*,\phi_2\right)}{a_{\mathrm{V}}\left(\phi_0,\phi_2\right)}
\int\dd\mathcal{N} \PfptV{\phi_0\to\phi_*}\left(\mathcal{N}\right)
\ee^{\left[\Lambda_0(\phi_1)+\Lambda_0(\phi_2)-6\right]\mathcal{N}}\, .
\eea
This generalises the structure observed in \Eq{eq:quantum:well:2pt:tail:exp}, which has important consequences. Indeed, the two-point distribution ``almost'' factorises, up to the factor $F(R_1,R_2,r)$. This implies that, although the black holes are correlated, these correlations are of a simple kind. For instance, using Baye's theorem, one has
\bea
P\left(\zeta_{R_2} \vert \zeta_{R_1}\right) = P\left(\zeta_{R_2}\right)F\left(R_1,R_2,r\right) ,
\eea 
so the conditional distribution of $\zeta_{R_2}$ at fixed $\zeta_{R_1}$ does not depend on $\zeta_{R_1}$.\footnote{Strictly speaking, the upper bound in the integral of \Eq{eq:F:def} is such that the arguments of the two other first-passage time distributions in \Eq{eq:2pt:single:clock}, $\zeta_{R_1}-\mathcal{N}_{\phi_0\to\phi_*}+\left\langle \mathcal{N}_{\phi_0} \right\rangle_{\mathrm{V}}-\left\langle \mathcal{N}_{\phi_1} \right\rangle_{\mathrm{V}}$ and $\zeta_{R_2}-\mathcal{N}_{\phi_0\to\phi_*}+\left\langle \mathcal{N}_{\phi_0} \right\rangle_{\mathrm{V}}-\left\langle \mathcal{N}_{\phi_2} \right\rangle_{\mathrm{V}}$, remain positive. Thus it implicitly depends on $\zeta_{R_1}$ and $\zeta_{R_2}$. However, at large distance $r$ that upper bound is always located in the tail of $\PfptV{\phi_0\to\phi_*}$, hence its contribution is exponentially suppressed.}
In other words, under the condition that $\zeta_{R_1}$ and $\zeta_{R_2}$ are  located on the tail, their values are uncorrelated. 

Another way to consider this result is to notice that, from \Eq{eq:2pt:almost:factorised}, the reduced correlation reads
\bea
\xi= F\left(R_1,R_2,r\right)-1\, ,
\eea 
which does not depend on the formation threshold $\zeta_\uc$. 
The reason is that, if $\zeta_\uc$ is located on the exponential tail, then $\zeta_{R_1}$ and $\zeta_{R_2}$ also have to lie on the exponential tail, hence their values are uncorrelated. As a consequence $\xi$ only measures the correlation between the two events ``$\zeta_{R_1}$ is in the exponential-tail regime'' and ``$\zeta_{R_2}$ is in the exponential-tail regime'', which depends on the value of $\zeta_{R_1}$ and $\zeta_{R_2}$ at which the exponential tail takes over, but not on $\zeta_\uc$.
This leads us to another important conclusion: any type of cosmological structure that requires a critical value $\zeta_\uc$ that lies within the exponential tail, features the same reduced correlation function. We have thus found a universal clustering behaviour, that applies to PBHs but also to all tail-born structures, independently of their formation threshold. 

Finally, the fact that the reduced correlation is independent of $\zeta_\uc$ implies that, in the large-peak limit, clustering is always larger when quantum diffusion is included. Indeed, as mentioned in \Sec{sec:Introduction}, in the large-threshold limit Gaussian clustering is suppressed by the ratio between the squared threshold and the field variance~\cite{Ali-Haimoud:2018dau, Desjacques:2018wuu}. In the stochastic picture however, it is independent of the threshold in that same limit, hence it is necessarily larger.
Note that the same conclusions would be drawn in the absence of volume weighting, \ie \Eqs{eq:2pt:almost:factorised} and~\eqref{eq:F:def} would still be valid, with non-volume-weighted quantities in \Eq{eq:F:def}.

However, when PBHs (or any structure of interest) form in the intermediate tail, \ie the region of the tail that is not yet exponential, the above limit does not always apply. This could be seen in the tilted-well model when parameters are chosen in the classical regime (the so-called wide-well limit), where the tail approximation is valid at large distances only. In this case different behaviours can be observed, and we found situations where the classical calculation overestimates the reduced correlation function.

Let us now mention a few directions that we think deserve further investigations.
First, a numerical analysis is required to go beyond the large-volume approximation and to test the validity of this approximation scheme. This requires to implement the recursive set of algorithms detailed in \App{app:recursivecode}. In this way, the exact mapping between distances measured by a local observer at the end of inflation and field values in the corresponding parent patches during inflation would be reconstructed. This would also allow us to test the assumption that the region emerging from a given patch is spherically symmetric. Note that a lattice code is not required to do so, since this approach only relies on sampling uncoupled Langevin equations. Lattice simulations would be required to include finite gradient effects, which may be necessary in models with sudden transitions~\cite{Jackson:2023obv}, but they are not  needed to reconstruct the real-space structure of the curvature perturbation. In stochastic inflation indeed, the physical distance between super-Hubble distant points is related to their past causal overlap. This provides inflating backgrounds with a graph structure that we have discussed and along which stochastic inflation should be solved.

Second, the volume-weighting procedure makes the eternal-inflation problem inevitable. If the first-passage-time distribution decays as $P(\mathcal{N})\propto \ee^{-\Lambda_0 \mathcal{N}}$ on its tail, its volume-weighted version scales as $P^{\mathrm{V}}(\mathcal{N})\propto \ee^{(3-\Lambda_0) \mathcal{N}}$, which diverges when $\Lambda_0\leq 3$. This occurs in a broad class of models. For instance, when field space is non compact, the tail of the first-passage-time distribution is monomial rather than exponential, and its volume-weighted version always diverges. This is the case for all single-field models currently favoured by the data (except for the sub-class of hilltop models that are still in agreement with CMB measurements). In this work, we have avoided this problem by considering toy models in which $\Lambda_0>3$, but it has to be addressed in order to extend our analysis to more realistic scenarios. Although there is no clear path for doing so at this stage, let us note that the quantity that is subject to divergences, \ie the first-passage time from a given field configuration, is not a quantity one can observe. Instead, a local observer only has access to a finite region around them, in which inflation did terminate, so by constructing the formalism only in terms of backward quantities one might be able to overcome these difficulties.

Third, in this work cosmological fluctuations have been discussed in terms of the comoving perturbation, since it is a natural outcome of the $\delta N$ formalism and given that it can be non-linearly defined at large scales. However, formation criteria are more often discussed in terms of the density contrast or the compaction function, since those have been showed to better capture the fate of local inhomogeneities. The calculation of these quantities in the present framework thus needs to be implemented, possibly along the lines of \Refa{Tada:2016pmk} where the compaction function is related to the coarse-shelled curvature perturbation. Another possibility is to investigate simple models where the full statistics of curvature perturbations can be extracted, such as in \Refa{Raatikainen:2023bzk}. Let us also mention the approach developed in \Refs{Kitajima:2021fpq, Gow:2022jfb, Ferrante:2022mui}, where a generalised-local ansatz is considered, $\zeta(r)=f[\zeta_{\mathrm{G}}(r),r]$, where $\zeta_{\mathrm{G}}$ is a Gaussian field, $r$ is the radial distance from the peak (assuming spherical symmetry) and $f$ is an arbitrary, possibly highly non-linear, function. The one-point statistics of the compaction function can be computed in this model, and one may consider extending the analysis to two-point distributions, but the local nature of the model implies that spatial correlations are mostly Gaussian, hence it is not clear that they can capture the statistics obtained from quantum diffusion.

Finally, the phenomenological consequences of our findings need to be further explored. In particular, we have found that the typical correlation length between PBHs is increased in the presence of quantum diffusion. At fixed PBH mean density, this implies that clusters are more likely to be formed; and if they form, they should tend to contain more black holes. This should have important implications for their subsequent merging rate, hence for the stochastic gravitational wave background to be expected from them.

\acknowledgments
It is a pleasure to thank Pierre Auclair and Baptiste Blachier for their feedback on this manuscript and for insightful discussions.

\appendix
\addtocontents{toc}{\protect\setcounter{tocdepth}{1}}

\section{Recursive algorithms for forward sampling}
\label{app:recursivecode}

In this appendix, we provide the algorithms for recursive codes that would sample the quantities introduced in \Sec{sec:forward:stat}.

\subsection{Final volume}

Starting from a patch with field value $\bm{\Phi}_*$, the following function produces one realisation of the final volume emerging from that patch, normalised to $V_*$ (see the discussion around \Fig{fig:sketch:volume}).
\\ $ $\\
\texttt{
function $V(\bm{\Phi}_*)$:
\begin{enumerate}
\item Set the initial volume $V=0$
\item From $\bm{\Phi}_*$, simulate two Langevin realisations with a duration of $\Delta N=\ln(2)/3$ \efolds \label{step:init:phi*}
\item For each realisation:
\subitem -- If it has ended inflation before the end of the simulation, after a time $N_\uend$, add $\ee^{3(N_\uend-\Delta N)}$ to $V$
\subitem -- Otherwise record the value $\bm{\Phi}'$ at the end of the simulation, and add $V(\bm{\Phi}')$ to $V$
\item Return $V$
\end{enumerate} 
}
This function is iterative since it calls itself in step \texttt{3}, if inflation does not end after $\Delta N$ \efolds. The factor $\ee^{3(N_\uend-\Delta N)}$ accounts for the fact that inflation may terminate on a branch, between two nodes of the tree. In this case, a leaf is created, but with a smaller size, reduced by $\ee^{3(N_\uend-\Delta N)}$.

Note also that in the case where $H$ depends on time, slow-roll corrections need to be added to $\Delta N=\ln(2)/3$, see footnote~\ref{footnote:slow:roll}. 

\subsection{Volume-averaged number of \efolds}

The above algorithm can be improved to also return the value of the volume-averaged number of \efolds realised along the wordlines emerging from a patch with field value $\mathcal{P}_*$. From \Eq{eq:volMean:N*}, this requires to compute
\bea
X\equiv \sum_{\mathrm{leaves}}  \mathcal{N} = \frac{V}{V_*} \mathbb{E}_{\mathcal{P}_*}\left[
\mathcal{N}_{\mathcal{P}_*}(\vecx)\right]\, ,
\eea
such that 
\bea
W=\mathbb{E}^{\mathrm{V}}_{\mathcal{P}_*} \left[ \mathcal{N}_{\mathcal{P}_*}(\vec{x})\right]  = \left(\frac{V_*}{V}\right)^2 X\, .
\eea
This can be done with the following function.
\\ $ $\\
\texttt{
function $XV(\bm{\Phi}_*,N)$
\begin{enumerate}
\item Set $X=0$, $V=0$
\item From $\bm{\Phi}_*$, simulate two Langevin realisations, with a duration of $\Delta N=\ln(2)/3$ \efolds \label{step:init:phi*}
\item For each realisation:
\subitem -- If it has ended inflation before the end of the simulation, after a time $N_\uend$, add $N_\uend$ to $N$, add $\ee^{3(N_\uend-\Delta N)}$ to $V$, and add $\ee^{3(N_\uend-\Delta N)} N$ to $X$
\subitem -- Otherwise add $N_\uend$ to $N$, record the value $\bm{\Phi}'$ at the end of the simulation, and call $XV(\bm{\Phi}',N)$. This returns a tuple $(X, V)$, the elements of which are added to $X$ and $V$ respectively
\item Return $(X, V)$
\end{enumerate} 
}

\section{Convolution of first-passage-time distributions}
\label{app:sub:vanishing:mean}

In this section, we show that first-passage times over separate portions of the stochastic dynamics are independent random variables. This is a natural consequence of the Markovian nature of the stochastic process, but it is instructive to check that it is indeed the case. For simplicity, we restrict the following considerations to one-dimensional processes, but they can be easily generalised to higher dimension. Let $\phi(N)$ satisfy a Langevin equation of the form~\eqref{eq:Langevin}, and let $\phi_1<\phi_2<\phi_3$ be three fixed values. Starting from $\phi_1$, $\mathcal{N}_{\phi_1\to \phi_3}$ denotes the first-passage time through $\phi_3$, and since realisations crossing $\phi_3$ must have crossed $\phi_2$ before, one must have
\bea
\mathcal{N}_{\phi_1\to\phi_3} = \mathcal{N}_{\phi_1\to\phi_2} +\mathcal{N}_{\phi_2\to\phi_3} \,.
\eea
Our goal is to show that this is true, and that $\mathcal{N}_{\phi_1\to\phi_2}$ and $\mathcal{N}_{\phi_2\to\phi_3}$ are independent.

Given a first-passage-time distribution $\Pfpt{\phi}(\mathcal{N})$, it can be associated with the characteristic function
\bea
\chi_{\calN}(t,\phi)=\langle e^{i t \calN} \rangle=\int_{-\infty}^{\infty} \dd \calN e^{i t \calN} \Pfpt{\phi}(\calN)\,,
\eea
where $t$ is a dummy variable and $\phi$ the initial field configuration. Namely, the characteristic function is nothing but the Fourier transform of the PDF of $\calN$, which satisfies the adjoint Fokker-Planck equation~\eqref{eq:Fokker-Planck}. This implies that $\Lfp^\dagger({\Phi})\cdot \chi_{\calN}(t,\phi) = -it \chi_{\calN}(t,\phi)$. The latter is a second-order differential equation in the variable $\phi$, so its solution is of the form
\bea
\chi_{\phi_\uend}(t,\phi) = A(t) f(\phi,t) + B(t) g(\phi,t)\, .
\eea  
Here, the two constants of integration $A(t)$ and $B(t)$ are set to satisfy some prescribed boundary conditions. One boundary condition is that the FPT distribution reduces to a Dirac delta distribution when the initial condition is set at the final value $\phi=\phi_\uend$, \ie $\chi(t,\phi_\uend)=1$. Another (\eg reflective) boundary condition may also be imposed, but it does not play an important role here. We have
\bea
\chi_{\phi_\uend}(t,\phi) = \frac{f(\phi,t) + \alpha(t) f(\phi,t)}{f(\phi_\uend,t) + \alpha(t) f(\phi_\uend,t)}\,,
\eea
where $\alpha$ is fixed by the additional boundary condition. This formula allows us to re-write the trivial identity
\bea
\frac{f(\phi_1,t) + \alpha(t) f(\phi_1,t)}{f(\phi_3,t) + \alpha(t) f(\phi_3,t)} = \frac{f(\phi_1,t) + \alpha(t) f(\phi_1,t)}{f(\phi_2,t) + \alpha(t) f(\phi_2,t)}  \frac{f(\phi_2,t) + \alpha(t) f(\phi_2,t)}{f(\phi_3,t) + \alpha(t) f(\phi_3,t)} 
\eea
as
\bea
\chi_{\phi_3}\left(t,\phi_1\right) = \chi_{\phi_2}\left(t,\phi_1\right)\chi_{\phi_3}\left(t,\phi_2\right)\, .
\eea
When characteristic functions get multiplied, their inverse Fourier transforms, \ie the first-passage time distributions, get convolved (since the convolution product is mapped to regular product in Fourier space), hence
\bea
\Pfpt{\phi_1\to\phi_3}\left(\mathcal{N}\right) = \int_0^\infty \dd \mathcal{N}_1 \Pfpt{\phi_1\to\phi_2}\left(\mathcal{N}_1\right) \Pfpt{\phi_2\to\phi_3}\left(\mathcal{N}-\mathcal{N}_1\right) \, .
\eea
This implies that 
\bea
\mathcal{N}_{\phi_1\to\phi_3} = \mathcal{N}_{\phi_1\to\phi_2} +\mathcal{N}_{\phi_2\to\phi_3}  
\eea
and that the two variables on the right-hand-side are indeed independent (and that they correspond to first-passage times since their PDFs derive from the characteristic function of the first-passage-time problem). 

\section{Power spectrum from the one-point distribution}
\label{app:power:spectrum:one:point}

In this appendix, we extract the power spectrum from the one-point distribution function of the coarse-grained curvature perturbation, in single-clock models and in the large-volume approximation. The same calculation is performed using the two-point distribution function in \Sec{sec:single:clock:models}, and our goal is to check that the same result is obtained from the one-point distribution function.

Let us start from
\bea
\left\langle \zeta_R^2\right\rangle = & \int  \zeta_R^2 P(\zeta_R) \dd\zeta_R\\ = &
\int \dd\zeta_R \PfptV{\phi_0\to\phi_*}\left(\zeta_R+\left\langle \mathcal{N}_{\phi_0}\right\rangle_{\mathrm{V}} - \left\langle \mathcal{N}_{\phi_*} \right\rangle_{\mathrm{V}}\right)\zeta_R^2\, ,
\eea
where \Eq{eq:1pt:single:clock} has been used in the second line. One can then perform the change of integration variable $\mathcal{N}_{\phi_0\to\phi_*} = \zeta_R+\langle \mathcal{N}_{\phi_0}\rangle_{\mathrm{V}} - \langle \mathcal{N}_{\phi_*} \rangle_{\mathrm{V}}$, leading to
\bea
\left\langle \zeta_R^2\right\rangle  = &
\int \dd\mathcal{N}_{\phi_0\to\phi_*}
\PfptV{\phi_0\to\phi_*}\left(\mathcal{N}_{\phi_0\to\phi_*}\right)\left(\mathcal{N}_{\phi_0\to\phi_*}-\left\langle \mathcal{N}_{\phi_0}\right\rangle_{\mathrm{V}} + \left\langle \mathcal{N}_{\phi_*} \right\rangle_{\mathrm{V}}\right)^2 \\
= & 
\left\langle \mathcal{N}_{\phi_0\to\phi_*}^2\right\rangle_{\mathrm{V}}
+2 \left(\left\langle \mathcal{N}_{\phi_*} \right\rangle_{\mathrm{V}}- \left\langle \mathcal{N}_{\phi_0}\right\rangle_{\mathrm{V}}   \right)\left\langle \mathcal{N}_{\phi_0\to\phi_*}\right\rangle_{\mathrm{V}}
+  \left(\left\langle \mathcal{N}_{\phi_*} \right\rangle_{\mathrm{V}}- \left\langle \mathcal{N}_{\phi_0}\right\rangle_{\mathrm{V}}   \right)^2\, .
\eea
Using the result established in \Eq{eq:convolv:vol:weighting}, one can replace $\langle\mathcal{N}_{\phi_0\to\phi_*}\rangle_{\mathrm{V}} = \langle\mathcal{N}_{\phi_0}\rangle_{\mathrm{V}}-\langle\mathcal{N}_{\phi_*}\rangle_{\mathrm{V}}$, which gives
\bea
\label{eq:zetaR2mean:inter:single:clock}
\left\langle \zeta_R^2\right\rangle  = &
\left\langle \mathcal{N}_{\phi_0\to\phi_*}^2\right\rangle
-  \left(\left\langle \mathcal{N}_{\phi_*} \right\rangle_{\mathrm{V}}- \left\langle \mathcal{N}_{\phi_0}\right\rangle_{\mathrm{V}}   \right)^2\, .
\eea
Let us then square the relation $\mathcal{N}_{\phi_0}=\mathcal{N}_{\phi_0\to\phi_*}+\mathcal{N}_{\phi_*}$ and take its (volume-weighted) mean value:
\bea
\left\langle\mathcal{N}_{\phi_0}^2\right\rangle_{\mathrm{V}} = \left\langle \mathcal{N}_{\phi_0\to\phi_*}^2\right\rangle_{\mathrm{V}}+\left\langle\mathcal{N}_{\phi_*} ^2\right\rangle _{\mathrm{V}}+ 2 \left\langle  \mathcal{N}_{\phi_0\to\phi_*} \mathcal{N}_{\phi_*}
\right\rangle_{\mathrm{V}}\,.
\label{eq:interm:mean2:indep}
\eea
Since $\mathcal{N}_{\phi_0\to\phi_*}$ and  $\mathcal{N}_{\phi_*}$ are independent variables, one has $\left\langle  \mathcal{N}_{\phi_0\to\phi_*} \mathcal{N}_{\phi_*}
\right\rangle_{\mathrm{V}}=\left\langle  \mathcal{N}_{\phi_0\to\phi_*}\right\rangle_{\mathrm{V}} \left\langle \mathcal{N}_{\phi_*}
\right\rangle_{\mathrm{V}}$ as shown around \Eq{eq:mean:product}. This leads to
\bea
\left\langle \mathcal{N}^2_{\phi_0\to\phi_*}\right\rangle_{\mathrm{V}} = &
\left\langle \mathcal{N}_{\phi_0}^2\right\rangle_{\mathrm{V}} - \left\langle \mathcal{N}_{\phi_*}^2\right\rangle_{\mathrm{V}} - 2 \left\langle \mathcal{N}_{\phi_0\to\phi_*}\right\rangle_{\mathrm{V}} \left\langle \mathcal{N}_{\phi_*}\right\rangle_{\mathrm{V}} \\
= &\left\langle \mathcal{N}_{\phi_0}^2\right\rangle_{\mathrm{V}} - \left\langle \mathcal{N}_{\phi_*}^2\right\rangle_{\mathrm{V}} - 2\left(\left\langle \mathcal{N}_{\phi_0}\right\rangle_{\mathrm{V}}-\left\langle \mathcal{N}_{\phi_*}\right\rangle_{\mathrm{V}}\right) \left\langle \mathcal{N}_{\phi_*}\right\rangle_{\mathrm{V}}\, ,
 \eea 
 where we have used the identity $\langle\mathcal{N}_{\phi_0\to\phi_*}\rangle_{\mathrm{V}} = \langle\mathcal{N}_{\phi_0}\rangle_{\mathrm{V}}-\langle\mathcal{N}_{\phi_*}\rangle_{\mathrm{V}}$. 
Inserting this formula into \Eq{eq:zetaR2mean:inter:single:clock} one finds
 \bea
\left\langle \zeta_R^2\right\rangle  = &
\left\langle \mathcal{N}_{\phi_0}^2\right\rangle_{\mathrm{V}} - \left\langle \mathcal{N}_{\phi_*}^2\right\rangle_{\mathrm{V}} - 2\left(\left\langle \mathcal{N}_{\phi_0}\right\rangle_{\mathrm{V}}-\left\langle \mathcal{N}_{\phi_*}\right\rangle_{\mathrm{V}}\right) \left\langle \mathcal{N}_{\phi_*}\right\rangle_{\mathrm{V}}-
\left(\left\langle \mathcal{N}_{\phi_*} \right\rangle_{\mathrm{V}}- \left\langle \mathcal{N}_{\phi_0}\right\rangle_{\mathrm{V}}   \right)^2
\\
= & 
\left\langle \mathcal{N}_{\phi_0}^2\right\rangle_{\mathrm{V}} - \left\langle \mathcal{N}_{\phi_*}^2\right\rangle_{\mathrm{V}} 
-\left\langle \mathcal{N}_{\phi_0}\right\rangle_{\mathrm{V}}^2 + \left\langle \mathcal{N}_{\phi_*}\right\rangle_{\mathrm{V}}^2\\
= & \left\langle \delta\mathcal{N}^2_{\phi_0}\right\rangle_{\mathrm{V}} - \left\langle \delta\mathcal{N}^2_{\phi_*}\right\rangle_{\mathrm{V}}\,.
\eea
This coincides with the expression~\eqref{eq:two:point:corr} obtained for the two-point correlation function in the coincident limit, which is consistent. Indeed, in the case where $R_1=R_2$ and $r=0$, $\phi_*$ plays the same role in both calculations. 

One can also compute $\langle \zeta_R^2\rangle $ in Fourier space using \Eq{eq:zetaR:Fourier} and this leads to
\bea
\left\langle \zeta_R^2\right\rangle = \int \calP_\zeta(k) \widetilde{W}^2\left(\frac{kR}{a}\right) \dd\ln(k)\, ,
\eea
where $\widetilde{W}$ is a Heaviside-type window function as explained below \Eq{eq:zetaR1:zetaR2:WWsinc}.
By differentiating both sides with respect to $R$ we obtain
\bea
\calP_\zeta(k) =- \left.\frac{\partial}{\partial \ln R}\left\langle \zeta_R^2\right\rangle  \right\vert_{R=a_\uend/k}=\left.\frac{\partial}{\partial \ln R} \left \langle \delta \calN_{\phi_*}^2 \right \rangle_{\mathrm{V}}\right\vert_{R=a_\uend/k}\,,
\eea
which coincides with the expression~\eqref{eq:pow:spec:interm} obtained from the two-point statistics. This confirms the consistency of our framework.

\bibliographystyle{JHEP}
\bibliography{biblio}

\providecommand{\href}[2]{#2}\begingroup\raggedright\begin{thebibliography}{10}

\bibitem{Escriva:2022duf}
A.~Escriv\`a, F.~Kuhnel and Y.~Tada, \emph{{Primordial Black Holes}},
  \href{http://arxiv.org/abs/2211.05767}{{\tt 2211.05767}}.

\bibitem{Carr:2023tpt}
B.~Carr, S.~Clesse, J.~Garcia-Bellido, M.~Hawkins and F.~Kuhnel,
  \emph{{Observational Evidence for Primordial Black Holes: A Positivist
  Perspective}},  \href{http://arxiv.org/abs/2306.03903}{{\tt 2306.03903}}.

\bibitem{LISACosmologyWorkingGroup:2023njw}
{\scshape LISA Cosmology Working Group} collaboration, E.~Bagui et~al.,
  \emph{{Primordial black holes and their gravitational-wave signatures}},
  \href{http://arxiv.org/abs/2310.19857}{{\tt 2310.19857}}.

\bibitem{Raidal:2017mfl}
M.~Raidal, V.~Vaskonen and H.~Veerm\"ae, \emph{{Gravitational Waves from
  Primordial Black Hole Mergers}},
  \href{http://dx.doi.org/10.1088/1475-7516/2017/09/037}{\emph{JCAP} {\bf 09}
  (2017) 037}, [\href{http://arxiv.org/abs/1707.01480}{{\tt 1707.01480}}].

\bibitem{Ballesteros:2018swv}
G.~Ballesteros, P.~D. Serpico and M.~Taoso, \emph{{On the merger rate of
  primordial black holes: effects of nearest neighbours distribution and
  clustering}},
  \href{http://dx.doi.org/10.1088/1475-7516/2018/10/043}{\emph{JCAP} {\bf 10}
  (2018) 043}, [\href{http://arxiv.org/abs/1807.02084}{{\tt 1807.02084}}].

\bibitem{Young:2019gfc}
S.~Young and C.~T. Byrnes, \emph{{Initial clustering and the primordial black
  hole merger rate}},
  \href{http://dx.doi.org/10.1088/1475-7516/2020/03/004}{\emph{JCAP} {\bf 03}
  (2020) 004}, [\href{http://arxiv.org/abs/1910.06077}{{\tt 1910.06077}}].

\bibitem{Atal:2020igj}
V.~Atal, A.~Sanglas and N.~Triantafyllou, \emph{{LIGO/Virgo black holes and
  dark matter: The effect of spatial clustering}},
  \href{http://dx.doi.org/10.1088/1475-7516/2020/11/036}{\emph{JCAP} {\bf 11}
  (2020) 036}, [\href{http://arxiv.org/abs/2007.07212}{{\tt 2007.07212}}].

\bibitem{Calcino:2018mwh}
J.~Calcino, J.~Garcia-Bellido and T.~M. Davis, \emph{{Updating the MACHO
  fraction of the Milky Way dark halo with improved mass models}},
  \href{http://dx.doi.org/10.1093/mnras/sty1368}{\emph{Mon. Not. Roy. Astron.
  Soc.} {\bf 479} (2018) 2889--2905},
  [\href{http://arxiv.org/abs/1803.09205}{{\tt 1803.09205}}].

\bibitem{Gorton:2022fyb}
M.~Gorton and A.~M. Green, \emph{{Effect of clustering on primordial black hole
  microlensing constraints}},
  \href{http://dx.doi.org/10.1088/1475-7516/2022/08/035}{\emph{JCAP} {\bf 08}
  (2022) 035}, [\href{http://arxiv.org/abs/2203.04209}{{\tt 2203.04209}}].

\bibitem{Petac:2022rio}
M.~Peta\v{c}, J.~Lavalle and K.~Jedamzik, \emph{{Microlensing constraints on
  clustered primordial black holes}},
  \href{http://dx.doi.org/10.1103/PhysRevD.105.083520}{\emph{Phys. Rev. D} {\bf
  105} (2022) 083520}, [\href{http://arxiv.org/abs/2201.02521}{{\tt
  2201.02521}}].

\bibitem{Bringmann:2018mxj}
T.~Bringmann, P.~F. Depta, V.~Domcke and K.~Schmidt-Hoberg, \emph{{Towards
  closing the window of primordial black holes as dark matter: The case of
  large clustering}},
  \href{http://dx.doi.org/10.1103/PhysRevD.99.063532}{\emph{Phys. Rev. D} {\bf
  99} (2019) 063532}, [\href{http://arxiv.org/abs/1808.05910}{{\tt
  1808.05910}}].

\bibitem{DeLuca:2022uvz}
V.~De~Luca, G.~Franciolini, A.~Riotto and H.~Veerm\"ae, \emph{{Ruling Out
  Initially Clustered Primordial Black Holes as Dark Matter}},
  \href{http://dx.doi.org/10.1103/PhysRevLett.129.191302}{\emph{Phys. Rev.
  Lett.} {\bf 129} (2022) 191302}, [\href{http://arxiv.org/abs/2208.01683}{{\tt
  2208.01683}}].

\bibitem{Dokuchaev:2004kr}
V.~Dokuchaev, Y.~Eroshenko and S.~Rubin, \emph{{Quasars formation around
  clusters of primordial black holes}}, {\emph{Grav. Cosmol.} {\bf 11} (2005)
  99--104}, [\href{http://arxiv.org/abs/astro-ph/0412418}{{\tt
  astro-ph/0412418}}].

\bibitem{Dokuchaev:2008hz}
V.~I. Dokuchaev, Y.~N. Eroshenko and S.~G. Rubin, \emph{{Early formation of
  galaxies initiated by clusters of primordial black holes}},
  \href{http://dx.doi.org/10.1134/S1063772908100016}{\emph{Astron. Rep.} {\bf
  52} (2008) 779--789}, [\href{http://arxiv.org/abs/0801.0885}{{\tt
  0801.0885}}].

\bibitem{Chisholm:2005vm}
J.~R. Chisholm, \emph{{Clustering of primordial black holes: basic results}},
  \href{http://dx.doi.org/10.1103/PhysRevD.73.083504}{\emph{Phys. Rev. D} {\bf
  73} (2006) 083504}, [\href{http://arxiv.org/abs/astro-ph/0509141}{{\tt
  astro-ph/0509141}}].

\bibitem{Belotsky:2018wph}
K.~M. Belotsky, V.~I. Dokuchaev, Y.~N. Eroshenko, E.~A. Esipova, M.~Y. Khlopov,
  L.~A. Khromykh et~al., \emph{{Clusters of primordial black holes}},
  \href{http://dx.doi.org/10.1140/epjc/s10052-019-6741-4}{\emph{Eur. Phys. J.
  C} {\bf 79} (2019) 246}, [\href{http://arxiv.org/abs/1807.06590}{{\tt
  1807.06590}}].

\bibitem{DeLuca:2020jug}
V.~De~Luca, V.~Desjacques, G.~Franciolini and A.~Riotto, \emph{{The clustering
  evolution of primordial black holes}},
  \href{http://dx.doi.org/10.1088/1475-7516/2020/11/028}{\emph{JCAP} {\bf 11}
  (2020) 028}, [\href{http://arxiv.org/abs/2009.04731}{{\tt 2009.04731}}].

\bibitem{Kaiser:1984sw}
N.~Kaiser, \emph{{On the Spatial correlations of Abell clusters}},
  \href{http://dx.doi.org/10.1086/184341}{\emph{Astrophys. J. Lett.} {\bf 284}
  (1984) L9--L12}.

\bibitem{Desjacques:2018wuu}
V.~Desjacques and A.~Riotto, \emph{{Spatial clustering of primordial black
  holes}}, \href{http://dx.doi.org/10.1103/PhysRevD.98.123533}{\emph{Phys. Rev.
  D} {\bf 98} (2018) 123533}, [\href{http://arxiv.org/abs/1806.10414}{{\tt
  1806.10414}}].

\bibitem{Shibata:1999zs}
M.~Shibata and M.~Sasaki, \emph{{Black hole formation in the Friedmann
  universe: Formulation and computation in numerical relativity}},
  \href{http://dx.doi.org/10.1103/PhysRevD.60.084002}{\emph{Phys. Rev. D} {\bf
  60} (1999) 084002}, [\href{http://arxiv.org/abs/gr-qc/9905064}{{\tt
  gr-qc/9905064}}].

\bibitem{Harada:2015yda}
T.~Harada, C.-M. Yoo, T.~Nakama and Y.~Koga, \emph{{Cosmological
  long-wavelength solutions and primordial black hole formation}},
  \href{http://dx.doi.org/10.1103/PhysRevD.91.084057}{\emph{Phys. Rev. D} {\bf
  91} (2015) 084057}, [\href{http://arxiv.org/abs/1503.03934}{{\tt
  1503.03934}}].

\bibitem{Musco:2018rwt}
I.~Musco, \emph{{Threshold for primordial black holes: Dependence on the shape
  of the cosmological perturbations}},
  \href{http://dx.doi.org/10.1103/PhysRevD.100.123524}{\emph{Phys. Rev. D} {\bf
  100} (2019) 123524}, [\href{http://arxiv.org/abs/1809.02127}{{\tt
  1809.02127}}].

\bibitem{Ali-Haimoud:2018dau}
Y.~Ali-Ha\"\i{}moud, \emph{{Correlation Function of High-Threshold Regions and
  Application to the Initial Small-Scale Clustering of Primordial Black
  Holes}}, \href{http://dx.doi.org/10.1103/PhysRevLett.121.081304}{\emph{Phys.
  Rev. Lett.} {\bf 121} (2018) 081304},
  [\href{http://arxiv.org/abs/1805.05912}{{\tt 1805.05912}}].

\bibitem{Auclair:2024jwj}
P.~Auclair and B.~Blachier, \emph{{Small-scale clustering of Primordial Black
  Holes: cloud-in-cloud and exclusion effects}},
  \href{http://arxiv.org/abs/2402.00600}{{\tt 2402.00600}}.

\bibitem{MoradinezhadDizgah:2019wjf}
A.~Moradinezhad~Dizgah, G.~Franciolini and A.~Riotto, \emph{{Primordial Black
  Holes from Broad Spectra: Abundance and Clustering}},
  \href{http://dx.doi.org/10.1088/1475-7516/2019/11/001}{\emph{JCAP} {\bf 11}
  (2019) 001}, [\href{http://arxiv.org/abs/1906.08978}{{\tt 1906.08978}}].

\bibitem{Suyama:2019cst}
T.~Suyama and S.~Yokoyama, \emph{{Clustering of primordial black holes with
  non-Gaussian initial fluctuations}},
  \href{http://dx.doi.org/10.1093/ptep/ptz105}{\emph{PTEP} {\bf 2019} (2019)
  103E02}, [\href{http://arxiv.org/abs/1906.04958}{{\tt 1906.04958}}].

\bibitem{DeLuca:2021hcf}
V.~De~Luca, G.~Franciolini and A.~Riotto, \emph{{Constraining the initial
  primordial black hole clustering with CMB distortion}},
  \href{http://dx.doi.org/10.1103/PhysRevD.104.063526}{\emph{Phys. Rev. D} {\bf
  104} (2021) 063526}, [\href{http://arxiv.org/abs/2103.16369}{{\tt
  2103.16369}}].

\bibitem{Franciolini:2018vbk}
G.~Franciolini, A.~Kehagias, S.~Matarrese and A.~Riotto, \emph{{Primordial
  Black Holes from Inflation and non-Gaussianity}},
  \href{http://dx.doi.org/10.1088/1475-7516/2018/03/016}{\emph{JCAP} {\bf 03}
  (2018) 016}, [\href{http://arxiv.org/abs/1801.09415}{{\tt 1801.09415}}].

\bibitem{Pattison:2017mbe}
C.~Pattison, V.~Vennin, H.~Assadullahi and D.~Wands, \emph{{Quantum diffusion
  during inflation and primordial black holes}},
  \href{http://dx.doi.org/10.1088/1475-7516/2017/10/046}{\emph{JCAP} {\bf 10}
  (2017) 046}, [\href{http://arxiv.org/abs/1707.00537}{{\tt 1707.00537}}].

\bibitem{Ezquiaga:2019ftu}
J.~M. Ezquiaga, J.~Garc\'\i{}a-Bellido and V.~Vennin, \emph{{The exponential
  tail of inflationary fluctuations: consequences for primordial black holes}},
  \href{http://dx.doi.org/10.1088/1475-7516/2020/03/029}{\emph{JCAP} {\bf 03}
  (2020) 029}, [\href{http://arxiv.org/abs/1912.05399}{{\tt 1912.05399}}].

\bibitem{Figueroa:2020jkf}
D.~G. Figueroa, S.~Raatikainen, S.~Rasanen and E.~Tomberg, \emph{{Non-Gaussian
  Tail of the Curvature Perturbation in Stochastic Ultraslow-Roll Inflation:
  Implications for Primordial Black Hole Production}},
  \href{http://dx.doi.org/10.1103/PhysRevLett.127.101302}{\emph{Phys. Rev.
  Lett.} {\bf 127} (2021) 101302}, [\href{http://arxiv.org/abs/2012.06551}{{\tt
  2012.06551}}].

\bibitem{Kitajima:2021fpq}
N.~Kitajima, Y.~Tada, S.~Yokoyama and C.-M. Yoo, \emph{{Primordial black holes
  in peak theory with a non-Gaussian tail}},
  \href{http://dx.doi.org/10.1088/1475-7516/2021/10/053}{\emph{JCAP} {\bf 10}
  (2021) 053}, [\href{http://arxiv.org/abs/2109.00791}{{\tt 2109.00791}}].

\bibitem{Ezquiaga:2022qpw}
J.~M. Ezquiaga, J.~Garc\'\i{}a-Bellido and V.~Vennin, \emph{{Massive Galaxy
  Clusters Like El Gordo Hint at Primordial Quantum Diffusion}},
  \href{http://dx.doi.org/10.1103/PhysRevLett.130.121003}{\emph{Phys. Rev.
  Lett.} {\bf 130} (2023) 121003}, [\href{http://arxiv.org/abs/2207.06317}{{\tt
  2207.06317}}].

\bibitem{Cai:2022erk}
Y.-F. Cai, X.-H. Ma, M.~Sasaki, D.-G. Wang and Z.~Zhou, \emph{{Highly
  non-Gaussian tails and primordial black holes from single-field inflation}},
  \href{http://dx.doi.org/10.1088/1475-7516/2022/12/034}{\emph{JCAP} {\bf 12}
  (2022) 034}, [\href{http://arxiv.org/abs/2207.11910}{{\tt 2207.11910}}].

\bibitem{Hooshangi:2023kss}
S.~Hooshangi, M.~H. Namjoo and M.~Noorbala, \emph{{Tail diversity from
  inflation}},
  \href{http://dx.doi.org/10.1088/1475-7516/2023/09/023}{\emph{JCAP} {\bf 09}
  (2023) 023}, [\href{http://arxiv.org/abs/2305.19257}{{\tt 2305.19257}}].

\bibitem{Kawaguchi:2023mgk}
R.~Kawaguchi, T.~Fujita and M.~Sasaki, \emph{{Highly asymmetric probability
  distribution from a finite-width upward step during inflation}},
  \href{http://dx.doi.org/10.1088/1475-7516/2023/11/021}{\emph{JCAP} {\bf 11}
  (2023) 021}, [\href{http://arxiv.org/abs/2305.18140}{{\tt 2305.18140}}].

\bibitem{Pi:2022ysn}
S.~Pi and M.~Sasaki, \emph{{Logarithmic Duality of the Curvature
  Perturbation}},
  \href{http://dx.doi.org/10.1103/PhysRevLett.131.011002}{\emph{Phys. Rev.
  Lett.} {\bf 131} (2023) 011002}, [\href{http://arxiv.org/abs/2211.13932}{{\tt
  2211.13932}}].

\bibitem{Enqvist:2008kt}
K.~Enqvist, S.~Nurmi, D.~Podolsky and G.~I. Rigopoulos, \emph{{On the
  divergences of inflationary superhorizon perturbations}},
  \href{http://dx.doi.org/10.1088/1475-7516/2008/04/025}{\emph{JCAP} {\bf 04}
  (2008) 025}, [\href{http://arxiv.org/abs/0802.0395}{{\tt 0802.0395}}].

\bibitem{Fujita:2013cna}
T.~Fujita, M.~Kawasaki, Y.~Tada and T.~Takesako, \emph{{A new algorithm for
  calculating the curvature perturbations in stochastic inflation}},
  \href{http://dx.doi.org/10.1088/1475-7516/2013/12/036}{\emph{JCAP} {\bf 12}
  (2013) 036}, [\href{http://arxiv.org/abs/1308.4754}{{\tt 1308.4754}}].

\bibitem{Vennin:2015hra}
V.~Vennin and A.~A. Starobinsky, \emph{{Correlation Functions in Stochastic
  Inflation}},
  \href{http://dx.doi.org/10.1140/epjc/s10052-015-3643-y}{\emph{Eur. Phys. J.
  C} {\bf 75} (2015) 413}, [\href{http://arxiv.org/abs/1506.04732}{{\tt
  1506.04732}}].

\bibitem{Ando:2020fjm}
K.~Ando and V.~Vennin, \emph{{Power spectrum in stochastic inflation}},
  \href{http://dx.doi.org/10.1088/1475-7516/2021/04/057}{\emph{JCAP} {\bf 04}
  (2021) 057}, [\href{http://arxiv.org/abs/2012.02031}{{\tt 2012.02031}}].

\bibitem{Tada:2021zzj}
Y.~Tada and V.~Vennin, \emph{{Statistics of coarse-grained cosmological fields
  in stochastic inflation}},
  \href{http://dx.doi.org/10.1088/1475-7516/2022/02/021}{\emph{JCAP} {\bf 02}
  (2022) 021}, [\href{http://arxiv.org/abs/2111.15280}{{\tt 2111.15280}}].

\bibitem{Starobinsky:1982ee}
A.~A. Starobinsky, \emph{{Dynamics of Phase Transition in the New Inflationary
  Universe Scenario and Generation of Perturbations}},
  \href{http://dx.doi.org/10.1016/0370-2693(82)90541-X}{\emph{Phys. Lett. B}
  {\bf 117} (1982) 175--178}.

\bibitem{Starobinsky:1985ibc}
A.~A. Starobinsky, \emph{{Multicomponent de Sitter (Inflationary) Stages and
  the Generation of Perturbations}}, {\emph{JETP Lett.} {\bf 42} (1985)
  152--155}.

\bibitem{Sasaki:1995aw}
M.~Sasaki and E.~D. Stewart, \emph{{A General analytic formula for the spectral
  index of the density perturbations produced during inflation}},
  \href{http://dx.doi.org/10.1143/PTP.95.71}{\emph{Prog. Theor. Phys.} {\bf 95}
  (1996) 71--78}, [\href{http://arxiv.org/abs/astro-ph/9507001}{{\tt
  astro-ph/9507001}}].

\bibitem{Sasaki:1998ug}
M.~Sasaki and T.~Tanaka, \emph{{Superhorizon scale dynamics of multiscalar
  inflation}}, \href{http://dx.doi.org/10.1143/PTP.99.763}{\emph{Prog. Theor.
  Phys.} {\bf 99} (1998) 763--782},
  [\href{http://arxiv.org/abs/gr-qc/9801017}{{\tt gr-qc/9801017}}].

\bibitem{Lyth:2004gb}
D.~H. Lyth, K.~A. Malik and M.~Sasaki, \emph{{A General proof of the
  conservation of the curvature perturbation}},
  \href{http://dx.doi.org/10.1088/1475-7516/2005/05/004}{\emph{JCAP} {\bf 05}
  (2005) 004}, [\href{http://arxiv.org/abs/astro-ph/0411220}{{\tt
  astro-ph/0411220}}].

\bibitem{Lyth:2005fi}
D.~H. Lyth and Y.~Rodriguez, \emph{{The Inflationary prediction for primordial
  non-Gaussianity}},
  \href{http://dx.doi.org/10.1103/PhysRevLett.95.121302}{\emph{Phys. Rev.
  Lett.} {\bf 95} (2005) 121302},
  [\href{http://arxiv.org/abs/astro-ph/0504045}{{\tt astro-ph/0504045}}].

\bibitem{Salopek:1990jq}
D.~S. Salopek and J.~R. Bond, \emph{{Nonlinear evolution of long wavelength
  metric fluctuations in inflationary models}},
  \href{http://dx.doi.org/10.1103/PhysRevD.42.3936}{\emph{Phys. Rev. D} {\bf
  42} (1990) 3936--3962}.

\bibitem{Wands:2000dp}
D.~Wands, K.~A. Malik, D.~H. Lyth and A.~R. Liddle, \emph{{A New approach to
  the evolution of cosmological perturbations on large scales}},
  \href{http://dx.doi.org/10.1103/PhysRevD.62.043527}{\emph{Phys. Rev. D} {\bf
  62} (2000) 043527}, [\href{http://arxiv.org/abs/astro-ph/0003278}{{\tt
  astro-ph/0003278}}].

\bibitem{Lyth:2003im}
D.~H. Lyth and D.~Wands, \emph{{Conserved cosmological perturbations}},
  \href{http://dx.doi.org/10.1103/PhysRevD.68.103515}{\emph{Phys. Rev. D} {\bf
  68} (2003) 103515}, [\href{http://arxiv.org/abs/astro-ph/0306498}{{\tt
  astro-ph/0306498}}].

\bibitem{Rigopoulos:2003ak}
G.~I. Rigopoulos and E.~P.~S. Shellard, \emph{{The separate universe approach
  and the evolution of nonlinear superhorizon cosmological perturbations}},
  \href{http://dx.doi.org/10.1103/PhysRevD.68.123518}{\emph{Phys. Rev. D} {\bf
  68} (2003) 123518}, [\href{http://arxiv.org/abs/astro-ph/0306620}{{\tt
  astro-ph/0306620}}].

\bibitem{Artigas:2021zdk}
D.~Artigas, J.~Grain and V.~Vennin, \emph{{Hamiltonian formalism for
  cosmological perturbations: the~separate-universe approach}},
  \href{http://dx.doi.org/10.1088/1475-7516/2022/02/001}{\emph{JCAP} {\bf 02}
  (2022) 001}, [\href{http://arxiv.org/abs/2110.11720}{{\tt 2110.11720}}].

\bibitem{Jackson:2023obv}
J.~H.~P. Jackson, H.~Assadullahi, A.~D. Gow, K.~Koyama, V.~Vennin and D.~Wands,
  \emph{{The separate-universe approach and sudden transitions during
  inflation}},  \href{http://arxiv.org/abs/2311.03281}{{\tt 2311.03281}}.

\bibitem{Starobinsky:1986fx}
A.~A. Starobinsky, \emph{{STOCHASTIC DE SITTER (INFLATIONARY) STAGE IN THE
  EARLY UNIVERSE}},
  \href{http://dx.doi.org/10.1007/3-540-16452-9_6}{\emph{Lect. Notes Phys.}
  {\bf 246} (1986) 107--126}.

\bibitem{Finelli:2010sh}
F.~Finelli, G.~Marozzi, A.~A. Starobinsky, G.~P. Vacca and G.~Venturi,
  \emph{{Stochastic growth of quantum fluctuations during slow-roll
  inflation}}, \href{http://dx.doi.org/10.1103/PhysRevD.82.064020}{\emph{Phys.
  Rev. D} {\bf 82} (2010) 064020}, [\href{http://arxiv.org/abs/1003.1327}{{\tt
  1003.1327}}].

\bibitem{risken1989fpe}
H.~Risken and H.~Haken, \emph{{The Fokker-Planck Equation: Methods of Solution
  and Applications Second Edition}}.
\newblock Springer, 1989.

\bibitem{Assadullahi:2016gkk}
H.~Assadullahi, H.~Firouzjahi, M.~Noorbala, V.~Vennin and D.~Wands,
  \emph{{Multiple Fields in Stochastic Inflation}},
  \href{http://dx.doi.org/10.1088/1475-7516/2016/06/043}{\emph{JCAP} {\bf 06}
  (2016) 043}, [\href{http://arxiv.org/abs/1604.04502}{{\tt 1604.04502}}].

\bibitem{Vennin:2016wnk}
V.~Vennin, H.~Assadullahi, H.~Firouzjahi, M.~Noorbala and D.~Wands,
  \emph{{Critical Number of Fields in Stochastic Inflation}},
  \href{http://dx.doi.org/10.1103/PhysRevLett.118.031301}{\emph{Phys. Rev.
  Lett.} {\bf 118} (2017) 031301}, [\href{http://arxiv.org/abs/1604.06017}{{\tt
  1604.06017}}].

\bibitem{Panagopoulos:2019ail}
G.~Panagopoulos and E.~Silverstein, \emph{{Primordial Black Holes from
  non-Gaussian tails}},  \href{http://arxiv.org/abs/1906.02827}{{\tt
  1906.02827}}.

\bibitem{Pattison:2021oen}
C.~Pattison, V.~Vennin, D.~Wands and H.~Assadullahi, \emph{{Ultra-slow-roll
  inflation with quantum diffusion}},
  \href{http://dx.doi.org/10.1088/1475-7516/2021/04/080}{\emph{JCAP} {\bf 04}
  (2021) 080}, [\href{http://arxiv.org/abs/2101.05741}{{\tt 2101.05741}}].

\bibitem{Vennin:2020kng}
V.~Vennin, \emph{{Stochastic inflation and primordial black holes}}.
\newblock PhD thesis, U. Paris-Saclay, 6, 2020.
\newblock \href{http://arxiv.org/abs/2009.08715}{{\tt 2009.08715}}.

\bibitem{Achucarro:2021pdh}
A.~Achucarro, S.~Cespedes, A.-C. Davis and G.~A. Palma, \emph{{The hand-made
  tail: non-perturbative tails from multifield inflation}},
  \href{http://dx.doi.org/10.1007/JHEP05(2022)052}{\emph{JHEP} {\bf 05} (2022)
  052}, [\href{http://arxiv.org/abs/2112.14712}{{\tt 2112.14712}}].

\bibitem{Animali:2022otk}
C.~Animali and V.~Vennin, \emph{{Primordial black holes from stochastic
  tunnelling}},
  \href{http://dx.doi.org/10.1088/1475-7516/2023/02/043}{\emph{JCAP} {\bf 02}
  (2023) 043}, [\href{http://arxiv.org/abs/2210.03812}{{\tt 2210.03812}}].

\bibitem{Jackson:2022unc}
J.~H.~P. Jackson, H.~Assadullahi, K.~Koyama, V.~Vennin and D.~Wands,
  \emph{{Numerical simulations of stochastic inflation using importance
  sampling}},
  \href{http://dx.doi.org/10.1088/1475-7516/2022/10/067}{\emph{JCAP} {\bf 10}
  (2022) 067}, [\href{http://arxiv.org/abs/2206.11234}{{\tt 2206.11234}}].

\bibitem{Briaud:2023eae}
V.~Briaud and V.~Vennin, \emph{{Uphill inflation}},
  \href{http://dx.doi.org/10.1088/1475-7516/2023/06/029}{\emph{JCAP} {\bf 06}
  (2023) 029}, [\href{http://arxiv.org/abs/2301.09336}{{\tt 2301.09336}}].

\bibitem{Tada:2016pmk}
Y.~Tada and V.~Vennin, \emph{{Squeezed bispectrum in the $\delta N$ formalism:
  local observer effect in field space}},
  \href{http://dx.doi.org/10.1088/1475-7516/2017/02/021}{\emph{JCAP} {\bf 02}
  (2017) 021}, [\href{http://arxiv.org/abs/1609.08876}{{\tt 1609.08876}}].

\bibitem{Noorbala:2018zlv}
M.~Noorbala, V.~Vennin, H.~Assadullahi, H.~Firouzjahi and D.~Wands,
  \emph{{Tunneling in Stochastic Inflation}},
  \href{http://dx.doi.org/10.1088/1475-7516/2018/09/032}{\emph{JCAP} {\bf 09}
  (2018) 032}, [\href{http://arxiv.org/abs/1806.09634}{{\tt 1806.09634}}].

\bibitem{Gratton:2005bi}
S.~Gratton and N.~Turok, \emph{{Langevin analysis of eternal inflation}},
  \href{http://dx.doi.org/10.1103/PhysRevD.72.043507}{\emph{Phys. Rev. D} {\bf
  72} (2005) 043507}, [\href{http://arxiv.org/abs/hep-th/0503063}{{\tt
  hep-th/0503063}}].

\bibitem{Winitzki:2008yb}
S.~Winitzki, \emph{{A Volume-weighted measure for eternal inflation}},
  \href{http://dx.doi.org/10.1103/PhysRevD.78.043501}{\emph{Phys. Rev. D} {\bf
  78} (2008) 043501}, [\href{http://arxiv.org/abs/0803.1300}{{\tt 0803.1300}}].

\bibitem{Steinhardt:1982kg}
P.~J. Steinhardt, \emph{{NATURAL INFLATION}},  in \emph{{Nuffield Workshop on
  the Very Early Universe}}, 7, 1982.

\bibitem{Vilenkin:1983xq}
A.~Vilenkin, \emph{{The Birth of Inflationary Universes}},
  \href{http://dx.doi.org/10.1103/PhysRevD.27.2848}{\emph{Phys. Rev. D} {\bf
  27} (1983) 2848}.

\bibitem{Guth:1985ya}
A.~H. Guth and S.-Y. Pi, \emph{{The Quantum Mechanics of the Scalar Field in
  the New Inflationary Universe}},
  \href{http://dx.doi.org/10.1103/PhysRevD.32.1899}{\emph{Phys. Rev. D} {\bf
  32} (1985) 1899--1920}.

\bibitem{Linde:1986fc}
A.~D. Linde, \emph{{ETERNAL CHAOTIC INFLATION}},
  \href{http://dx.doi.org/10.1142/S0217732386000129}{\emph{Mod. Phys. Lett. A}
  {\bf 1} (1986) 81}.

\bibitem{Garcia-Bellido:1994gng}
J.~Garcia-Bellido and A.~D. Linde, \emph{{Stationarity of inflation and
  predictions of quantum cosmology}},
  \href{http://dx.doi.org/10.1103/PhysRevD.51.429}{\emph{Phys. Rev. D} {\bf 51}
  (1995) 429--443}, [\href{http://arxiv.org/abs/hep-th/9408023}{{\tt
  hep-th/9408023}}].

\bibitem{Bousso:2006ev}
R.~Bousso, \emph{{Holographic probabilities in eternal inflation}},
  \href{http://dx.doi.org/10.1103/PhysRevLett.97.191302}{\emph{Phys. Rev.
  Lett.} {\bf 97} (2006) 191302},
  [\href{http://arxiv.org/abs/hep-th/0605263}{{\tt hep-th/0605263}}].

\bibitem{Linde:2008xf}
A.~D. Linde, V.~Vanchurin and S.~Winitzki, \emph{{Stationary Measure in the
  Multiverse}},
  \href{http://dx.doi.org/10.1088/1475-7516/2009/01/031}{\emph{JCAP} {\bf 01}
  (2009) 031}, [\href{http://arxiv.org/abs/0812.0005}{{\tt 0812.0005}}].

\bibitem{DeSimone:2008if}
A.~De~Simone, A.~H. Guth, A.~D. Linde, M.~Noorbala, M.~P. Salem and
  A.~Vilenkin, \emph{{Boltzmann brains and the scale-factor cutoff measure of
  the multiverse}},
  \href{http://dx.doi.org/10.1103/PhysRevD.82.063520}{\emph{Phys. Rev. D} {\bf
  82} (2010) 063520}, [\href{http://arxiv.org/abs/0808.3778}{{\tt 0808.3778}}].

\bibitem{Linde:2010xz}
A.~Linde and M.~Noorbala, \emph{{Measure Problem for Eternal and Non-Eternal
  Inflation}},
  \href{http://dx.doi.org/10.1088/1475-7516/2010/09/008}{\emph{JCAP} {\bf 09}
  (2010) 008}, [\href{http://arxiv.org/abs/1006.2170}{{\tt 1006.2170}}].

\bibitem{Guth:2011ie}
A.~H. Guth and V.~Vanchurin, \emph{{Eternal Inflation, Global Time Cutoff
  Measures, and a Probability Paradox}},
  \href{http://arxiv.org/abs/1108.0665}{{\tt 1108.0665}}.

\bibitem{Freivogel:2011eg}
B.~Freivogel, \emph{{Making predictions in the multiverse}},
  \href{http://dx.doi.org/10.1088/0264-9381/28/20/204007}{\emph{Class. Quant.
  Grav.} {\bf 28} (2011) 204007}, [\href{http://arxiv.org/abs/1105.0244}{{\tt
  1105.0244}}].

\bibitem{Garriga:2012bc}
J.~Garriga and A.~Vilenkin, \emph{{Watchers of the multiverse}},
  \href{http://dx.doi.org/10.1088/1475-7516/2013/05/037}{\emph{JCAP} {\bf 05}
  (2013) 037}, [\href{http://arxiv.org/abs/1210.7540}{{\tt 1210.7540}}].

\bibitem{10.1215/S0012-7094-48-01568-3}
W.~Hoeffding and H.~Robbins, \emph{{The central limit theorem for dependent
  random variables}},
  \href{http://dx.doi.org/10.1215/S0012-7094-48-01568-3}{\emph{Duke
  Mathematical Journal} {\bf 15} (1948) 773 -- 780}.

\bibitem{Tokeshi:2023swe}
K.~Tokeshi and V.~Vennin, \emph{{Why does inflation look single field to us?}},
   \href{http://arxiv.org/abs/2310.16649}{{\tt 2310.16649}}.

\bibitem{Grain:2017dqa}
J.~Grain and V.~Vennin, \emph{{Stochastic inflation in phase space: Is slow
  roll a stochastic attractor?}},
  \href{http://dx.doi.org/10.1088/1475-7516/2017/05/045}{\emph{JCAP} {\bf 05}
  (2017) 045}, [\href{http://arxiv.org/abs/1703.00447}{{\tt 1703.00447}}].

\bibitem{Efron:1979bxm}
B.~Efron, \emph{{Bootstrap Methods: Another Look at the Jackknife}},
  \href{http://dx.doi.org/10.1214/aos/1176344552}{\emph{Annals Statist.} {\bf
  7} (1979) 1--26}.

\bibitem{NIST:DLMF}
``{\it NIST Digital Library of Mathematical Functions}.''
  \url{https://dlmf.nist.gov/}, Release 1.1.12 of 2023-12-15.

\bibitem{Raatikainen:2023bzk}
S.~Raatikainen, S.~Rasanen and E.~Tomberg, \emph{{Primordial black hole
  compaction function from stochastic fluctuations in ultra-slow-roll
  inflation}},  \href{http://arxiv.org/abs/2312.12911}{{\tt 2312.12911}}.

\bibitem{Gow:2022jfb}
A.~D. Gow, H.~Assadullahi, J.~H.~P. Jackson, K.~Koyama, V.~Vennin and D.~Wands,
  \emph{{Non-perturbative non-Gaussianity and primordial black holes}},
  \href{http://dx.doi.org/10.1209/0295-5075/acd417}{\emph{EPL} {\bf 142} (2023)
  49001}, [\href{http://arxiv.org/abs/2211.08348}{{\tt 2211.08348}}].

\bibitem{Ferrante:2022mui}
G.~Ferrante, G.~Franciolini, A.~Iovino, Junior. and A.~Urbano,
  \emph{{Primordial non-Gaussianity up to all orders: Theoretical aspects and
  implications for primordial black hole models}},
  \href{http://dx.doi.org/10.1103/PhysRevD.107.043520}{\emph{Phys. Rev. D} {\bf
  107} (2023) 043520}, [\href{http://arxiv.org/abs/2211.01728}{{\tt
  2211.01728}}].

\end{thebibliography}\endgroup

\end{document}